\documentclass[preprint,12pt,numbers,sort&compress]{elsarticle}

\usepackage{amsmath}
\usepackage{amssymb}
\usepackage{mathrsfs}
\usepackage{stackrel}

\usepackage{graphicx}
\usepackage{float}
\usepackage{geometry}
\usepackage{subcaption}
\usepackage{microtype}
\usepackage{xcolor}
\usepackage{units}
\usepackage{booktabs}
\usepackage{makecell}

\usepackage{rotating}
\usepackage[colorlinks=true, linkcolor=blue, citecolor=blue, urlcolor=blue]{hyperref}


\newcommand{\modelcaption}[1]{%
  \par\vspace{1pt}%
  {\centering\scriptsize #1\par}%
  \vspace{3pt}%
}

\makeatletter
\def\ps@pprintTitle{%
  \let\@oddhead\@empty
  \let\@evenhead\@empty
  \def\@oddfoot{\hfill\footnotesize\itshape\today}
  \let\@evenfoot\@oddfoot
}
\makeatother

\usepackage{array}
\usepackage{multirow}

\usepackage{babel}
\usepackage{lineno}
\usepackage{setspace}
\usepackage[symbol]{footmisc}

\usepackage[ruled,vlined]{algorithm2e}

\usepackage{etoolbox}
\makeatletter
\patchcmd{\pprintMaketitle}{\par\vskip36pt}{\par\vskip6pt}{}{}
\makeatother

\begin{document}
\begin{frontmatter}

\title{A Comparative Study of Deep Learning Models for Geological Carbon Sequestration}

\author[inst1]{Giovanni Zingaro}

\author[inst1]{Robert Gracie\corref{cor1}}
\ead{rgracie@uwaterloo.ca}

\cortext[cor1]{Corresponding author.}

\author[inst2,inst3]{Yuri Leonenko}

\affiliation[inst1]{organization={Department of Civil and Environmental Engineering, University of Waterloo},%Department and Organization
            city={Waterloo},
            state={Ontario},
            country={Canada}}

\affiliation[inst2]{organization={Department of Earth and Environmental Sciences, University of Waterloo},%Department and Organization
            city={Waterloo},
            state={Ontario},
            country={Canada}}

\affiliation[inst3]{organization={Geography and Environmental Management, University of Waterloo},%Department and Organization
            city={Waterloo},
            state={Ontario},
            country={Canada}}

%% --- ABSTRACT ---
\begin{abstract}
Numerical reservoir simulations are extremely computationally expensive, as they require the repeated solution of large nonlinear algebraic systems derived from the discretized governing equations. With growing demand for real-time optimization, uncertainty quantification, and history matching in digital twin applications, reducing computatio\-nal cost has become essential. Deep learning (DL)--based surrogate models have emerged as an effective approach for accelerating subsurface flow simulations. Here, we seek to determine which DL architectures are best suited for high-dimensional, transient subsurface flow problems. In this study, we examine the advantages and relative costs associated with training such models, including memory requirements, training speed, accuracy, robustness, and generalization. We conduct a comparative study of several DL architectures commonly used as surrogate models for subsurface flow problems, including U-Net, V-Net, Temporal Convolutional Networks, Fourier Neural Operators (FNO), and a U-Net--enhanced FNO (U-FNO). As a benchmark, we compare the performance of the studied models for geological carbon sequestration to predict transient pressure build-up and CO$_2$ saturation fields. We study the problem of CO$_2$ injection into a single wellbore in a two-dimensional domain, which is parameterized by anisotropic, heterogeneous permeability and porosity fields, injection configurations, and reservoir properties. Results demonstrate that surrogate model performance is strongly dependent on the underlying PDE type (i.e., hyperbolic vs.\ elliptic). The U-FNO achieves the highest accuracy for predicting CO$_2$ saturation fields, while the FNO provides the best performance for pressure build-up prediction.
\end{abstract}


\end{frontmatter}

\section{Introduction}
\label{sec:intro}
Subsurface flow simulations are important in many geoscience applications, this includes, geological environmental remediation \cite{bear2010modeling}, hydrology \cite{xu2024forward}, carbon sequestration \cite{ladubec2025vertically}, geothermal \cite{partl2026computational}, oil and gas extraction (e.g., Steam-Assisted Gravity Drainage \cite{hatefi2025parameterized, akin2006mathematical}, and Hydraulic Fracturing \cite{parchei2020dynamic, parchei2019undrained}). In all of these applications, the accurate prediction of fluid migration, pressure evolution, and geomechanical response is critical for ensuring safe and effective subsurface management. However, the biggest challenge associated with subsurface simulations is that their governing physics are often described by a set of strongly coupled, nonlinear Partial Differential Equations (PDEs), which require the solution of a system with many Degrees of Freedom (DoFs). This expense becomes prohibitive in workflows that require many forward evaluations, such as uncertainty quantification, history matching, and real-time operational control. In this work, we focus on benchmarking surrogate models for Geological Carbon Sequestration (GCS) due to the complexity associated with multi-phase subsurface advection-dominated flow.

Surrogate models have been widely used to reduce the computational cost associated with high-fidelity numerical simulations. The main idea is to construct a mathematical proxy model that is orders of magnitude more computationally efficient than the high-fidelity model using a nonlinear function approximator \cite{brunton2022data}. Modern surrogate modeling has leveraged Deep Neural Networks (DNNs) as nonlinear function approximators due to their ability to learn accurate and efficient approximations of high-dimensional systems. These Deep Learning (DL) architectures have included Artificial Neural Networks (ANNs) \cite{lu2021learning}, Convolutional Neural Networks (CNNs) \cite{vspetlik2026convolutional, taccari2022attention}, Long Short-Term Memory Networks (LSTMs) \cite{hajisharifi2024lstm, conti2023multi}, and Graph Neural Networks (GNNs) \cite{ju2024learning}, depending on the type, structure, and complexity of the data (e.g., static vs.\ dynamic inputs, spatial vs.\ temporal dependencies, structured vs.\ unstructured meshes). Another emerging class of DL surrogates, specifically designed for PDEs, is operator learning methods, which aim to learn mappings between infinite-dimensional function spaces, such as mapping input fields (e.g., initial or boundary conditions) to solution fields of PDEs. Notable operator-based architectures include DeepONet, Graph Neural Operators, and Fourier Neural Operators.

Methodologies for DL-based surrogate modeling are generally categorized into two different approaches: (1) finite-dimensional surrogates, which learn mappings between vector spaces (e.g., Euclidean-to-Euclidean) \cite{san2019artificial, thuerey2020deep, maulik2021reduced, hajisharifi2024lstm, li2019deep, jahangir2023quantile, geneva2022transformers, nguyen2024scaling, hadizadeh2025graph, ju2024learning}, and (2) operator-based surrogates, which learn function-to-function mappings \cite{lu2021learning, li2020neural, li2020fourier}. Finite-dimensional models often use a combination of ANNs/CNNs and RNNs to capture spatial and temporal dependencies (e.g., when inputs are time-dependent), but they typically require large datasets and are prone to overfitting. In contrast, operator-based approaches aim to learn mappings between function spaces and have shown promising generalization, particularly across discretizations. Fourier Neural Operators (FNOs) are among the most successful examples, performing spectral convolutions in Fourier space using the Fast Fourier Transform (FFT). By operating on a truncated set of Fourier modes, they are more computationally efficient compared to standard global operators \cite{li2020fourier}.

A number of studies have explored the use of surrogate modeling as an alternative to traditional reservoir simulations of GCS processes. One of the first studies to investigate the use of DL-based surrogates in GCS modeling was Mo et al.\ \cite{mo2019deep} developed an encoder-decoder CNN to predict both saturation and pressure fields. This work adopted a training strategy combining both a regression loss and a segmentation loss to improve the approximation of the saturation field. Zhong et al.\ \cite{zhong2019predicting} explored the use of a deep convolutional generative adversarial network (cDC-GAN) to predict the spatio-temporal evolution of $\text{CO}_2$ saturation, where high-resolution permeability fields and time were inputs to the model. Yan et al.\ \cite{yan2022robust} and Badawi and Gildin \cite{badawi2025neural} developed FNO-based surrogate models to predict the spatio-temporal evolution of CO$_2$ saturation and pressure build-up fields. Both works considered two-dimensional planar domains parameterized by heterogeneous permeability fields, well locations, and well control (i.e., constant injection rate). Feng et al.\ \cite{feng2024encoder} developed a hybrid CNN-LSTM network to predict the pressure and $\text{CO}_2$ saturation fields for dynamic injection schedules (i.e., time-discontinuous). Other works by Zingaro et al.\ \cite{zingaro2025deep} demonstrated the motivation for developing DL-based surrogates by using them to solve the history-matching problem (i.e., monitoring), successfully recovering full-field reservoir state variables from sparse geophysical measurements.

Among the most influential works in this area are the series of studies by Wen et al.\ \cite{wen2019multiphase, wen2021ccsnet, wen2022u}. Wen et al.\ \cite{wen2019multiphase} developed a two-dimensional U-Net to predict the CO$_2$ saturation field, parameterized by a heterogeneous permeability field and a constant injection rate. This was followed by CCSNet \cite{wen2021ccsnet}, a three-dimensional Temporal Convolutional Neural Network that predicts CO$_2$ saturation and pressure build-up fields over fixed time steps for inputs parameterized by heterogeneous permeability and porosity fields, injection duration, rate, and perforation. Wen et al.\ \cite{wen2022u} further extended this to a single-wellbore domain parameterized by anisotropic heterogeneous permeability and porosity fields, injection rate and perforation, reservoir thickness, initial pressure, temperature, capillary pressure, Van Genuchten scaling factor, and irreducible water saturation.

While DL-based surrogates have demonstrated effectiveness across multiple publica\-tions, several gaps remain. Existing studies do not compare multiple architectures on a common dataset, making it difficult to assess the relative performance of different surrogate models. Moreover, the number of trainable parameters varies greatly across architectures and studies. Architectural choices are also not analyzed in the context of the underlying PDE type (i.e., hyperbolic vs.\ elliptic). Finally, most studies train surrogate models on different full-physics simulators with different parameterizations, making direct comparison of cost-accuracy tradeoffs across studies challenging. Thus, the contributions of this work are as follows:

\begin{enumerate}
    \item We present a comprehensive benchmark comparison of the U-Net, V-Net, Temporal CNN, FNO, and U-FNO architectures on a complex anisotropic heterogeneous multi-phase flow dataset.
    \item The predictive accuracy, offline training time, and peak memory usage during offline training are compared.
    \item Surrogate models are constructed for both CO$_2$ saturation and pressure build-up fields. We discuss their performance and behavior in the context of their governing PDEs (i.e., hyperbolic vs.\ elliptic) and provide insight into why architectural differences lead to performance differences.
\end{enumerate}

We acknowledge that the dataset used is limited to a static field-to-field mapping and does not capture the most complex parameterizations (e.g., wellbore placement, discontinuous injection schedules). However, it is designed to be sufficiently complex to yield meaningful architectural comparisons, while remaining computationally feasible for training multiple DL models under a reasonable high-performance computing (HPC) resource allocation.

\section{Problem Formulation}
\begin{figure}[!htbp]
    \centering
    \includegraphics[width=0.65\linewidth]{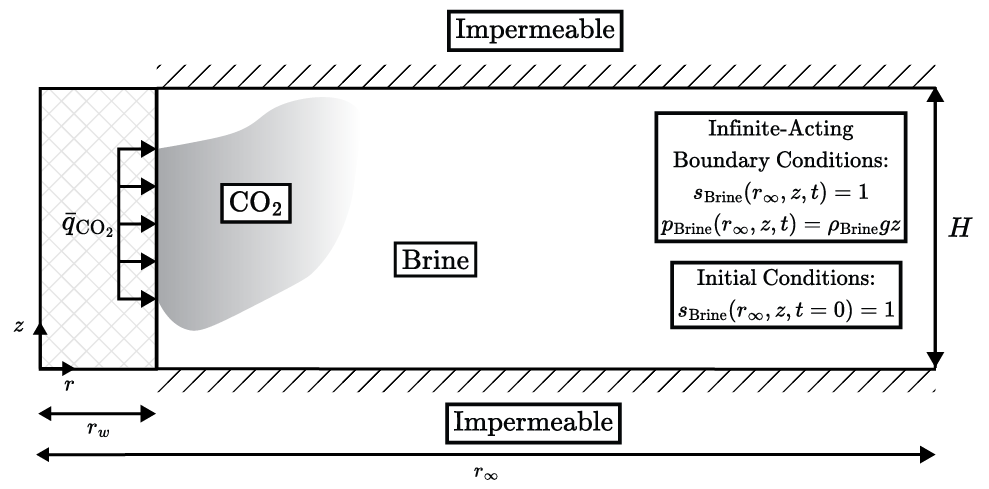}
    \caption{Schematic of the radially symmetric single-wellbore GCS domain, including domain geometry, initial conditions, and boundary conditions.}
    \label{fig:Radial_Wellbore_Schematic}
\end{figure}

In this work, we consider the injection of supercritical CO$_2$ into a radially symmetric vertical wellbore in a two-dimensional heterogeneous anisotropic reservoir. 
Specifically, we aim to predict the spatio-temporal evolution of the CO$_2$ saturation and pressure build-up fields over a 30-year injection period. In this section, we define the forward problem governing the transient reservoir response for geological carbon sequestration operations, which involves predicting CO$_2$ saturation and pressure fields given a set of geological properties and injection perforation configurations, as shown in Figure~\ref{fig:Radial_Wellbore_Schematic}. The forward problem is defined as follows:

\begin{equation}
    [\mathbf{s},\mathbf{p}] = f(\boldsymbol{\theta}, \boldsymbol{\beta})
\end{equation}

\noindent where, $f$ denotes the forward model of the numerical simulator, and $\mathbf{s}\in\mathbb{R}^{n_r \times n_z \times n_t}$, $\mathbf{p} \in\mathbb{R}^{n_r \times n_z \times n_t}$ represent the CO$_2$ saturation and pressure fields. Here, $n_r$, $n_z$ and $n_t$ denote the total number of grid blocks used at discrete points in (r, z) space and in time (t). Additionally, $\boldsymbol{\theta} = [\mathbf{k_r}, \mathbf{k_z}, \boldsymbol{\phi}, \mathbf{b}, \mathbf{b_{perf}}]$ denotes anisotropic permeability, porosity, reservoir thickness and perforation fields, respectively. Scalar variables are denoted by $\boldsymbol{\beta} = [Q, P_{init}, T, S_{wi}, \lambda]$ which includes injection rate, initial pressure, iso-thermal reservoir temperature, irreducible water saturation and van Genuchten scaling factor.

\subsection{Governing Equations}
The governing equations for coupled multi-constituent flow of CO$_2$ and brine in the context of geological CO$_2$ storage consist of statements of mass conservation for the brine and CO$_2$ phases, combined with the multi-phase fluxes of each phase. The mass continuity equations are expressed as follows:

\begin{equation}
    \nabla \cdot \left( \sum_{j} \rho_{j} x_{j}^{\text{CO}_2} \mathbf{v}_{j} \right) + q^{\text{CO}_2} = \frac{\partial}{\partial t}\left(\sum_{j} \phi \rho_{j} s_j x_{j}^{\text{CO}_2}\right)
\end{equation}

\begin{equation}
    \nabla \cdot \left( \sum_{j} \rho_{j} x_{j}^{\text{Brine}} \mathbf{v}_{j} \right) + q^{\text{Brine}} = \frac{\partial}{\partial t}\left(\sum_{j} \phi \rho_{j} s_j x_{j}^{\text{Brine}}\right)
\end{equation}

in which, the subscript $j$ denotes the wetting and non-wetting phases, which are brine and CO$_2$, respectively. The parameter $\phi$ is the porosity, $s_j$ is the saturation of phase $j$, and $x_j^{\mu}$ is the mass fraction of component $\mu$ in phase $j$.

The Darcy velocity of each phase $j$, $\mathbf{v}_j$, is given by:

\begin{equation}
    \mathbf{v}_j = \frac{\mathbf{k} \, k_{rj}(s_j)}{\mu_j} \left( -\nabla p_{j} + \rho_{j} g \nabla z \right)
\end{equation}

\noindent where, $\mathbf{k}$ is the permeability tensor, $k_{rj}$ is the relative permeability, $\mu_j$ is the viscosity, and $p_j$ is the fluid pressure of phase $j$. The variable $g$ denotes gravitational acceleration, and $z$ is the depth within the reservoir. The phase pressures are related through the capillary pressure $p_c$ as follows:

\begin{equation}
	p_c = p_{\text{CO}_2} - p_{\text{Brine}}
\end{equation}

\noindent where, $p_c(s_{\text{CO}_2})$ is an empirical relation. Conservation of the system is enforced by:

\begin{equation}
	s_{\text{CO}_2} + s_{\text{Brine}} = 1
\end{equation}

For more details on the high-fidelity simulations used to generate the synthetic dataset, refer to the work of Wen et al. \cite{wen2022u}.

\subsection{Numerical Simulation Configuration}
In this dataset, CO$_2$ is injected into a radially symmetric vertical wellbore with a radius of 100 mm. The modeled reservoir is radially symmetric, with no-flow boundaries at the top and bottom. It is treated as an infinite-acting reservoir with a radius of 100 km. The injection well can be perforated over the entire thickness or a partially selected depth interval of the reservoir. Constant CO$_2$ injection rates are considered over a 30 year period, ranging from 0.2 to 2 MT/year. Additionally, the height of the reservoir is parameterized with values ranging from 12.5 to 200 m. The dataset used in this study was previously developed by Wen et al.\ \cite{wen2022u}. The dataset was generated using the ECLIPSE (e300) numerical software, which is based on the Finite Difference Method with upstream weighting for the spatial discretization and adaptive implicit method for the temporal discretization.

The numerical simulation models the spatio-temporal evolution of both the CO$_2$ saturation and pressure fields during injection operations. The initial and boundary conditions for the infinite-acting reservoir multi-phase flow boundary value problem were previously shown in Figure \ref{fig:Radial_Wellbore_Schematic}. Here, the numerical model saves the CO$_2$ saturation and pressure fields at 24 time-steps: 1, 2, 4, 7, 11, 17, 25, 37, 53, 77, 111, 158, 226, and 323 days, and 1.3, 1.8, 2.6, 3.6, 5.2, 7.3, 10.4, 14.8, 21.1, and 30.0 years, which are later used to train the presented surrogate models.

\subsection{Parameterization of Dataset}
This dataset considers the injection of CO$_2$ into a radially symmetric infinite-acting reservoir, in which both field and scalar variables were 
parameterized and randomly sampled for each numerical simulation. The field variables include horizontal permeability ($k_x$), vertical permeability ($k_z$), porosity ($\phi$), and injection perforation depth. The reservoir thickness ($b$) is uniformly sampled per realization and used to define an active mask over the field variables, accounting for varying simulation domains. The scalar parameters sampled for each case are initial reservoir pressure ($P_\text{init}$), reservoir temperature ($T$), injection rate ($Q$), Van Genuchten capillary pressure scaling factor ($\lambda$), and irreducible water saturation ($S_{wi}$). The parameter space sampling distributions are shown in Table~\ref{table:ParameterSpace}.

\begin{table}[!htbp]
\scriptsize
\centering
\begin{tabular}{llccc}
\toprule
\textbf{Variable} & \textbf{Sampling Parameter} & \textbf{Notation} & \textbf{Distribution} & \textbf{Unit} \\
\midrule
\multirow{7}{*}{Field}
    & Radial Permeability Field    & $k_r$             & Heterogeneous                         & --    \\
    & No. of Anisotropic Materials & $n_\text{aniso}$  & $X \sim U'\{1,6\}$                    & --    \\
    & Material Anisotropy Ratio    & $k_r/k_z$         & $X \sim U'[1,150]$                    & --    \\
    & Porosity (Perturbation)      & $\phi$            & $\epsilon \sim \mathcal{N}(0, 0.005)$ & --    \\
    & Reservoir Thickness          & $b$               & $X \sim U[12.5, 200]$                 & m     \\
    & Perforation Thickness        & $b_\text{perf}$   & $X \sim U'[12, b]$                    & m     \\
    & Perforation Location         & --                & Randomly Placed                       & --    \\
\midrule
\multirow{5}{*}{Scalar}
    & Injection Rate               & $Q$               & $X \sim U'[0.2, 2]$                   & MT/yr \\
    & Initial Pressure             & $P_\text{init}$   & $X \sim U'[100, 300]$                 & bar   \\
    & Reservoir Temperature        & $T$               & $X \sim U'[35, 170]$                  & °C    \\
    & Irreducible Water Saturation & $S_\text{wc}$     & $X \sim U'[0.1, 0.3]$                 & --    \\
    & Van Genuchten Scaling Factor & $\lambda$         & $X \sim U'[0.3, 0.7]$                 & --    \\
\bottomrule
\end{tabular}
\caption{Summary of parameter space for input variables, sampling range and distribution. We refer to Wen et al.\ \cite{wen2022u} for more detail on the construction of the dataset.}
\label{table:ParameterSpace}
\end{table}

The Geostatistical Modeling Software (SGeMS) \cite{remy2009applied} was used to generate stochastic spatially correlated radial permeability ($k_r$) maps based on correlation lengths, medium appearances, and permeability mean and standard deviation. The vertical permeability field ($k_z$) was computed by scaling $k_r$ with an independently generated anisotropy map, which is determined from a randomly sampled anisotropy ratio. Porosity values were determined from empirical relationships between porosity and permeability \cite{pape2000variation} and further perturbed by Gaussian noise to introduce spatial variability. The perforation zones were randomly placed within the thickness of the reservoir, and the top depth was sampled uniformly and constrained to ensure that the perforation thickness ($b_\text{perf}$) remained within the limits of the parameterized domain.

\subsection{Data Configuration}
In this section, we describe the format and data preprocessing techniques used for both the input and output data for all studied models. Each data sample predicts the full spatio-temporal solution over $r$, $z$, and $t$ through a single forward pass. The input and output tensors have dimensions of $(N, N_r, N_z, N_t, N_c)$ and $(N, N_r, N_z, N_t)$, respectively, where $N_c$ denotes the number of input channels. Each field variable is assigned its own channel, while scalar variables are broadcast into a constant-valued matrix matching the spatial dimensions of the field variables, providing global context to the model. Since the dataset contains reservoir configurations of variable thickness, zero-padding is applied in the spatial dimensions to fill cells outside the active reservoir with a neutral value that does not contaminate the loss landscape during training. An active cell mask is also constructed for each sample so that the loss is computed only within the true reservoir domain, consistent with the approach of Wen et al.~\cite{wen2022u}.

In addition to the inputs described in Table~\ref{table:ParameterSpace}, the model also receives spatial coordinate information as input. Specifically, two additional channels encode the radial ($r$) and vertical ($z$) cell coordinates, providing the model with spatial grid information. Temporal information is provided by broadcasting the current time value across the $N_t$ dimension as an additional channel. The full model input is constructed by concatenating the field, scalar, spatial, and temporal channels.

Similar to Wen et al.~\cite{wen2022u}, data preprocessing techniques are applied to improve the data efficiency of the developed networks; however, using a different approach. All input channels are preprocessed using techniques appropriate to the range and distribution of the data, as summarized in Table~\ref{table:Preprocessing}. Permeability fields are log-normalized prior to Z-Score normalization to compress the effect of large values and outliers inherent in their distributions, while scalar variables are normalized using MinMax normalization, as they are uniformly sampled over known bounded ranges. For the output variables, CO$_2$ saturation is normalized using MinMax normalization, as its values are inherently bounded between 0 and 1, while pressure build-up is first log-normalized prior to Z-Score normalization for the same reason.

\begin{table}[!htbp]
\scriptsize
\centering
\begin{tabular}{llll}
\toprule
\textbf{Type} & \textbf{Variable} & \textbf{Notation} & \textbf{Normalization} \\
\midrule
\multirow{5}{*}{Field Input}
    & Radial Permeability   & $k_r$            & $\ln(k_r) \rightarrow$ Z-Score    \\
    & Vertical Permeability & $k_z$            & $\ln(k_z) \rightarrow$ Z-Score    \\
    & Porosity              & $\phi$           & Z-Score                           \\
    & Radial Coordinate     & $r$              & $r/r_{\infty} \rightarrow$ Z-Score \\
    & Vertical Coordinate   & $z$              & Z-Score                           \\
\midrule
\multirow{1}{*}{Temporal Input}
    & Time                  & $t$              & $t/t_{\infty} \rightarrow$ Z-Score \\
\midrule
\multirow{6}{*}{Scalar Input}
    & Initial Pressure      & $P_\text{init}$  & MinMax \\
    & Reservoir Temperature & $T$              & MinMax \\
    & Injection Rate        & $Q$              & MinMax \\
    & Irred. Water Sat.     & $S_{wc}$         & MinMax \\
    & Van Genuchten Factor  & $\lambda$        & MinMax \\
    & Perforation Location  & $b$              & MinMax \\
\midrule
\multirow{2}{*}{Output}
    & CO$_2$ Saturation     & $s$              & MinMax                            \\
    & Pressure Build-up     & $p$              & $\ln(p) \rightarrow$ Z-Score      \\
\bottomrule
\end{tabular}
\caption{Summary of data preprocessing applied to input and output variables.}
\label{table:Preprocessing}
\end{table}

The dataset adopted from Wen et al.~\cite{wen2022u} consists of 5,500 unique input-to-output mappings. In this work, 4,500 samples are used for training, 500 for validation, and 500 samples are used as an unseen test set to assess model performance after training.

\section{Methodology}
\label{sec:Methodolgy}
In this section, we explain the Temporal Convolutional Neural Network (CNN), U-Net, V-Net, Fourier Neural Operator (FNO) and U-Net Enhanced FNO (U-FNO) architectures compared in this study. In addition, we discuss the training approach and evaluation metrics used to assess the performance of the networks studied. We should note that all these models are compared for both pressure build-up and CO$_2$ saturation field surrogate models.

\subsection{Temporal Convolutional Neural Network (CNN)}
Convolutional Neural Networks (CNNs) are neural networks designed to process grid-like topological data. In recent years, these models have demonstrated success in computer vision, particularly in image recognition tasks \cite{deng2009imagenet, girshick2014uc, simonyan2014very}, due to their ability to extract local correlations and hierarchical spatial dependencies. One of the main benefits of such architectures is their ability to learn efficient representations via parameter sharing through the convolution operation, which significantly reduces the number of trainable parameters in the model.

The CNN architecture used in this work is inspired by the Temporal CNN proposed by Wen et al.\ \cite{wen2021ccsnet}. The network consists of an encoder, a residual trunk, and a decoder. The encoder successively compresses the spatial dimensions through a series of $3\times3\times3$ convolutional layers with stride $2\times2\times2$, each followed by batch normalization and a ReLU activation. The residual trunk consists of eight sequential residual blocks applied at the bottleneck resolution, each of which adds the block input directly to the output of two convolutional layers. The main purposes of these bottleneck layers is to learn the relation between the input and output embeddings. The decoder restores the original spatial dimensions through nearest-neighbour upsampling followed by a reflection-padded convolution. A final $3\times3\times3$ convolutional layer maps the output to the number of output channels. Unlike the U-Net, this architecture employs strided convolutions rather than max pooling for downsampling, which learns an optimal compression. The detailed outline of the CNN model architecture used in this study is shown in Table~\ref{tab:cnn_arch}.

\begin{table}[!htbp]
\scriptsize
\centering
\begin{tabular}{lll}
\toprule
\textbf{Unit} & \textbf{Layer} & \textbf{Output Shape} \\
\midrule
Input       &                                                          & $(N, 96, 200, 24, 12)$   \\[2pt]
Encoder 1   & \texttt{Conv3D(c32k3s2)/BN/ReLU}                        & $(N, 48, 100, 12, 32)$   \\
Encoder 2   & \texttt{Conv3D(c64k3s1)/BN/ReLU}                        & $(N, 48, 100, 12, 64)$   \\
Encoder 3   & \texttt{Conv3D(c128k3s2)/BN/ReLU}                       & $(N, 24, 50, 6, 128)$    \\
Encoder 4   & \texttt{Conv3D(c128k3s1)/BN/ReLU}                       & $(N, 24, 50, 6, 128)$    \\
Encoder 5   & \texttt{Conv3D(c256k3s2)/BN/ReLU}                       & $(N, 12, 25, 3, 256)$    \\
Encoder 6   & \texttt{Conv3D(c256k3s1)/BN/ReLU}                       & $(N, 12, 25, 3, 256)$    \\[2pt]
ResConv 1   & \texttt{Conv3D(c256k3s1)/BN/ReLU/Conv3D(c256k3s1)/BN}  & $(N, 12, 25, 3, 256)$    \\
ResConv 2   & \texttt{Conv3D(c256k3s1)/BN/ReLU/Conv3D(c256k3s1)/BN}  & $(N, 12, 25, 3, 256)$    \\
ResConv 3   & \texttt{Conv3D(c256k3s1)/BN/ReLU/Conv3D(c256k3s1)/BN}  & $(N, 12, 25, 3, 256)$    \\
ResConv 4   & \texttt{Conv3D(c256k3s1)/BN/ReLU/Conv3D(c256k3s1)/BN}  & $(N, 12, 25, 3, 256)$    \\
ResConv 5   & \texttt{Conv3D(c256k3s1)/BN/ReLU/Conv3D(c256k3s1)/BN}  & $(N, 12, 25, 3, 256)$    \\
ResConv 6   & \texttt{Conv3D(c256k3s1)/BN/ReLU/Conv3D(c256k3s1)/BN}  & $(N, 12, 25, 3, 256)$    \\
ResConv 7   & \texttt{Conv3D(c256k3s1)/BN/ReLU/Conv3D(c256k3s1)/BN}  & $(N, 12, 25, 3, 256)$    \\
ResConv 8   & \texttt{Conv3D(c256k3s1)/BN/ReLU/Conv3D(c256k3s1)/BN}  & $(N, 12, 25, 3, 256)$    \\[2pt]
Decoder 6   & \texttt{Conv3D(c256k3s1)/BN/ReLU}                       & $(N, 12, 25, 3, 256)$    \\
Decoder 5   & \texttt{Upsample(s=2)/ReflPad/Conv3D(c256k3s1)/BN/ReLU} & $(N, 24, 50, 6, 256)$    \\
Decoder 4   & \texttt{Conv3D(c128k3s1)/BN/ReLU}                       & $(N, 24, 50, 6, 128)$    \\
Decoder 3   & \texttt{Upsample(s=2)/ReflPad/Conv3D(c128k3s1)/BN/ReLU} & $(N, 48, 100, 12, 128)$  \\
Decoder 2   & \texttt{Conv3D(c64k3s1)/BN/ReLU}                        & $(N, 48, 100, 12, 64)$   \\
Decoder 1   & \texttt{Upsample(s=2)/ReflPad/Conv3D(c32k3s1)/BN/ReLU}  & $(N, 96, 200, 24, 32)$   \\[2pt]
Output      & \texttt{Conv3D(c1k3s1)}                                  & $(N, 96, 200, 24, 1)$    \\
\bottomrule
\end{tabular}
\caption{Architecture summary of the Temporal CNN surrogate model.}
\label{tab:cnn_arch}
\end{table}

\subsection{U-Net and V-Net Architectures}
The U-Net architecture, originally introduced by Ronneberger et al.\ \cite{ronneberger2015u} is a type of Convolutional Neural Network that was specifically designed for biomedical image segmentation tasks. The network consists of an encoder-decoder architecture that uses convolutional layers to progressively reduce the spatial dimensions of the input in the encoder while extracting important features. The decoder uses transposed convolutions to restore the original spatial dimensions while retaining the features learned in the bottleneck. The most defining aspect of the U-Net architecture is its use of skip connections, where feature representations from the encoder layers are concatenated into the corresponding decoder layers, enabling the transfer of both low and high-resolution spatial information while preserving spatial context during reconstruction. This results in a model that has demonstrated strong performance and robustness across various segmentation tasks \cite{ronneberger2015u, oktay2018attention, cciccek20163d} and has shown some success in the context of subsurface flow surrogate modelling \cite{wen2019multiphase, tang2021deep}.

Previous studies have used the U-Net architecture for field-based PDE surrogate modeling, as it has demonstrated the ability to extract and learn both low- and high-frequency spatial features from grid-like topological data \cite{wen2019multiphase}. In subsurface flow, coarse global behaviour is governed by boundary conditions and large pressure gradients, while fine local features arise from heterogeneous permeability and porosity fields. This makes the U-Net architecture particularly suitable as the encoder downsamples the input to aggregate global features, while the skip connections restore fine-scale spatial details along the decoder path.

In this work, we study both the U-Net \cite{ronneberger2015u} and V-Net \cite{milletari2016v} architectures, each extended to a three-dimensional setting. Both models use double convolution blocks consisting of two $3\times3\times3$ convolutional layers, batch normalization, and LeakyReLU activation. The models differ as the U-Net applies $2\times2\times2$ max pooling for downsampling, while the V-Net uses $2\times2\times2$-strided convolutions. The bottleneck consists of a double convolution block followed by six residual blocks, each of which adds the block input to the output after two convolutions. The decoder restores spatial dimensions via transposed convolutions, concatenates the corresponding encoder skip connections, and then applies a double convolution block. A final $1\times1\times1$ convolution maps the output to the desired number of output channels. The detailed architecture of the U-Net and V-Net used in this study are summarized in Tables~\ref{tab:unet3d_arch} and~\ref{tab:vnet3d_arch}, respectively.

\begin{table}[!htbp]
\scriptsize
\centering
\begin{tabular}{lll}
\toprule
\textbf{Unit} & \textbf{Layer} & \textbf{Output Shape} \\
\midrule
Input           &                                                              & $(N, 96, 200, 24, 12)$ \\[2pt]
\multirow{2}{*}{Encoder 1}
                & \texttt{Conv3D(c16,k3,s1)/BN/LeakyReLU}                     & $(N, 96, 200, 24, 16)$ \\
                & \texttt{Conv3D(c16,k3,s1)/BN/LeakyReLU}                     & $(N, 96, 200, 24, 16)$ \\
MaxPool 1       & \texttt{MaxPool3D(k=2,\ s=2)}                                & $(N, 48, 100, 12, 16)$ \\
\multirow{2}{*}{Encoder 2}
                & \texttt{Conv3D(c32,k3,s1)/BN/LeakyReLU}                     & $(N, 48, 100, 12, 32)$ \\
                & \texttt{Conv3D(c32,k3,s1)/BN/LeakyReLU}                     & $(N, 48, 100, 12, 32)$ \\
MaxPool 2       & \texttt{MaxPool3D(k=2,\ s=2)}                                & $(N, 24, 50, 6, 32)$   \\
\multirow{2}{*}{Encoder 3}
                & \texttt{Conv3D(c64,k3,s1)/BN/LeakyReLU}                     & $(N, 24, 50, 6, 64)$   \\
                & \texttt{Conv3D(c64,k3,s1)/BN/LeakyReLU}                     & $(N, 24, 50, 6, 64)$   \\
MaxPool 3       & \texttt{MaxPool3D(k=2,\ s=2)}                                & $(N, 12, 25, 3, 64)$   \\
\multirow{2}{*}{Encoder 4}
                & \texttt{Conv3D(c128,k3,s1)/BN/LeakyReLU}                    & $(N, 12, 25, 3, 128)$  \\
                & \texttt{Conv3D(c128,k3,s1)/BN/LeakyReLU}                    & $(N, 12, 25, 3, 128)$  \\
MaxPool 4       & \texttt{MaxPool3D(k=2,\ s=2)}                                & $(N, 6, 12, 1, 128)$   \\[2pt]
\multirow{2}{*}{Bottleneck}
                & \texttt{Conv3D(c256,k3,s1)/BN/LeakyReLU}                    & $(N, 6, 12, 1, 256)$   \\
                & \texttt{Conv3D(c256,k3,s1)/BN/LeakyReLU}                    & $(N, 6, 12, 1, 256)$   \\
ResConv 1       & \texttt{Conv3D/BN/ReLU/Conv3D/BN} $+$ skip                  & $(N, 6, 12, 1, 256)$   \\
ResConv 2       & \texttt{Conv3D/BN/ReLU/Conv3D/BN} $+$ skip                  & $(N, 6, 12, 1, 256)$   \\
ResConv 3       & \texttt{Conv3D/BN/ReLU/Conv3D/BN} $+$ skip                  & $(N, 6, 12, 1, 256)$   \\
ResConv 4       & \texttt{Conv3D/BN/ReLU/Conv3D/BN} $+$ skip                  & $(N, 6, 12, 1, 256)$   \\
ResConv 5       & \texttt{Conv3D/BN/ReLU/Conv3D/BN} $+$ skip                  & $(N, 6, 12, 1, 256)$   \\
ResConv 6       & \texttt{Conv3D/BN/ReLU/Conv3D/BN} $+$ skip                  & $(N, 6, 12, 1, 256)$   \\[2pt]
\multirow{3}{*}{Decoder 3}
                & \texttt{ConvTranspose3D(c128,k2,s2) + Concat(skip4)}         & $(N, 12, 25, 3, 256)$  \\
                & \texttt{Conv3D(c128,k3,s1)/BN/LeakyReLU}                    & $(N, 12, 25, 3, 128)$  \\
                & \texttt{Conv3D(c128,k3,s1)/BN/LeakyReLU}                    & $(N, 12, 25, 3, 128)$  \\
\multirow{3}{*}{Decoder 2}
                & \texttt{ConvTranspose3D(c64,k2,s2) + Concat(skip3)}          & $(N, 24, 50, 6, 128)$  \\
                & \texttt{Conv3D(c64,k3,s1)/BN/LeakyReLU}                     & $(N, 24, 50, 6, 64)$   \\
                & \texttt{Conv3D(c64,k3,s1)/BN/LeakyReLU}                     & $(N, 24, 50, 6, 64)$   \\
\multirow{3}{*}{Decoder 1}
                & \texttt{ConvTranspose3D(c32,k2,s2) + Concat(skip2)}          & $(N, 48, 100, 12, 128)$\\
                & \texttt{Conv3D(c32,k3,s1)/BN/LeakyReLU}                     & $(N, 48, 100, 12, 32)$ \\
                & \texttt{Conv3D(c32,k3,s1)/BN/LeakyReLU}                     & $(N, 48, 100, 12, 32)$ \\
\multirow{3}{*}{Decoder 0}
                & \texttt{ConvTranspose3D(c16,k2,s2) + Concat(skip1)}          & $(N, 96, 200, 24, 32)$ \\
                & \texttt{Conv3D(c16,k3,s1)/BN/LeakyReLU}                     & $(N, 96, 200, 24, 16)$ \\
                & \texttt{Conv3D(c16,k3,s1)/BN/LeakyReLU}                     & $(N, 96, 200, 24, 16)$ \\[2pt]
Output          & \texttt{Conv3D(c1,k1,s1)} + squeeze                         & $(N, 96, 200, 24)$     \\
\bottomrule
\end{tabular}
\caption{Architecture summary of the U-Net surrogate model.}
\label{tab:unet3d_arch}
\end{table}

\begin{table}[!htbp]
\scriptsize
\centering
\begin{tabular}{lll}
\toprule
\textbf{Unit} & \textbf{Layer} & \textbf{Output Shape} \\
\midrule
Input           &                                                              & $(N, 96, 200, 24, 12)$ \\[2pt]
\multirow{2}{*}{Encoder 0}
                & \texttt{Conv3D(c16,k3,s1)/BN/LeakyReLU}                     & $(N, 16, 96, 200, 24)$ \\
                & \texttt{Conv3D(c16,k3,s1)/BN/LeakyReLU}                     & $(N, 16, 96, 200, 24)$ \\[2pt]
\multirow{3}{*}{Encoder 1}
                & \texttt{Conv3D(c32,k2,s2)/BN/LeakyReLU}                     & $(N, 32, 48, 100, 12)$ \\
                & \texttt{Conv3D(c32,k3,s1)/BN/LeakyReLU}                     & $(N, 32, 48, 100, 12)$ \\
                & \texttt{Conv3D(c32,k3,s1)/BN/LeakyReLU}                     & $(N, 32, 48, 100, 12)$ \\[2pt]
\multirow{3}{*}{Encoder 2}
                & \texttt{Conv3D(c64,k2,s2)/BN/LeakyReLU}                     & $(N, 64, 24, 50, 6)$   \\
                & \texttt{Conv3D(c64,k3,s1)/BN/LeakyReLU}                     & $(N, 64, 24, 50, 6)$   \\
                & \texttt{Conv3D(c64,k3,s1)/BN/LeakyReLU}                     & $(N, 64, 24, 50, 6)$   \\[2pt]
\multirow{3}{*}{Encoder 3}
                & \texttt{Conv3D(c128,k2,s2)/BN/LeakyReLU}                    & $(N, 128, 12, 25, 3)$  \\
                & \texttt{Conv3D(c128,k3,s1)/BN/LeakyReLU}                    & $(N, 128, 12, 25, 3)$  \\
                & \texttt{Conv3D(c128,k3,s1)/BN/LeakyReLU}                    & $(N, 128, 12, 25, 3)$  \\[2pt]
\multirow{2}{*}{Bottleneck}
                & \texttt{Conv3D(c256,k3,s1)/BN/LeakyReLU}                    & $(N, 256, 12, 25, 3)$  \\
                & \texttt{Conv3D(c256,k3,s1)/BN/LeakyReLU}                    & $(N, 256, 12, 25, 3)$  \\
ResConv 1       & \texttt{Conv3D/BN/ReLU/Conv3D/BN} $+$ skip                  & $(N, 256, 12, 25, 3)$  \\
ResConv 2       & \texttt{Conv3D/BN/ReLU/Conv3D/BN} $+$ skip                  & $(N, 256, 12, 25, 3)$  \\
ResConv 3       & \texttt{Conv3D/BN/ReLU/Conv3D/BN} $+$ skip                  & $(N, 256, 12, 25, 3)$  \\
ResConv 4       & \texttt{Conv3D/BN/ReLU/Conv3D/BN} $+$ skip                  & $(N, 256, 12, 25, 3)$  \\
ResConv 5       & \texttt{Conv3D/BN/ReLU/Conv3D/BN} $+$ skip                  & $(N, 256, 12, 25, 3)$  \\
ResConv 6       & \texttt{Conv3D/BN/ReLU/Conv3D/BN} $+$ skip                  & $(N, 256, 12, 25, 3)$  \\[2pt]
\multirow{3}{*}{Decoder 3}
                & \texttt{ConvTranspose3D(c128,k2,s2) + Concat(x3)}           & $(N, 256, 12, 25, 3)$  \\
                & \texttt{Conv3D(c128,k3,s1)/BN/LeakyReLU}                    & $(N, 128, 12, 25, 3)$  \\
                & \texttt{Conv3D(c128,k3,s1)/BN/LeakyReLU}                    & $(N, 128, 12, 25, 3)$  \\
\multirow{3}{*}{Decoder 2}
                & \texttt{ConvTranspose3D(c64,k2,s2) + Concat(x2)}            & $(N, 128, 24, 50, 6)$  \\
                & \texttt{Conv3D(c64,k3,s1)/BN/LeakyReLU}                     & $(N, 64, 24, 50, 6)$   \\
                & \texttt{Conv3D(c64,k3,s1)/BN/LeakyReLU}                     & $(N, 64, 24, 50, 6)$   \\
\multirow{3}{*}{Decoder 1}
                & \texttt{ConvTranspose3D(c32,k2,s2) + Concat(x1)}            & $(N, 64, 48, 100, 12)$ \\
                & \texttt{Conv3D(c32,k3,s1)/BN/LeakyReLU}                     & $(N, 32, 48, 100, 12)$ \\
                & \texttt{Conv3D(c32,k3,s1)/BN/LeakyReLU}                     & $(N, 32, 48, 100, 12)$ \\
\multirow{3}{*}{Decoder 0}
                & \texttt{ConvTranspose3D(c16,k2,s2) + Concat(x0)}            & $(N, 32, 96, 200, 24)$ \\
                & \texttt{Conv3D(c16,k3,s1)/BN/LeakyReLU}                     & $(N, 16, 96, 200, 24)$ \\
                & \texttt{Conv3D(c16,k3,s1)/BN/LeakyReLU}                     & $(N, 16, 96, 200, 24)$ \\[2pt]
Output          & \texttt{Conv3D(c1,k1,s1)} + squeeze                         & $(N, 96, 200, 24)$     \\
\bottomrule
\end{tabular}
\caption{Architecture summary of the V-Net surrogate model.}
\label{tab:vnet3d_arch}
\end{table}

\subsection{Fourier Neural Operator (FNO)}
In recent years, neural operators have emerged as a new paradigm for learning mesh-free, infinite-dimensional operators for solving PDEs using neural networks \cite{lu2021learning, li2020fourier, kovachki2023neural}. The objective of these approaches is to learn a solution operator that maps input functions (e.g., scalar coefficients, forcing terms, and boundary conditions) to the PDE solution, defined as:

\begin{equation}
    \mathcal{G}: \mathcal{A} \rightarrow \mathcal{U} \equiv \mathcal{G}_\theta: \mathcal{A} \rightarrow \mathcal{U}
\end{equation}

\noindent where $\mathcal{A}$ and $\mathcal{U}$ are infinite-dimensional function spaces consisting of the inputs and solutions, respectively. The operator learning problem involves approximating $\mathcal{G}$ with a parameterised neural network $\mathcal{G}_\theta$. This is fundamentally different from classical neural network surrogates, which have primarily focused on learning mappings between finite-dimensional Euclidean spaces ($f: \mathbb{R}^n \rightarrow \mathbb{R}^m$).

Neural operators are iterative architectures in which the input $a(x) \in \mathcal{A}$ is lifted to a higher-dimensional representation $v_0(x) = P(a(x))$ via a local pointwise transformation $P$. This representation is then transformed through a sequence of nonlocal and local nonlinear operations, where each update step is written as:

\begin{equation}
    v_{t+1}(x) = \sigma\!\left( (\mathcal{K}(a;\phi)v_t)(x) + W v_t(x) \right), \quad \forall x \in D
    \label{eqn:neural_operator}
\end{equation}

\noindent where $\mathcal{K}$ is a parameterised nonlocal integral operator, $W$ is a linear pointwise transformation, and $\sigma$ is a nonlinear activation function applied component-wise. The discrete input-output pairs $\{a_j, u_j\}$ are defined on the discretisation $D_j = \{x_1, \dots, x_n\} \subset D$ of the domain $D$.

To obtain an efficient kernel representation, the kernel integral operator in Equation~(\ref{eqn:neural_operator}) is restricted to the translation-invariant form $\kappa(x, y; \phi) = \kappa(x - y;\, \phi)$, which removes the dependence on the input function $a(x)$. Under this assumption, the integral operator reduces to a convolution operator, and can be evaluated efficiently in the Fourier domain as:

\begin{equation}
    (\mathcal{K}(\phi)\, v_t)(x) = \mathcal{F}^{-1}\!\big( R_\phi \cdot \mathcal{F}(v_t) \big)(x), \quad x \in D
    \label{eqn:FNO}
\end{equation}

\noindent in which, $\mathcal{F}$ and $\mathcal{F}^{-1}$ denote the Fourier transform and its inverse, respectively, and $R_\phi$ is a trainable weight tensor parameterised in the Fourier domain and truncated to a finite number of modes $k_{\max}$. This truncation acts as a filter, retaining only the dominant low-frequency modes of the solution. The convolution can thereby be computed in quasi-linear complexity $\mathcal{O}(n \log n)$ using the Fast Fourier Transform (FFT). After the final Fourier layer, the output is projected back to the output space by a pointwise transformation $Q$, which maps the hidden representation to a higher-dimensional intermediate space, followed by a nonlinear activation and a final feedforward layer $L$ that maps to the number of output channels.

\begin{figure}[!htbp]
    \centering
    \includegraphics[width=0.8\linewidth, trim={0 0 0 77pt}, clip]{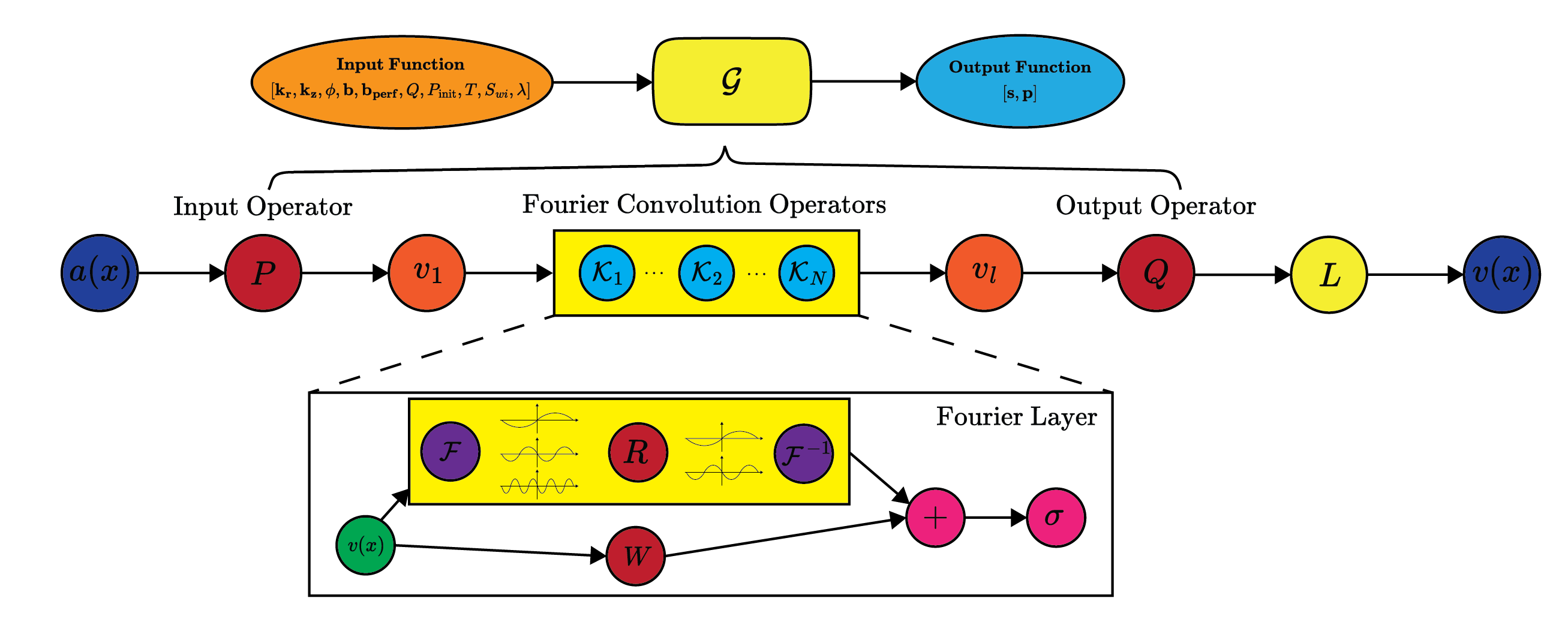}
    \caption{Fourier Neural Operator (FNO) architecture schematic.}
    \label{fig:FNO_Arch}
\end{figure}

The FNO architecture used in this work follows the formulation of Li et al.\ \cite{li2020fourier}, consisting of a lifting layer, $T$ sequential Fourier layers as defined in Equation~(\ref{eqn:FNO}), and a projection layer. The detailed outline of the FNO model architecture used in this study is shown in Table~\ref{tab:fno_arch} and illustrated in Figure~\ref{fig:FNO_Arch}.

\begin{table}[!htbp]
\scriptsize
\centering
\begin{tabular}{llll}
\toprule
\textbf{Unit} & \textbf{Layer} & \textbf{Operation} & \textbf{Output Shape} \\
\midrule
Input        &               &                                                    & $(N, 96, 200, 24, 12)$   \\[2pt]
Lifting      & Linear        & Pointwise Linear Projection ($P$)                  & $(N, 96, 200, 24, 36)$   \\[2pt]
Fourier 1    & Spectral Conv & FFT / $R_\phi \cdot \mathcal{F}(v)$ / IFFT         & $(N, 96, 200, 24, 36)$   \\
             & Residual      & Pointwise Linear Projection ($W$)                  & $(N, 96, 200, 24, 36)$   \\
             & Activation    & ReLU                                               & $(N, 96, 200, 24, 36)$   \\
Fourier 2    & Spectral Conv & FFT / $R_\phi \cdot \mathcal{F}(v)$ / IFFT         & $(N, 96, 200, 24, 36)$   \\
             & Residual      & Pointwise Linear Projection ($W$)                  & $(N, 96, 200, 24, 36)$   \\
             & Activation    & ReLU                                               & $(N, 96, 200, 24, 36)$   \\
Fourier 3    & Spectral Conv & FFT / $R_\phi \cdot \mathcal{F}(v)$ / IFFT         & $(N, 96, 200, 24, 36)$   \\
             & Residual      & Pointwise Linear Projection ($W$)                  & $(N, 96, 200, 24, 36)$   \\
             & Activation    & ReLU                                               & $(N, 96, 200, 24, 36)$   \\
Fourier 4    & Spectral Conv & FFT / $R_\phi \cdot \mathcal{F}(v)$ / IFFT         & $(N, 96, 200, 24, 36)$   \\
             & Residual      & Pointwise Linear Projection ($W$)                  & $(N, 96, 200, 24, 36)$   \\
             & Activation    & ReLU                                               & $(N, 96, 200, 24, 36)$   \\
Fourier 5    & Spectral Conv & FFT / $R_\phi \cdot \mathcal{F}(v)$ / IFFT         & $(N, 96, 200, 24, 36)$   \\
             & Residual      & Pointwise Linear Projection ($W$)                  & $(N, 96, 200, 24, 36)$   \\
             & Activation    & ReLU                                               & $(N, 96, 200, 24, 36)$   \\[2pt]
Feedforward  & Linear        & Pointwise Linear Projection ($Q$)                  & $(N, 96, 200, 24, 128)$  \\
             & Activation    & ReLU                                               & $(N, 96, 200, 24, 128)$  \\
Output       & Linear        & Pointwise Linear Projection ($L$)                  & $(N, 96, 200, 24, 1)$    \\
\bottomrule
\end{tabular}
\caption{Architecture summary of the Fourier Neural Operator (FNO) surrogate model. The Fourier modes are set to $k_{\max} = 10$ in all spatial and temporal dimensions, with a channel width of $N_W = 36$.}
\label{tab:fno_arch}
\end{table}

\subsection{U-Net Enhanced Fourier Neural Operator (U-FNO)}
The U-Net Enhanced Fourier Neural Operator (U-FNO) was originally developed by Wen et al.\ \cite{wen2022u} to improve performance on surrogate models for hyperbolic PDE problems with highly heterogeneous fields that often contain high-frequency modes that can potentially dominate the solution. The motivation behind their work was that the standard FNO struggles to retain high-frequency modes from data because a large number of modes must be retained during truncation. To address this, the U-FNO architecture adds a U-Net operator into each Fourier layer, which appends/captures high-frequency modes to the low-frequency modes captured by the standard FNO layers, thereby enriching the overall approximation.

It should be noted that including the U-Net operator removes the mesh-invariance property of the standard FNO; however, this trade-off is considered acceptable given the potential performance gains. This choice was made by the original authors, as the underlying multi-phase numerical method used in this context depends on the grid resolution; thus, numerical dispersion and dissolution effects are tied to the training grid resolution.

Mathematically, the U-FNO layer update extends the standard FNO layer by incorporating an additional U-Net branch, and can be formally expressed as the following:

\begin{equation}
    v_{t+1}(x) = \sigma\!\left( (\mathcal{K}(a;\phi)v_t)(x) + (\mathcal{U}v_{t})(x) + W v_t(x) \right), \quad \forall x \in D
\end{equation}

\noindent where $\mathcal{K}$ is a parameterized nonlocal integral operator, $\mathcal{U}$ is the U-Net CNN operator, $W$ is a linear pointwise transformation, and $\sigma$ is a nonlinear activation function applied component-wise.

\begin{figure}[!htbp]
    \centering
    \includegraphics[width=0.7\linewidth, trim={0 0 0 77pt}, clip]{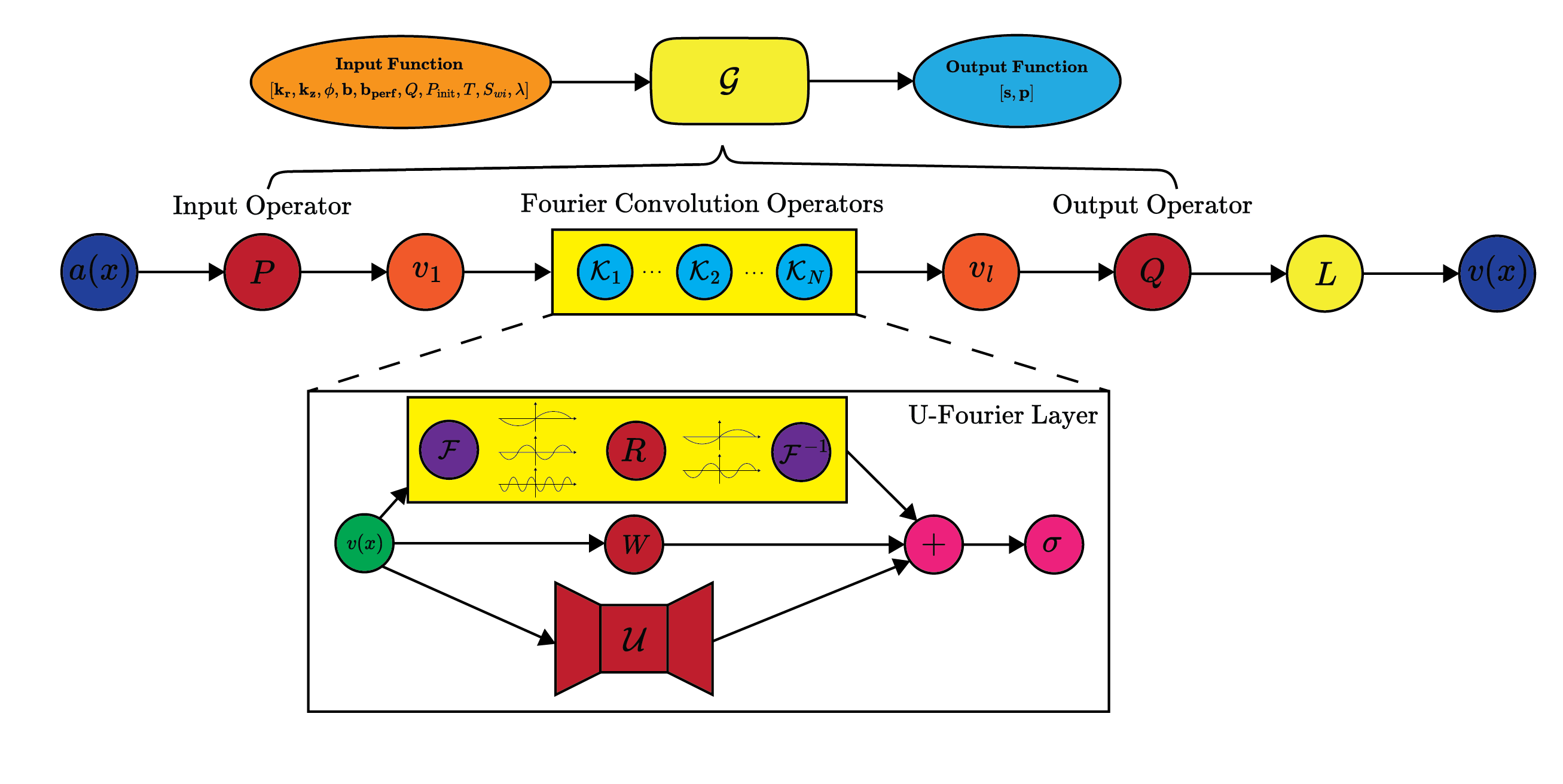}
    \caption{U-Net Enhanced Fourier Neural Operator (U-FNO) architecture schematic.}
    \label{fig:UFNO_Arch}
\end{figure}

The U-FNO architecture used in this work follows that of Wen et al.\ \cite{wen2022u}. The architecture is illustrated in Figure~\ref{fig:UFNO_Arch}, and a detailed summary are shown in Tables~\ref{tab:ufno_arch} and~\ref{tab:ufno_unet}.

\begin{table}[!htbp]
\scriptsize
\centering
\begin{tabular}{llll}
\toprule
\textbf{Unit} & \textbf{Layer} & \textbf{Operation} & \textbf{Output Shape} \\
\midrule
Input        &               &                                                    & $(N, 96, 200, 24, 12)$   \\[2pt]
Lifting      & Linear        & Pointwise Linear Projection ($P$)                  & $(N, 96, 200, 24, 36)$   \\[2pt]
Fourier 1    & Spectral Conv & FFT / $R_\phi \cdot \mathcal{F}(v)$ / IFFT         & $(N, 96, 200, 24, 36)$   \\
             & Residual      & Pointwise Linear Projection ($W$)                  & $(N, 96, 200, 24, 36)$   \\
             & Activation    & ReLU                                               & $(N, 96, 200, 24, 36)$   \\[2pt]
Fourier 2    & Spectral Conv & FFT / $R_\phi \cdot \mathcal{F}(v)$ / IFFT         & $(N, 96, 200, 24, 36)$   \\
             & Residual      & Pointwise Linear Projection ($W$)                  & $(N, 96, 200, 24, 36)$   \\
             & Activation    & ReLU                                               & $(N, 96, 200, 24, 36)$   \\[2pt]
Fourier 3    & Spectral Conv & FFT / $R_\phi \cdot \mathcal{F}(v)$ / IFFT         & $(N, 96, 200, 24, 36)$   \\
             & Residual      & Pointwise Linear Projection ($W$)                  & $(N, 96, 200, 24, 36)$   \\
             & Activation    & ReLU                                               & $(N, 96, 200, 24, 36)$   \\[2pt]
Fourier 4    & Spectral Conv & FFT / $R_\phi \cdot \mathcal{F}(v)$ / IFFT         & $(N, 96, 200, 24, 36)$   \\
             & Residual      & Pointwise Linear Projection ($W$)                  & $(N, 96, 200, 24, 36)$   \\
             & U-Net         & UNet                                               & $(N, 96, 200, 24, 36)$   \\
             & Activation    & ReLU                                               & $(N, 96, 200, 24, 36)$   \\[2pt]
Fourier 5    & Spectral Conv & FFT / $R_\phi \cdot \mathcal{F}(v)$ / IFFT         & $(N, 96, 200, 24, 36)$   \\
             & Residual      & Pointwise Linear Projection ($W$)                  & $(N, 96, 200, 24, 36)$   \\
             & U-Net         & UNet                                               & $(N, 96, 200, 24, 36)$   \\
             & Activation    & ReLU                                               & $(N, 96, 200, 24, 36)$   \\[2pt]
Fourier 6    & Spectral Conv & FFT / $R_\phi \cdot \mathcal{F}(v)$ / IFFT         & $(N, 96, 200, 24, 36)$   \\
             & Residual      & Pointwise Linear Projection ($W$)                  & $(N, 96, 200, 24, 36)$   \\
             & U-Net         & UNet                                               & $(N, 96, 200, 24, 36)$   \\
             & Activation    & ReLU                                               & $(N, 96, 200, 24, 36)$   \\[2pt]
Feedforward  & Linear        & Pointwise Linear Projection ($Q$)                  & $(N, 96, 200, 24, 128)$  \\
             & Activation    & ReLU                                               & $(N, 96, 200, 24, 128)$  \\
Output       & Linear        & Pointwise Linear Projection ($L$)                  & $(N, 96, 200, 24, 1)$    \\
\bottomrule
\end{tabular}
\caption{Architecture summary of the U-FNO surrogate model. The Fourier modes are set to $k_{\max} = 10$ in all spatial and temporal dimensions, with a channel width of $N_W = 36$.}
\label{tab:ufno_arch}
\end{table}

\begin{table}[!htbp]
\scriptsize
\centering
\begin{tabular}{lll}
\toprule
\textbf{Unit} & \textbf{Operation} & \textbf{Output Shape} \\
\midrule
Input      &                                                    & $(N, N_W, 104, 208, 32)$ \\[2pt]
Encoder 1  & \texttt{Conv3D(k3,s2)/BN/LeakyReLU/Drop}           & $(N, N_W, 52, 104, 16)$  \\
Encoder 2  & \texttt{Conv3D(k3,s2)/BN/LeakyReLU/Drop}           & $(N, N_W, 26, 52, 8)$    \\
           & \texttt{Conv3D(k3,s1)/BN/LeakyReLU/Drop}           & $(N, N_W, 26, 52, 8)$    \\
Encoder 3  & \texttt{Conv3D(k3,s2)/BN/LeakyReLU/Drop}           & $(N, N_W, 13, 26, 4)$    \\
           & \texttt{Conv3D(k3,s1)/BN/LeakyReLU/Drop}           & $(N, N_W, 13, 26, 4)$    \\[2pt]
Decoder 2  & \texttt{ConvTranspose3D(k4,s2)} $+$ Concat(Enc 2) & $(N, 2N_W, 26, 52, 8)$   \\
Decoder 1  & \texttt{ConvTranspose3D(k4,s2)} $+$ Concat(Enc 1) & $(N, 2N_W, 52, 104, 16)$ \\
Decoder 0  & \texttt{ConvTranspose3D(k4,s2)} $+$ Concat(Input) & $(N, 2N_W, 104, 208, 32)$\\[2pt]
Output     & \texttt{Conv3D(k3,s1)}                             & $(N, N_W, 104, 208, 32)$ \\
\bottomrule
\end{tabular}
\caption{U-Net block used in Fourier Layers 4--6 of the U-FNO (Table~\ref{tab:ufno_arch}).}
\label{tab:ufno_unet}
\end{table}

\subsection{Loss Function and Training Design}
Similar to Wen et al.\ \cite{wen2022u}, we employ a multi-component loss function for both the CO$_2$ saturation and pressure build-up fields. However, we extend their formulation by replacing the first-derivative term with the full spatial gradient of the predicted field. The original loss function is defined as:
\begin{equation}
    \mathcal{L}_{r}(\hat{\mathbf{y}}, \mathbf{y}) =
    \frac{\|\mathbf{y}-\hat{\mathbf{y}}\|_2}{\|\mathbf{y}\|_2}
    + \lambda_r
    \frac{\|\partial_r \mathbf{y} - \partial_r \hat{\mathbf{y}}\|_2}
    {\|\partial_r \mathbf{y}\|_2}
    \label{eqn:loss_radial}
\end{equation}

while our modified full-gradient loss is defined as:
\begin{equation}
    \mathcal{L}_{rz}(\hat{\mathbf{y}}, \mathbf{y}) =
    \frac{\|\mathbf{y} - \hat{\mathbf{y}}\|_2}{\|\mathbf{y}\|_2}
    + \lambda_r
    \frac{\|\partial_r \mathbf{y} - \partial_r \hat{\mathbf{y}}\|_2}
    {\|\partial_r \mathbf{y}\|_2}
    + \lambda_z
    \frac{\|\partial_z \mathbf{y} - \partial_z \hat{\mathbf{y}}\|_2}
    {\|\partial_z \mathbf{y}\|_2}
    \label{eqn:loss_full}
\end{equation}

\noindent where $\hat{\mathbf{y}}$ is the predicted output, $\mathbf{y}$ is the ground truth, and $\lambda_r$ and $\lambda_z$ are hyperparameters that control the relative contributions of the radial and vertical gradient terms, respectively.

In contrast to Wen et al.\ \cite{wen2022u}, who found that the vertical gradient penalty term did not improve predictive performance, our results indicate a marginal improvement for select architectures. The inclusion of the gradient term improved the ability of select models to capture the sharp discontinuity associated with the CO$_2$ plume front in the saturation field, as well as the buoyancy effects present in the vertical direction. Since the dataset consists of reservoirs with varying thicknesses, a variable active cell mask is applied to each sample during the computation of the loss, ensuring that the loss is evaluated only over the active region of the domain. Zero-padding is applied to both the input and output tensors to fill cells outside the active reservoir with a neutral value that does not contaminate the loss landscape.

All models presented in this study are trained using the Adam optimizer with an initial learning rate of $1\times10^{-3}$ and $\mathcal{L}_2$-regularization with a weight decay of $\lambda = 1\times10^{-4}$. A Cosine Annealing Warm Restart learning rate scheduler is used for all models with $T_0 = 50$ and $T_{\text{mult}} = 2$, where $T_0$ is the number of iterations for the first restart and $T_{\text{mult}}$ is the multiplication factor for the cycle length after each restart. The FNO-based models are trained for 350 epochs, while the CNN-based models are trained for 750 epochs. In all cases, the model with the highest validation $R^2$ over the training process is saved as the best model.

\section{Results}
This section presents and compares the predictive accuracy, training cost, and memory requirements of the five surrogate model architectures: the Temporal CNN, U-Net, V-Net, FNO, and U-FNO. The Temporal CNN and U-FNO architectures follow those presented by Wen et al.\ \cite{wen2021ccsnet, wen2022u}, and the same dataset is used to enable a consistent and fair comparison. All models are trained using the proposed loss function (Equ.~\ref{eqn:loss_full}) and predict both the CO$_2$ saturation and pressure build-up fields. Their performance is discussed in the context of the predictive behavior in terms of PDE type (i.e., elliptic vs.\ hyperbolic). Model architectures are described in detail in Section~\ref{sec:Methodolgy}.

\subsection{CO$_2$ Saturation Field}
\label{sec:CO2_sat}
This section compares the performance of the five surrogate model architectures using the loss functions defined in Equations~\eqref{eqn:loss_radial} and~\eqref{eqn:loss_full} for the CO$_2$ saturation field (i.e., hyperbolic PDE). A parametric study is performed to investigate the sensitivity of model performance to the gradient penalty hyperparameter $\lambda$, with values of 0.01, 0.1, and 1.0 for both the radial-only and full gradient loss formulations. The overall accuracy of each model is assessed using the Root Mean Square Error (RMSE) and $R^2$ metrics. Representative test cases are then presented to illustrate the prediction errors, followed by a frequency-domain analysis using the Power Spectral Density (PSD). Finally, the sensitivity of model performance to permeability heterogeneity is analyzed.

\subsubsection{Model Performance}
Tables~\ref{tab:results_asymmetric} and~\ref{tab:results_symmetric} summarize the performance of all trained surrogate models for the CO$_2$ saturation field. The results demonstrate that the U-FNO achieves the highest predictive accuracy, followed by the U-Net and V-Net, and then the FNO and Temporal CNN. All models are discussed and evaluated in terms of testing performance. The results are consistent with the expected behaviour of each architecture. The Temporal CNN produces smoother, more diffuse predictions due to its lack of skip connections, while the FNO exhibits a similar behavior that can be mainly attributed to the low-frequency bias introduced by spectral truncation. These limitations are significant for the highly localized features present in the anisotropic heterogeneous hyperbolic saturation field. In contrast, the U-Net and V-Net benefit from skip connections that preserve high-frequency spatial information along the decoder path. The U-FNO achieves the best overall performance as the FNO component captures the global low-frequency structure of the solution, while the embedded U-Net branch enriches the prediction with high-frequency features.

The results also demonstrate that including the full spatial gradient error term ($\lambda_z > 0$) improves predictive accuracy for both the Temporal CNN and the U-FNO. This improvement can be attributed to the sharp, localized features of the CO$_2$ saturation field, where gradient information in multiple directions provides additional context on the direction of flow, as Darcy fluxes are driven by pressure gradients, which more effectively guides the optimization process. In particular, the vertical gradient error term ($\lambda_z$, Equ.~\ref{eqn:loss_full}) captures flow phenomena such as buoyancy-driven flow and viscous fingering, both of which are critically influenced by vertical permeability anisotropy and heterogeneity in the reservoir.

The full gradient loss function does not benefit the U-Net and V-Net architectures. This can be attributed to their highly localized predictions driven by skip connections, where the additional gradient information does not meaningfully improve learning. The FNO similarly struggles with the addition of the gradient error term, as the high-gradient regions of the saturation field are predominantly associated with high-frequency modes that spectral truncation prevents the model from capturing, introducing noise into the optimization process. The U-FNO benefits from the full gradient loss as the FNO component captures the global low-frequency structure of the solution, while the embedded U-Net branch retains the high-frequency features associated with the sharp gradient regions. The combination of the primary and gradient error terms in the loss function results in convergence to a better local minimum. Overall, the U-FNO trained with the full gradient loss function (Equ.~\ref{eqn:loss_full}) and $\lambda_r = \lambda_z = 0.01$ achieves the best predictive accuracy on the CO$_2$ saturation field, with a test $R^2$ of 0.967.

Table~\ref{tab:computational_cost_SG} summarizes the computational efficiency and cost of all surrogate models. All models were trained on the Compute Canada Trillium cluster using a single NVIDIA H100 GPU with 80 GB of memory. It should be mentioned that all models were constructed to have a similar number of learnable parameters to enable a fair comparison. The Temporal CNN has the largest number of trainable parameters at 36.9M; however, it achieves the shortest training time of 15.9 hours due to the computational efficiency of convolutional operations. The U-Net and V-Net are the most parameter-efficient models at approximately 26.9M and 27.3M parameters, respectively, and also require the least memory during training. The FNO requires significantly more training time (27.8 hours) despite a comparable parameter count, due to the computational overhead of FFT operations performed at full spatial resolution. The U-FNO is the most computationally expensive model, requiring 63.3 hours of training time and 5,960~MB of memory per forward/backward pass, illustrating the additional cost of the embedded U-Net branches within Fourier layers.

\begin{table}[!htbp]
\scriptsize
\centering
\begin{tabular}{llccc}
\toprule
\multirow{2}{*}{\textbf{Model}} & \multirow{2}{*}{\textbf{Parameters}} &
\multirow{2}{*}{\textbf{\makecell{Forward/Backward \\ Pass Memory (MB)}}} &
\multicolumn{2}{c}{\textbf{Training Cost}} \\
\cmidrule(lr){4-5}
& & & \textbf{GPU Type} & \textbf{Time (hr)} \\
\midrule
Temporal CNN & 36,850,017 & 1,562.09 & NVIDIA H100 -- 80 GB & 15.9 \\
U-Net        & 26,894,017 & 1,152.69 & NVIDIA H100 -- 80 GB & 10.5 \\
V-Net        & 27,270,785 & 1,198.80 & NVIDIA H100 -- 80 GB & 11.0 \\
FNO          & 31,117,325 & 1,576.46 & NVIDIA H100 -- 80 GB & 27.8 \\
U-FNO        & 33,097,829 & 2,980.11 & NVIDIA H100 -- 80 GB & 63.3 \\
\bottomrule
\end{tabular}
\caption{Summary of model architecture and training cost for CO$_2$ saturation field.}
\label{tab:computational_cost_SG}
\end{table}

\begin{figure}[!htbp]
    \centering
    \begin{subfigure}[t]{0.45\textwidth}
        \centering
        \includegraphics[width=\textwidth]{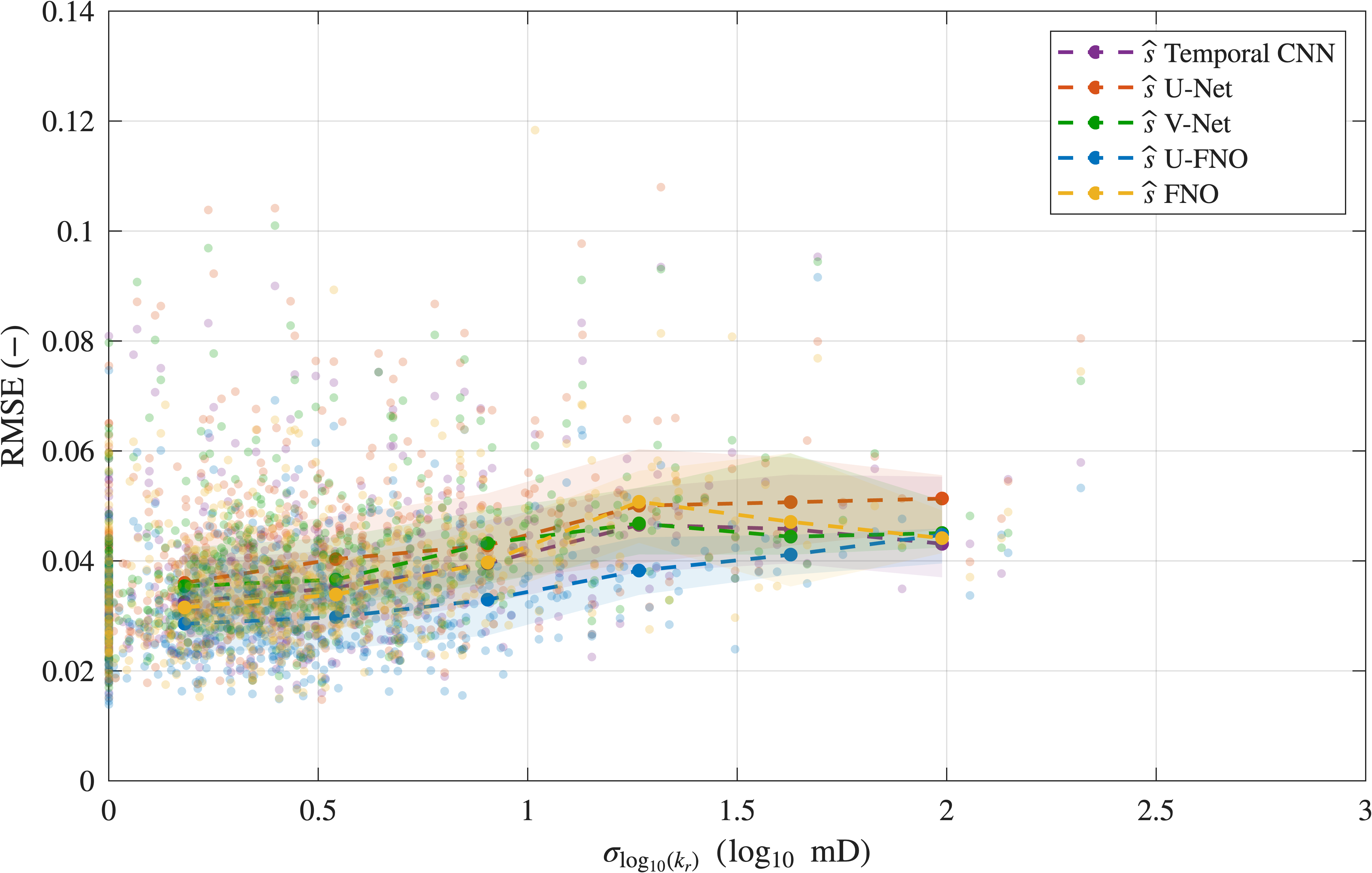}
        \caption{RMSE vs $\sigma_{\log_{10}(k_r)}$}
        \label{fig:perm_r_std}
    \end{subfigure}
    \hfill
    \begin{subfigure}[t]{0.45\textwidth}
        \centering
        \includegraphics[width=\textwidth]{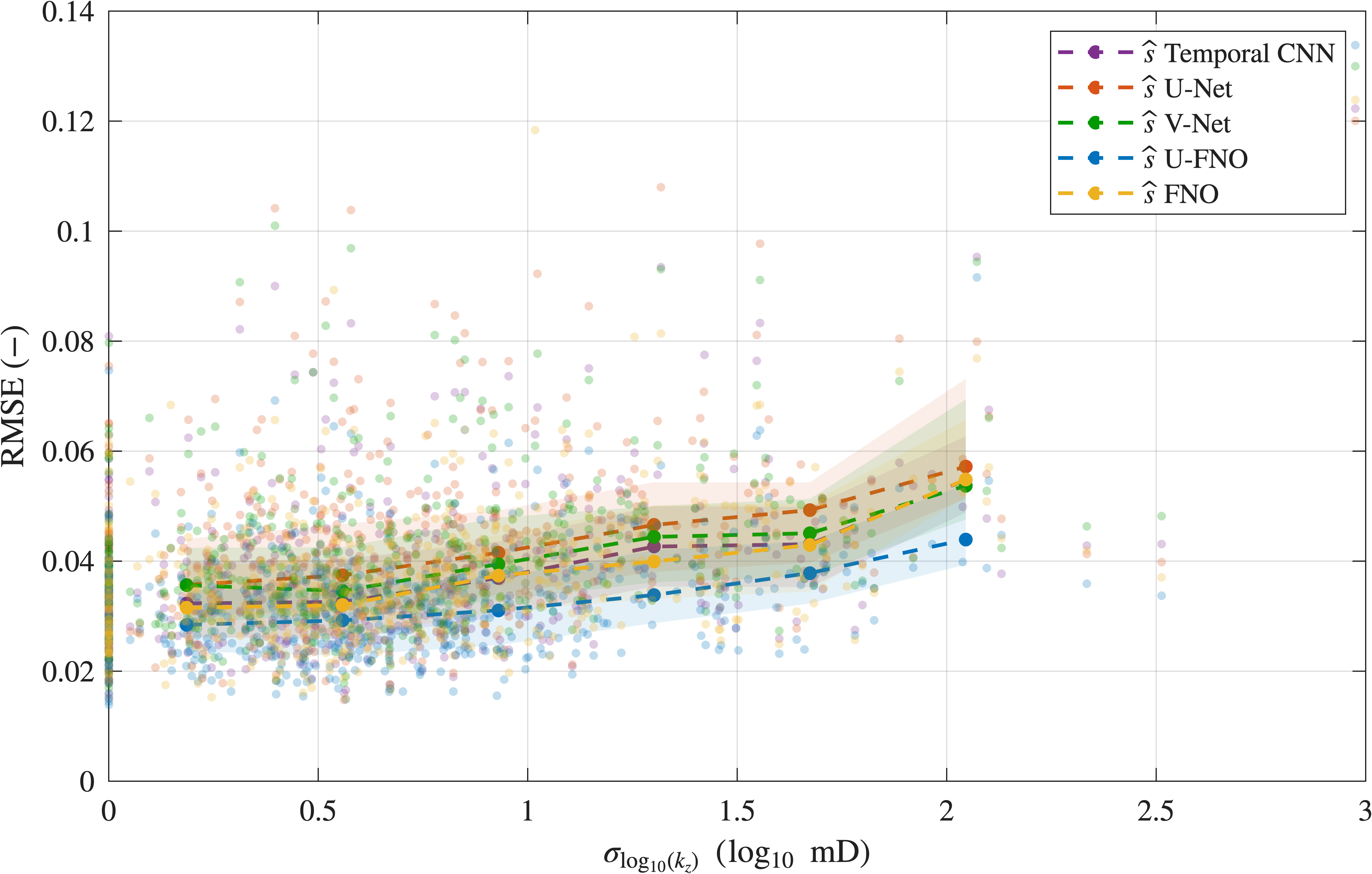}
        \caption{RMSE vs $\sigma_{\log_{10}(k_z)}$}
        \label{fig:perm_z_std}
    \end{subfigure}
    \caption{Test dataset root mean square error (RMSE) vs.\ standard deviation in permeability field for all models for CO$_2$ saturation field.}
    \label{fig:SG_perm_all}
\end{figure}

The sensitivity of model performance to permeability heterogeneity is shown in Figure~\ref{fig:SG_perm_all}. As discussed earlier, highly anisotropic and heterogeneous reservoirs inherently exhibit spectral compositions dominated by high-frequency modes. The U-FNO achieves the highest accuracy across the full range of permeability variability. Although the U-Net and V-Net architectures achieve lower overall RMSE relative to the Temporal CNN and FNO from Tables~\ref{tab:results_asymmetric} and~\ref{tab:results_symmetric}, both models show lower performance for highly heterogeneous permeability fields. This behavior can be attributed to their skip connections, which preserve high-frequency spatial detail along the decoder path and improve overall predictive accuracy. However, for highly heterogeneous permeability fields, the localized nature of the skip connection features becomes insufficient to capture the broader global spectral composition of the saturation field. It is also observed that this performance degradation is more significant in the $z$-direction, as the vertical permeability field exhibits a slower decay in its spatial modes.

\subsubsection{Representative Test Cases}
\begin{figure}[!htbp]
    \centering
    \begin{subfigure}[b]{0.24\linewidth}
        \centering
        \includegraphics[width=\linewidth]{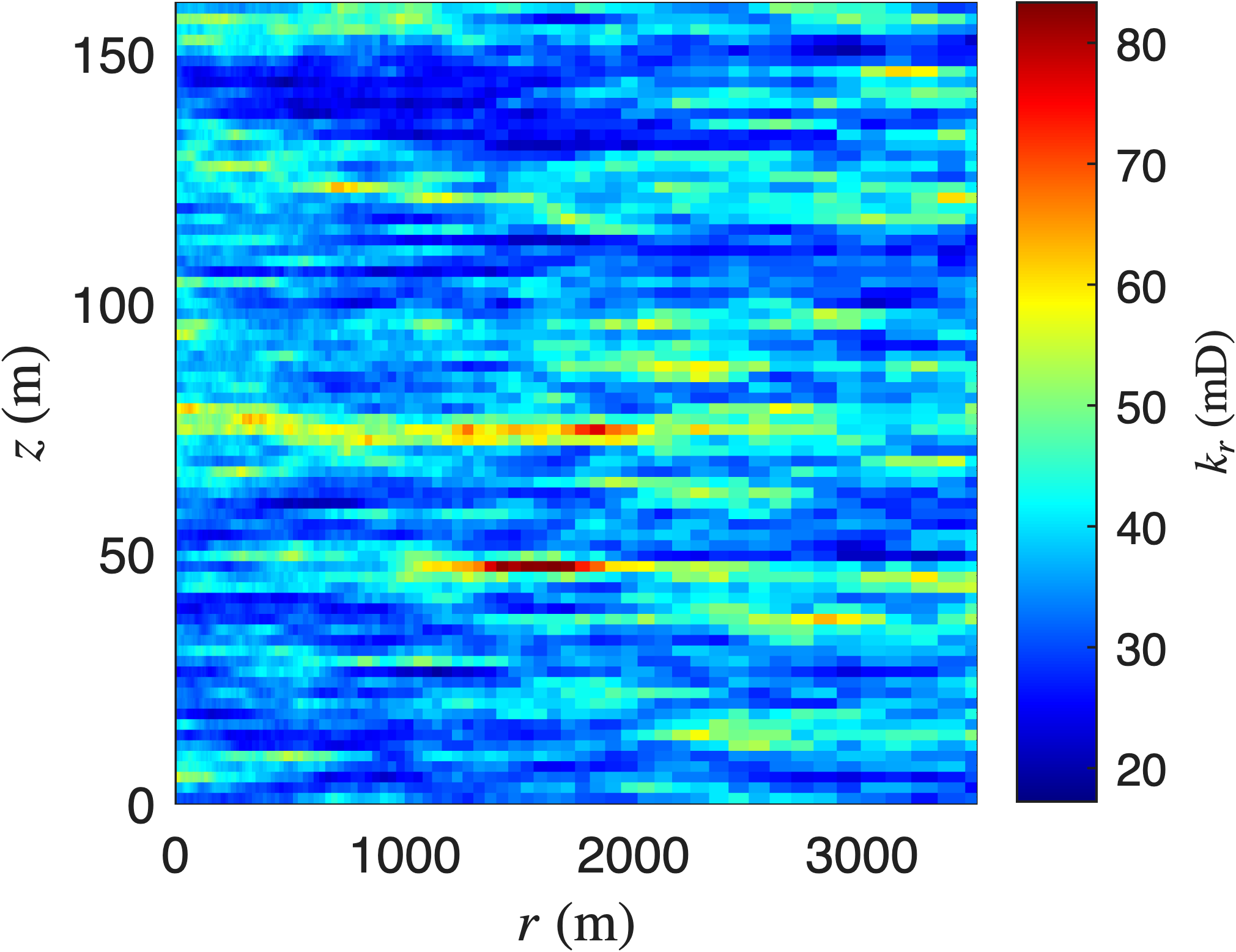}
        \caption{$k_r$ (mD)}
    \end{subfigure}
    \hfill
    \begin{subfigure}[b]{0.24\linewidth}
        \centering
        \includegraphics[width=\linewidth]{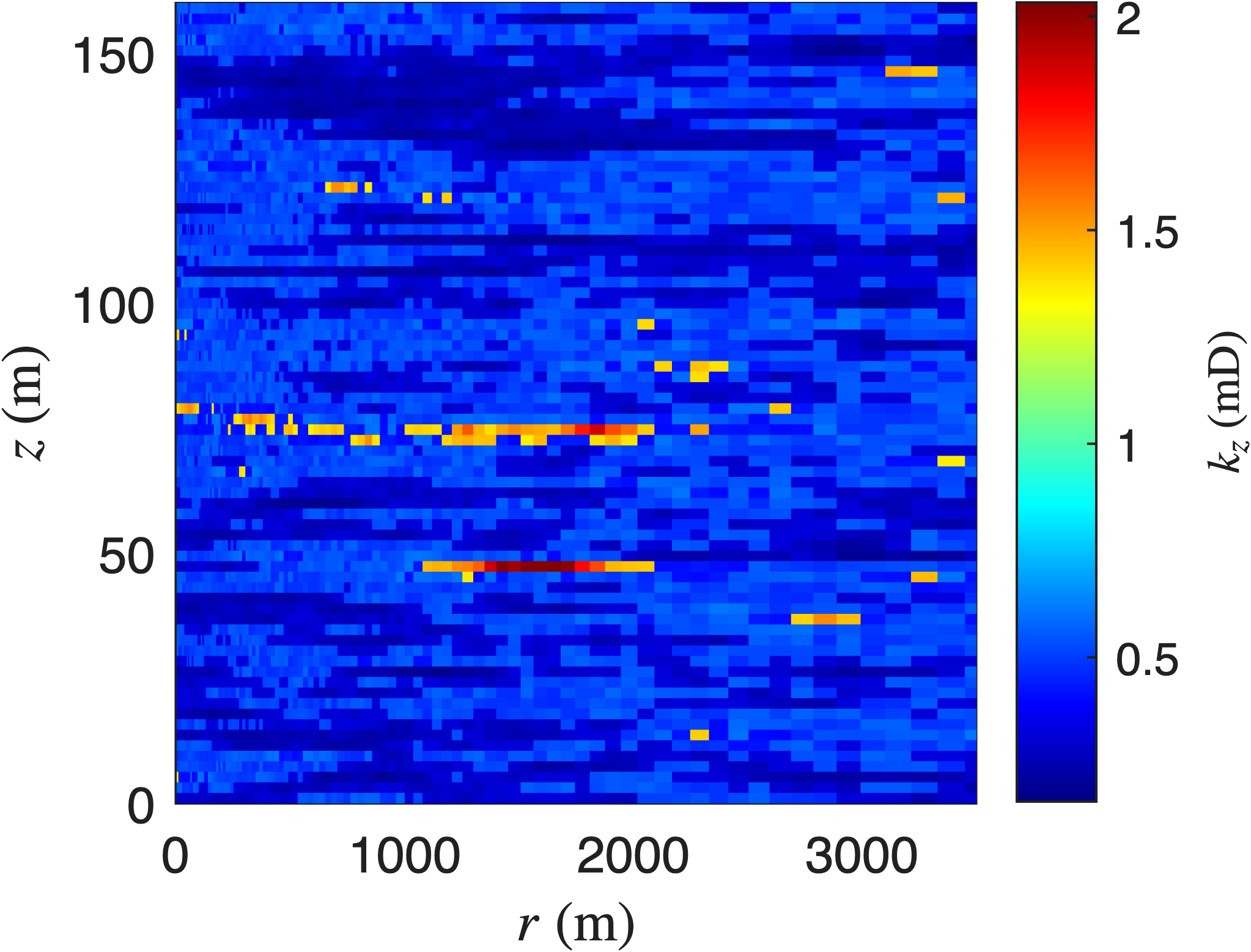}
        \caption{$k_z$ (mD)}
    \end{subfigure}
    \hfill
    \begin{subfigure}[b]{0.25\linewidth}
        \centering
        \includegraphics[width=\linewidth]{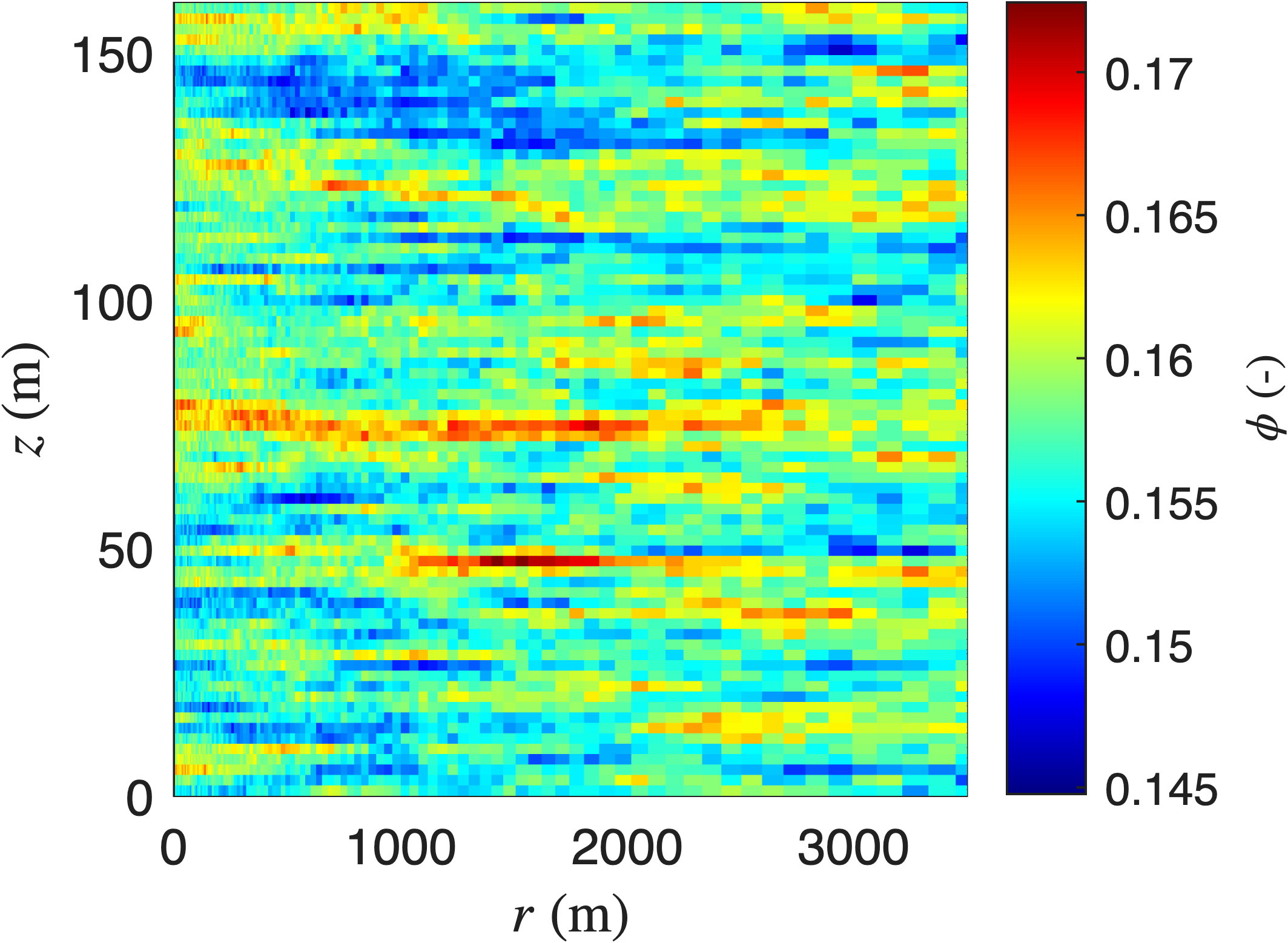}
        \caption{$\phi$ (-)}
    \end{subfigure}
    \hfill
    \begin{subfigure}[b]{0.24\linewidth}
        \centering
        \includegraphics[width=\linewidth]{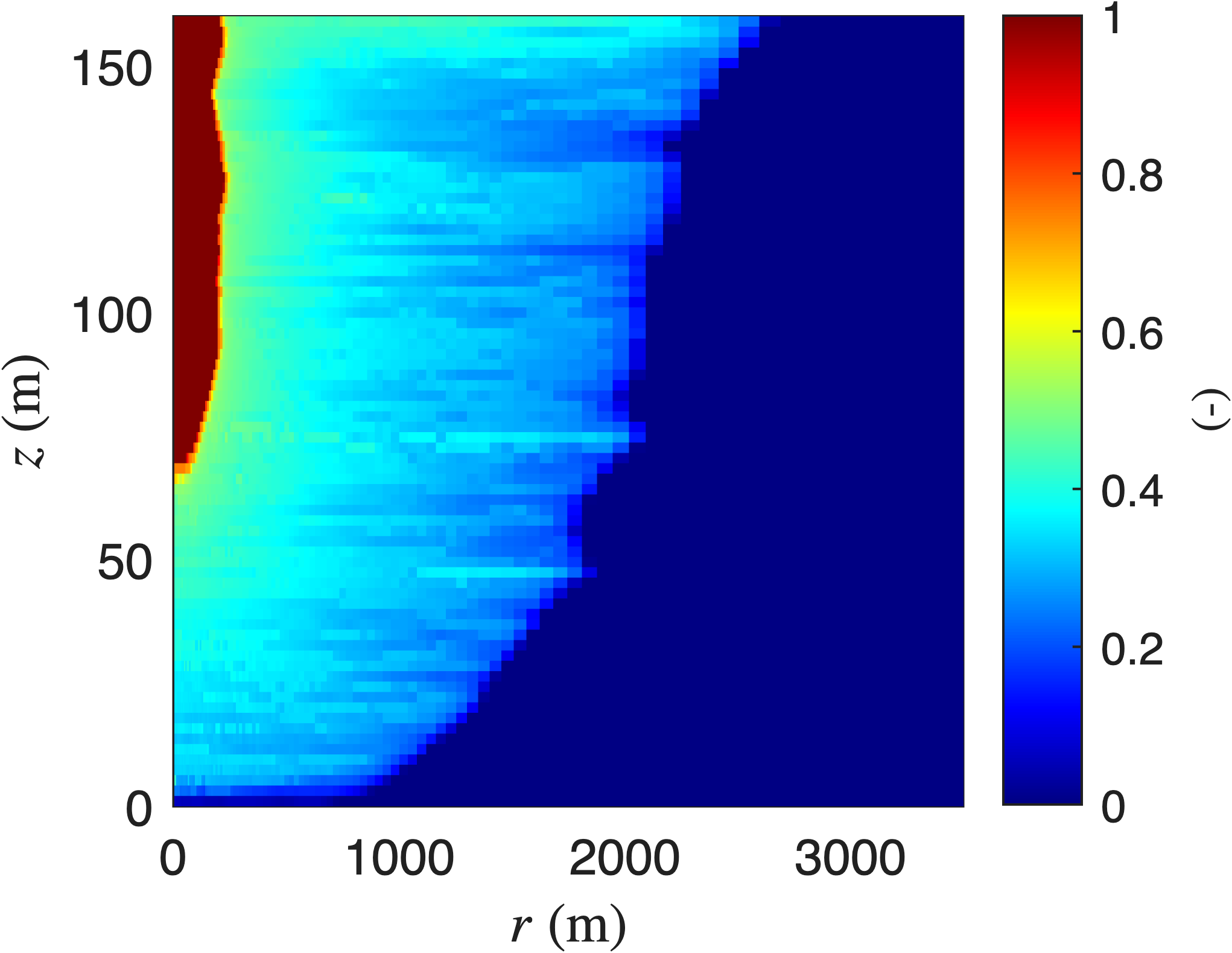}
        \caption{$s$ (-), $t = 30.0$ yr}
    \end{subfigure}
    \caption{Input fields and ground truth CO$_2$ saturation field for Test Case 1. Scalar inputs: $Q_\text{inj} = 1.99$ MT/yr,\; $T = 118.9\,^\circ$C,\; $P_\text{init} = 213.3$ bar,\; $S_{wi} = 0.24$,\; $\lambda = 0.52$.}
    \label{fig:Inputs_GT_Test_1}
\end{figure}

\begin{figure}[!htbp]
    \centering
    \begin{subfigure}[b]{0.24\linewidth}
        \centering
        \includegraphics[width=\linewidth]{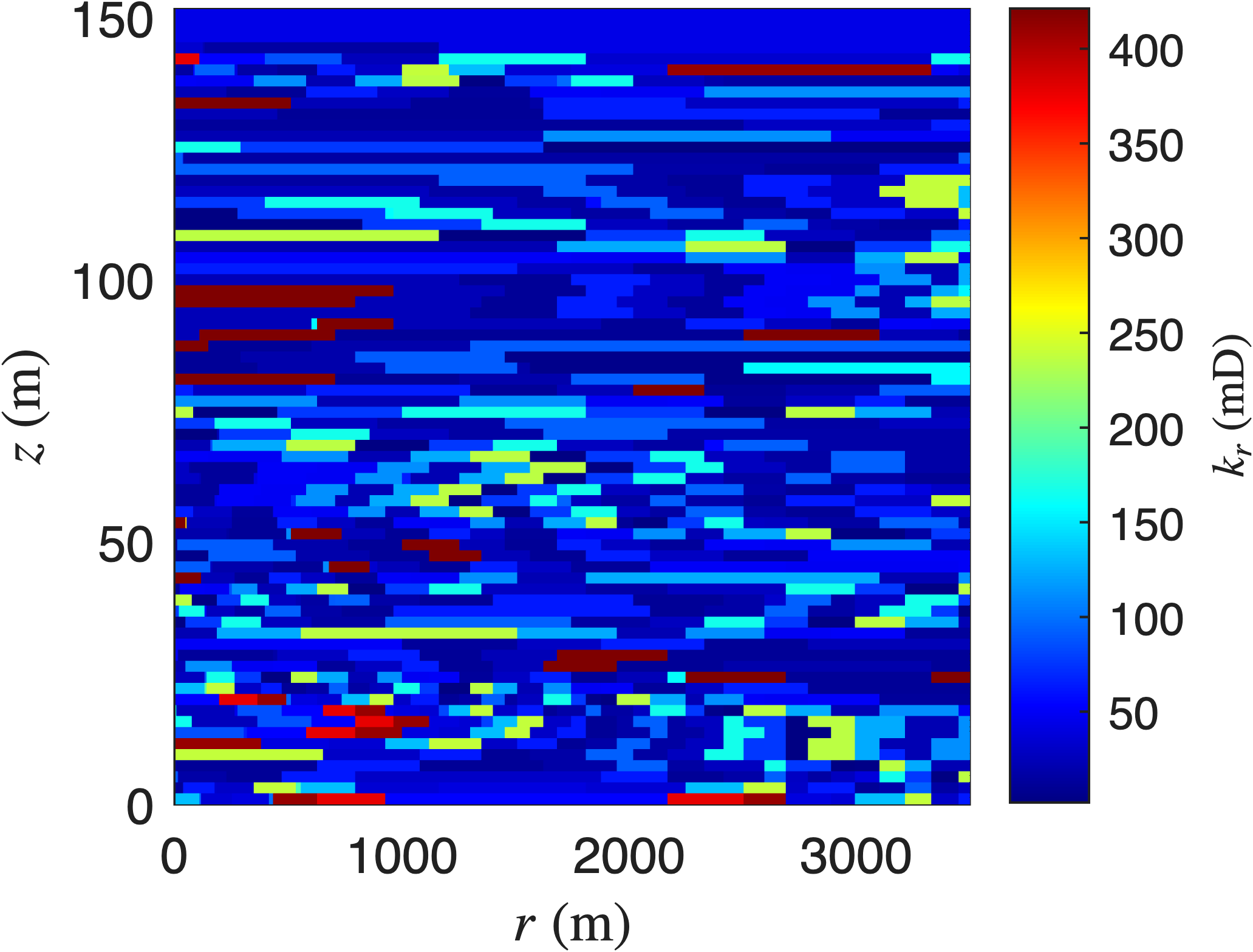}
        \caption{$k_r$ (mD)}
    \end{subfigure}
    \hfill
    \begin{subfigure}[b]{0.24\linewidth}
        \centering
        \includegraphics[width=\linewidth]{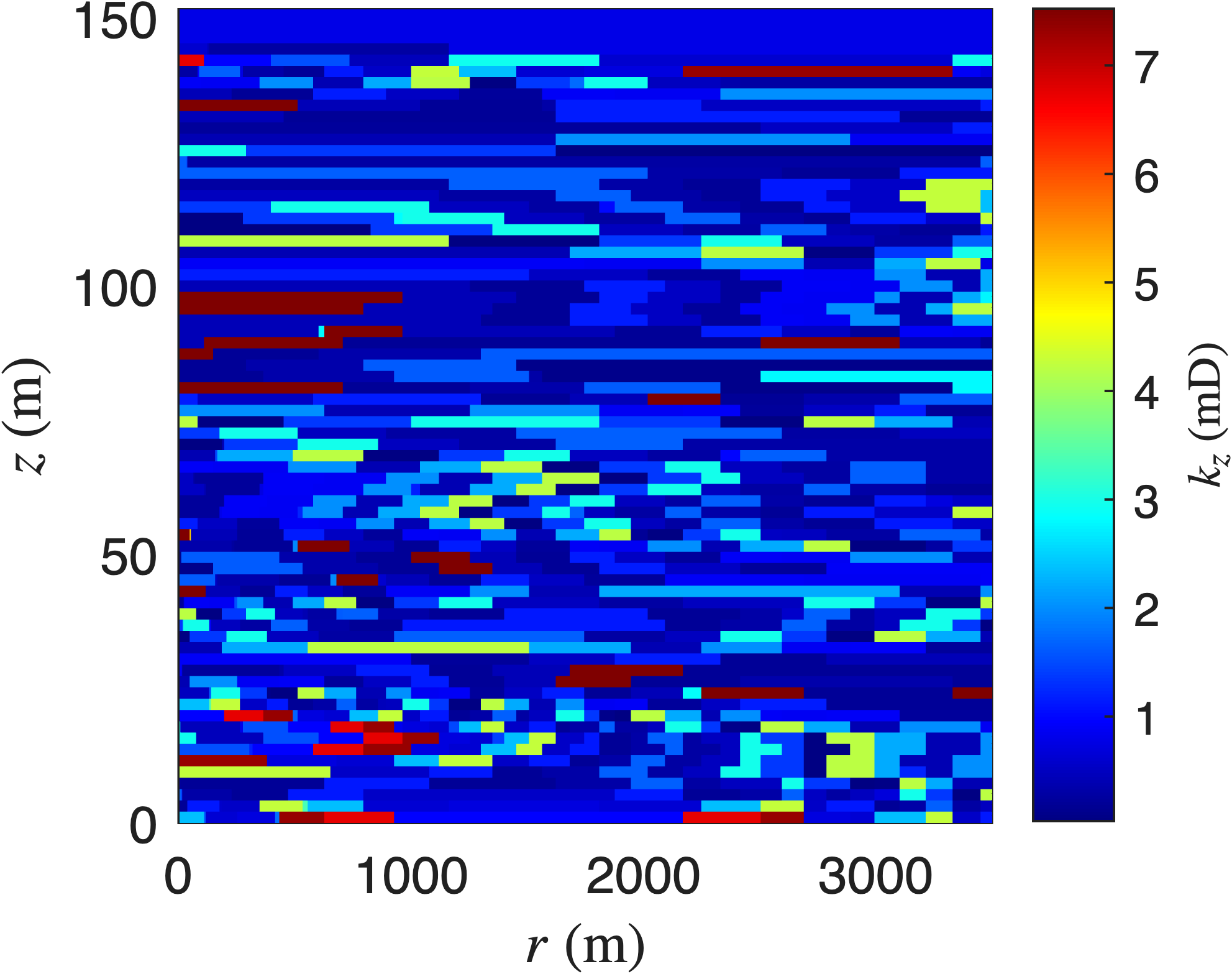}
        \caption{$k_z$ (mD)}
    \end{subfigure}
    \hfill
    \begin{subfigure}[b]{0.25\linewidth}
        \centering
        \includegraphics[width=\linewidth]{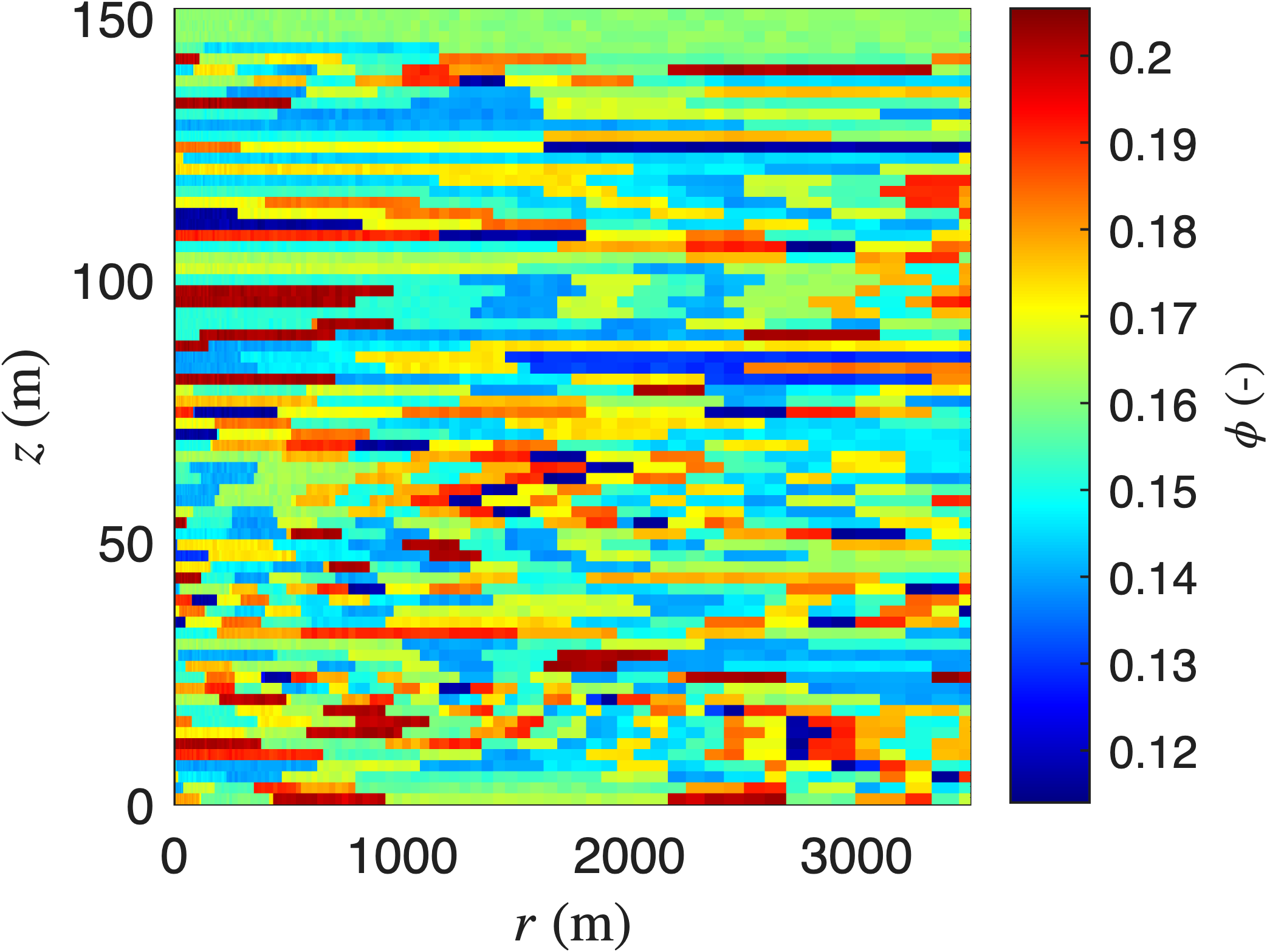}
        \caption{$\phi$ (-)}
    \end{subfigure}
    \hfill
    \begin{subfigure}[b]{0.24\linewidth}
        \centering
        \includegraphics[width=\linewidth]{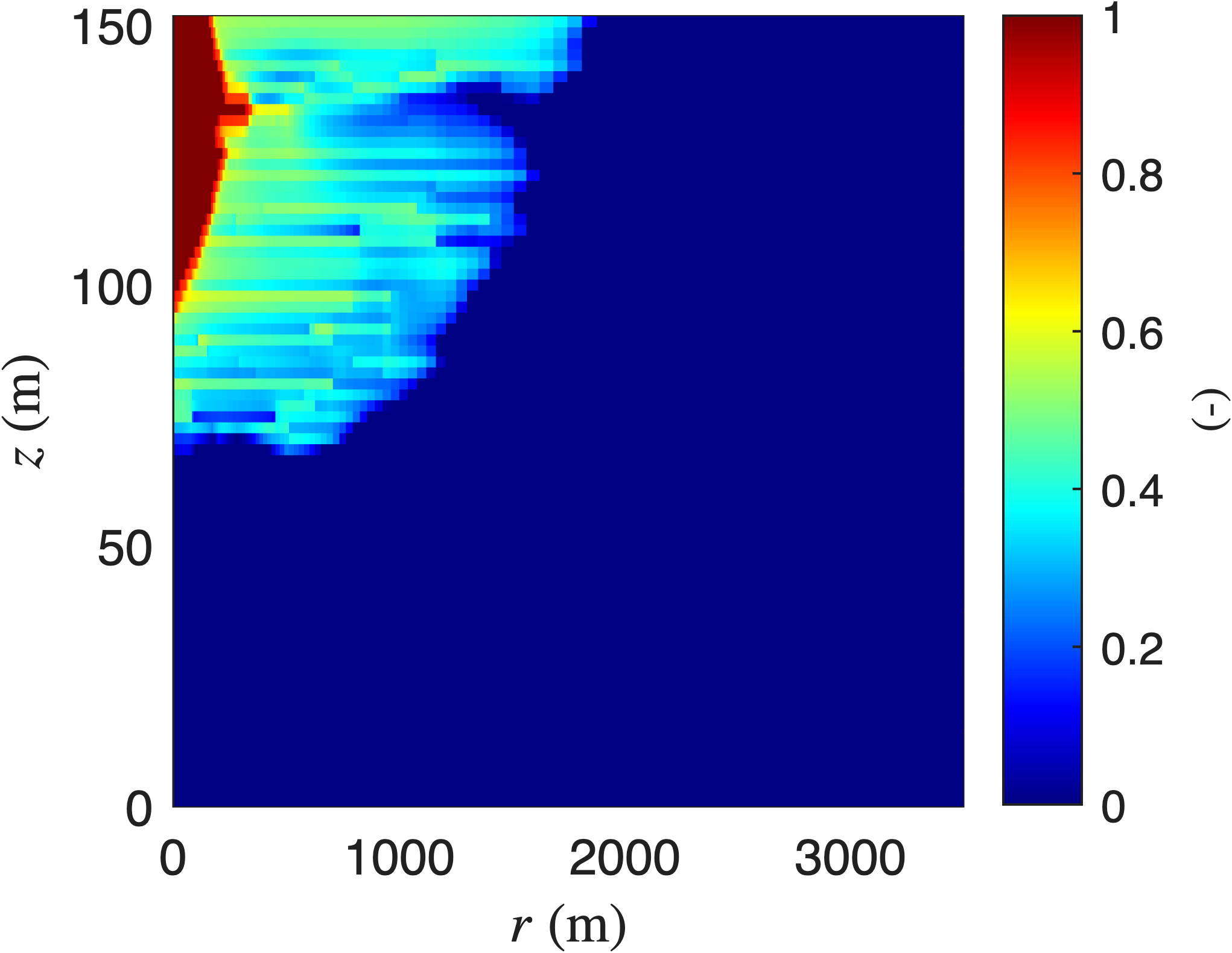}
        \caption{$s$ (-), $t = 30.0$ yr}
    \end{subfigure}
    \caption{Input fields and ground truth CO$_2$ saturation field for Test Case 2. Scalar inputs: $Q_\text{inj} = 0.66$ MT/yr,\; $T = 148.6\,^\circ$C,\; $P_\text{init} = 279.4$ bar,\; $S_{wi} = 0.15$,\; $\lambda = 0.45$.}
    \label{fig:Inputs_GT_Test_2}
\end{figure}

\begin{figure}[!tbp]
    \centering
    \includegraphics[width=0.74\linewidth]{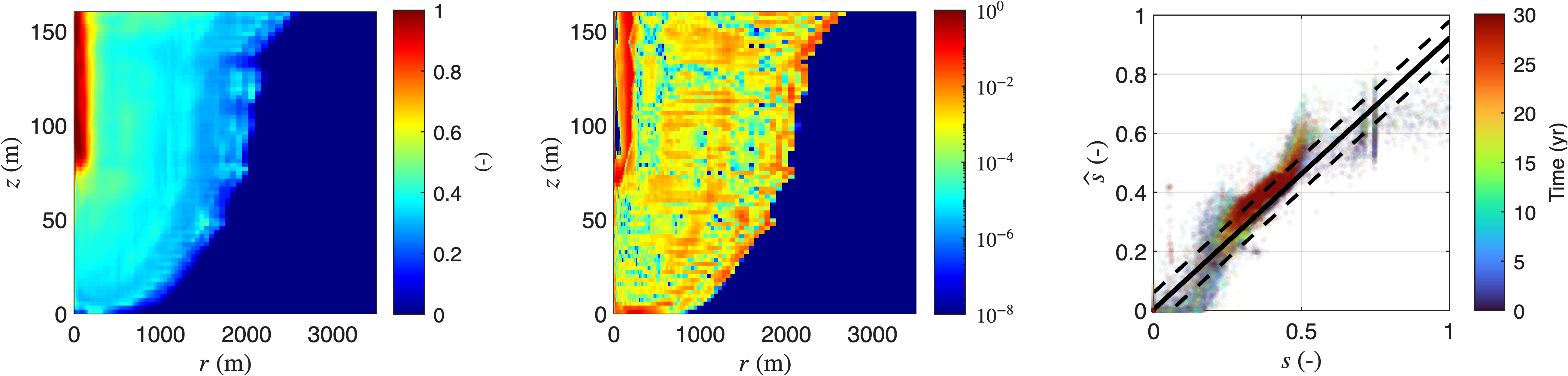}
    \modelcaption{Temporal CNN ($R^2 = 0.9699$, RMSE $= 3.3 \times 10^{-2}$)}
    \includegraphics[width=0.74\linewidth]{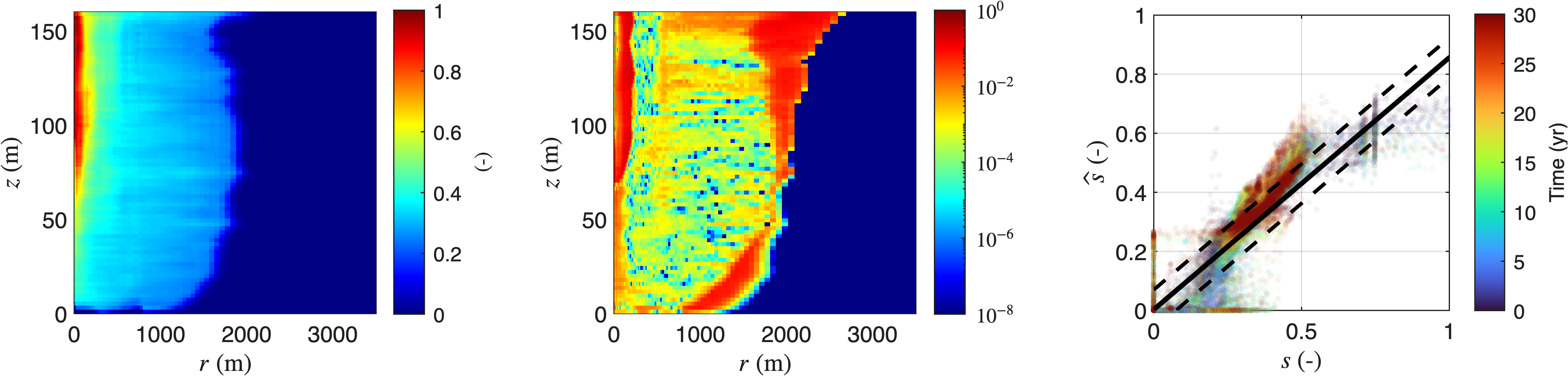}
    \modelcaption{U-Net ($R^2 = 0.9448$, RMSE $= 4.4 \times 10^{-2}$)}
    \includegraphics[width=0.74\linewidth]{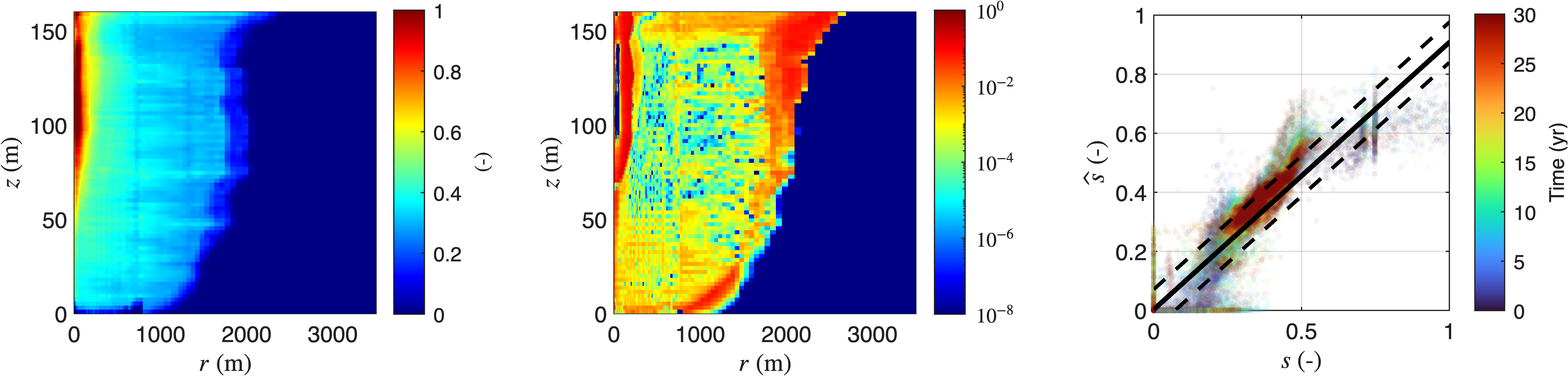}
    \modelcaption{V-Net ($R^2 = 0.9574$, RMSE $= 3.9 \times 10^{-2}$)}
    \includegraphics[width=0.74\linewidth]{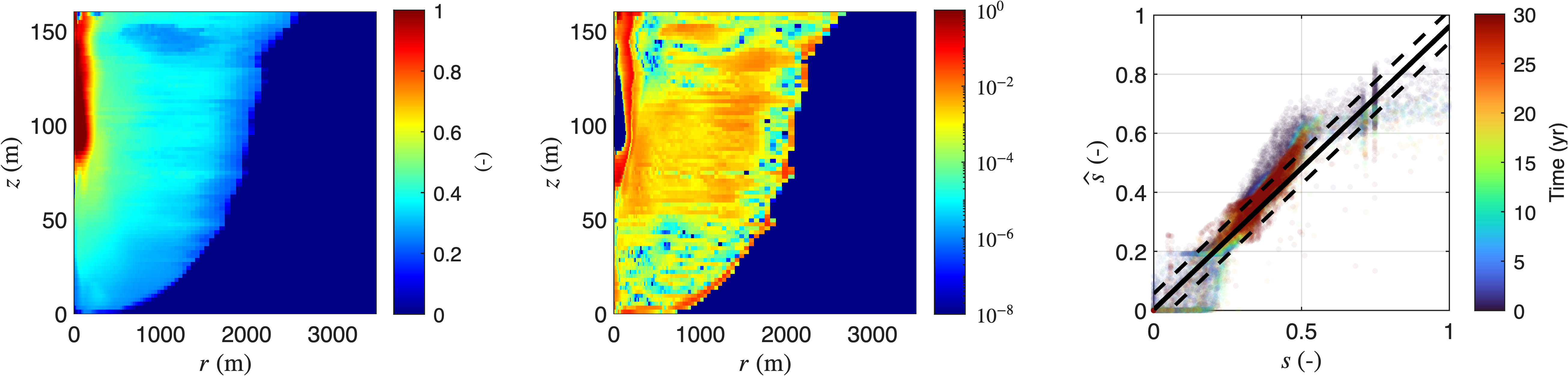}
    \modelcaption{FNO ($R^2 = 0.9772$, RMSE $= 2.9 \times 10^{-2}$)}
    \includegraphics[width=0.74\linewidth]{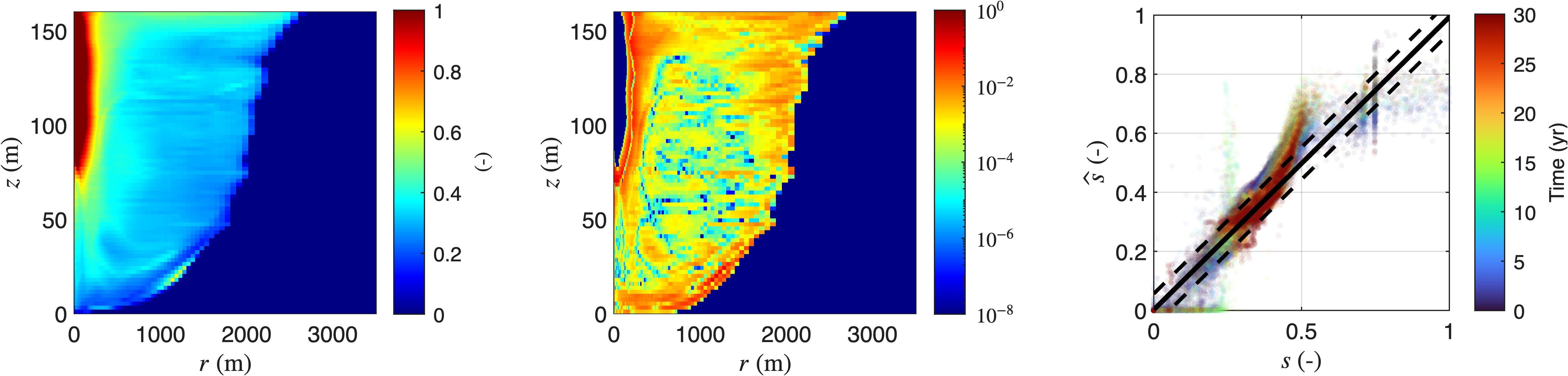}
    \modelcaption{U-FNO ($R^2 = 0.9787$, RMSE $= 2.8 \times 10^{-2}$)}
    \caption{CO$_2$ saturation field predictions for Test Case 1: (a) predicted field $\hat{s}$ at $t = 30.0$ yr, (b) normalized absolute error $|s - \hat{s}|/|s|_{\max}$ at $t = 30.0$ yr, and (c) parity plot over the full 30-year injection period.}
    \label{fig:sat_test_1_results}
\end{figure}

\begin{figure}[!tbp]
    \centering
    \includegraphics[width=0.74\linewidth]{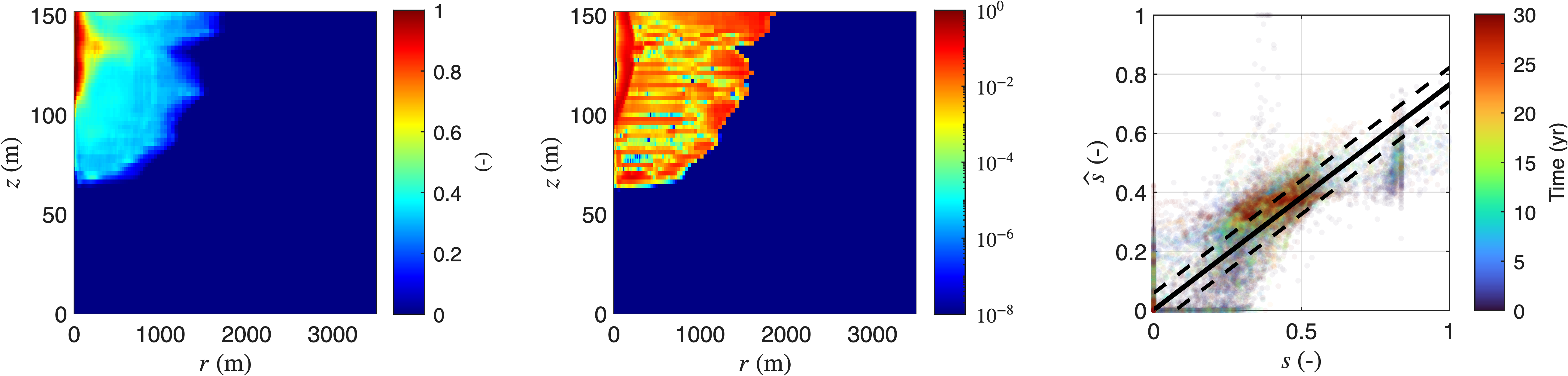}
    \modelcaption{Temporal CNN ($R^2 = 0.9025$, RMSE $= 4.6 \times 10^{-2}$)}
    \includegraphics[width=0.74\linewidth]{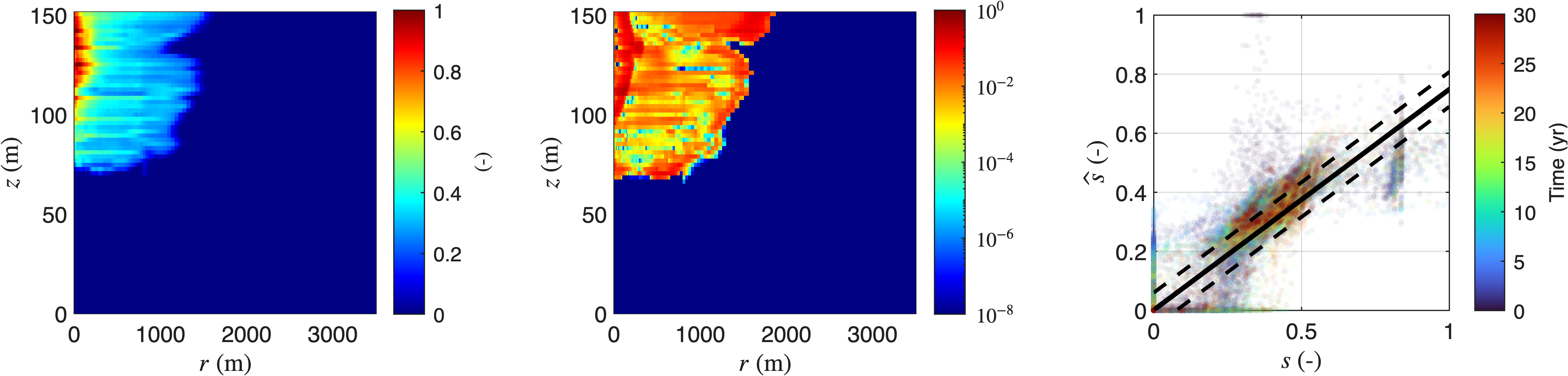}
    \modelcaption{U-Net ($R^2 = 0.8922$, RMSE $= 4.8 \times 10^{-2}$)}
    \includegraphics[width=0.74\linewidth]{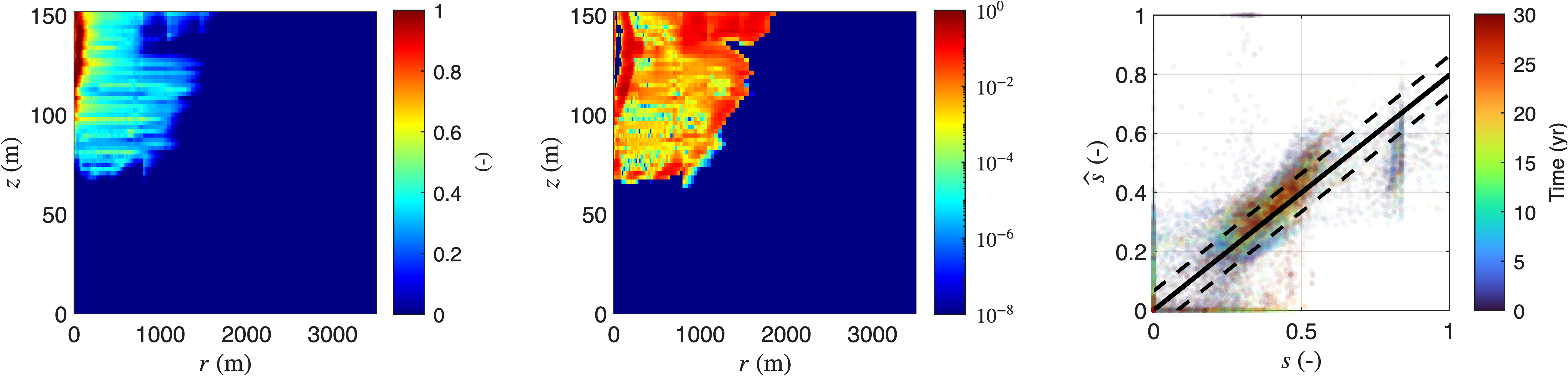}
    \modelcaption{V-Net ($R^2 = 0.9062$, RMSE $= 4.5 \times 10^{-2}$)}
    \includegraphics[width=0.74\linewidth]{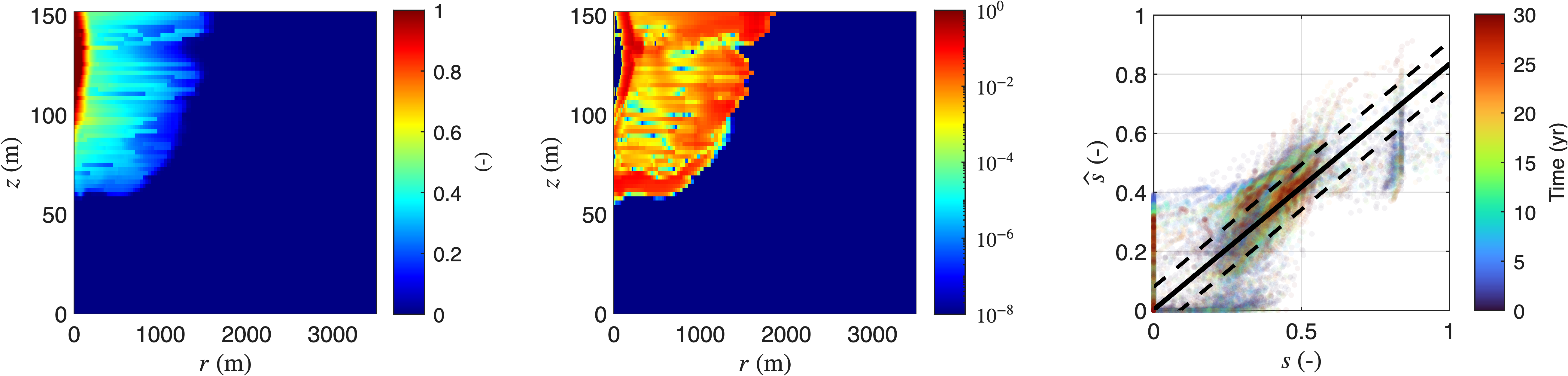}
    \modelcaption{FNO ($R^2 = 0.9007$, RMSE $= 4.6 \times 10^{-2}$)}
    \includegraphics[width=0.74\linewidth]{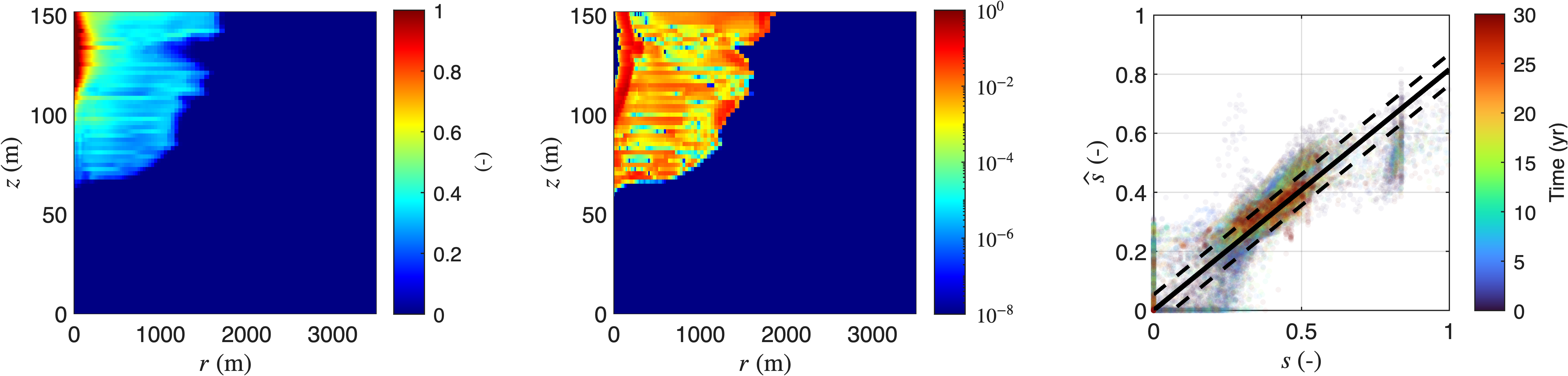}
    \modelcaption{U-FNO ($R^2 = 0.9312$, RMSE $= 3.9 \times 10^{-2}$)}
    \caption{CO$_2$ saturation field predictions for Test Case 2: (a) predicted field $\hat{s}$ at $t = 30.0$ yr, (b) normalized absolute error $|s - \hat{s}|/|s|_{\max}$ at $t = 30.0$ yr, and (c) parity plot over the full 30-year injection period.}
    \label{fig:sat_test_2_results}
\end{figure}

The predictive performance of each surrogate model is further evaluated on two representative test cases. The input fields and ground truth CO$_2$ saturation fields for Test Cases 1 and 2 are shown in Figures~\ref{fig:Inputs_GT_Test_1} and~\ref{fig:Inputs_GT_Test_2}, respectively. The predicted and true CO$_2$ saturation fields are compared over the full 30-year injection period in Figures~\ref{fig:sat_test_1_results} and~\ref{fig:sat_test_2_results}.

All models perform reasonably well with $R^2 > 0.94$ for Test Case 1. The U-FNO produces the sharpest plume fronts, while the U-Net and V-Net show slightly more diffuse predictions. The Temporal CNN, FNO, and U-FNO tend to have the largest errors at the fully saturated plume front; in contrast, the U-Net and V-Net architectures tend to have larger errors in the partially saturated region. This differing behavior can be explained by the lack of skip connections in the Temporal CNN and FNO and their ability to resolve sharp discontinuities in the fully saturated front, while the localized nature of skip connections introduces errors in more diffuse areas of the partially saturated region. From Test Case 2, it is clear that performance drops across all models with $R^2$ values between 0.89 and 0.93, with the U-FNO maintaining the best performance.

\subsubsection{Frequency-Domain Analysis}
\begin{figure}[!htbp]
    \centering
    \begin{subfigure}[b]{0.45\linewidth}
        \centering
        \includegraphics[width=\linewidth]{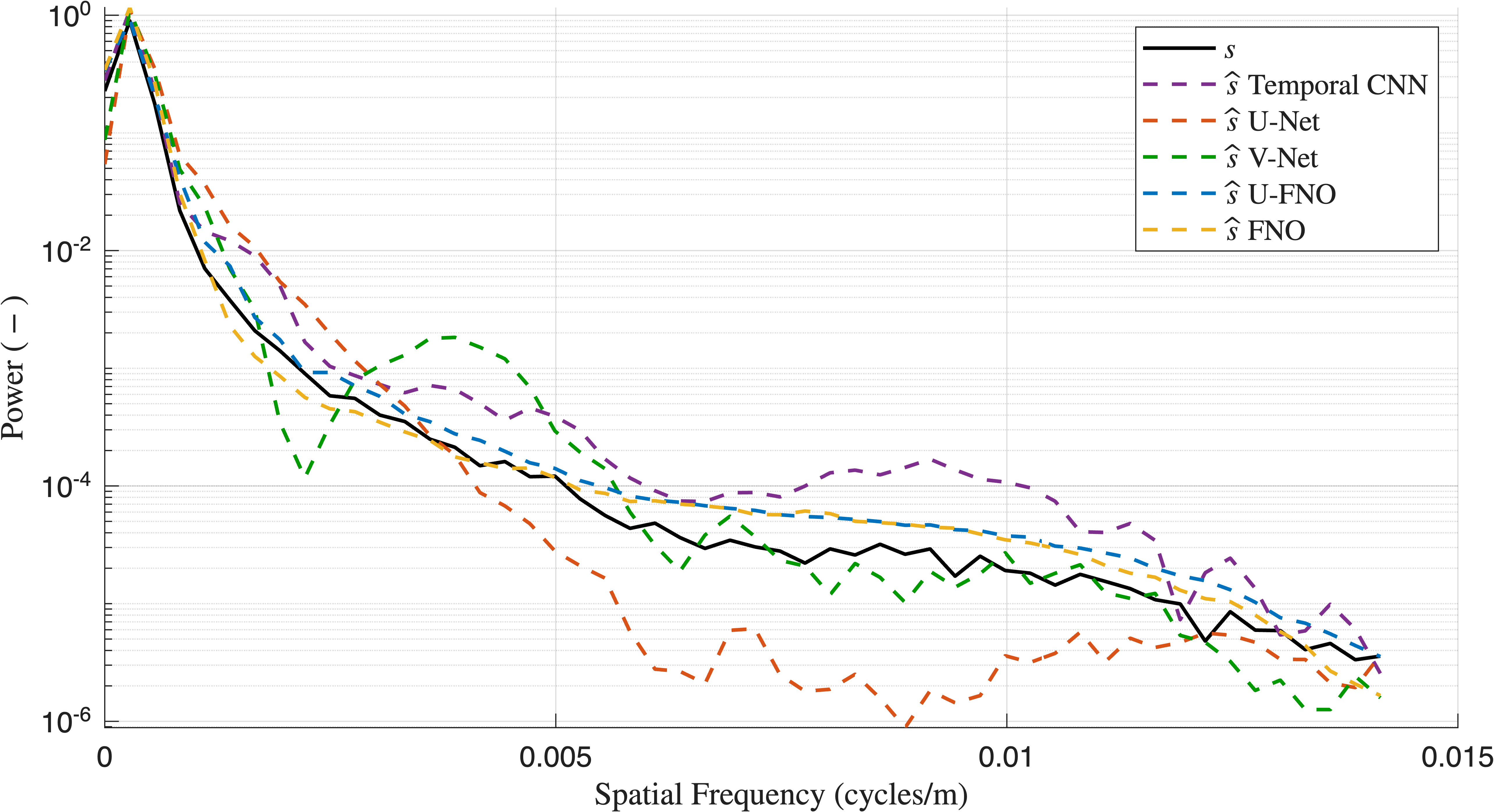}
        \caption{$r$-direction, $t = 30.0$ yr}
    \end{subfigure}
    \hfill
    \begin{subfigure}[b]{0.45\linewidth}
        \centering
        \includegraphics[width=\linewidth]{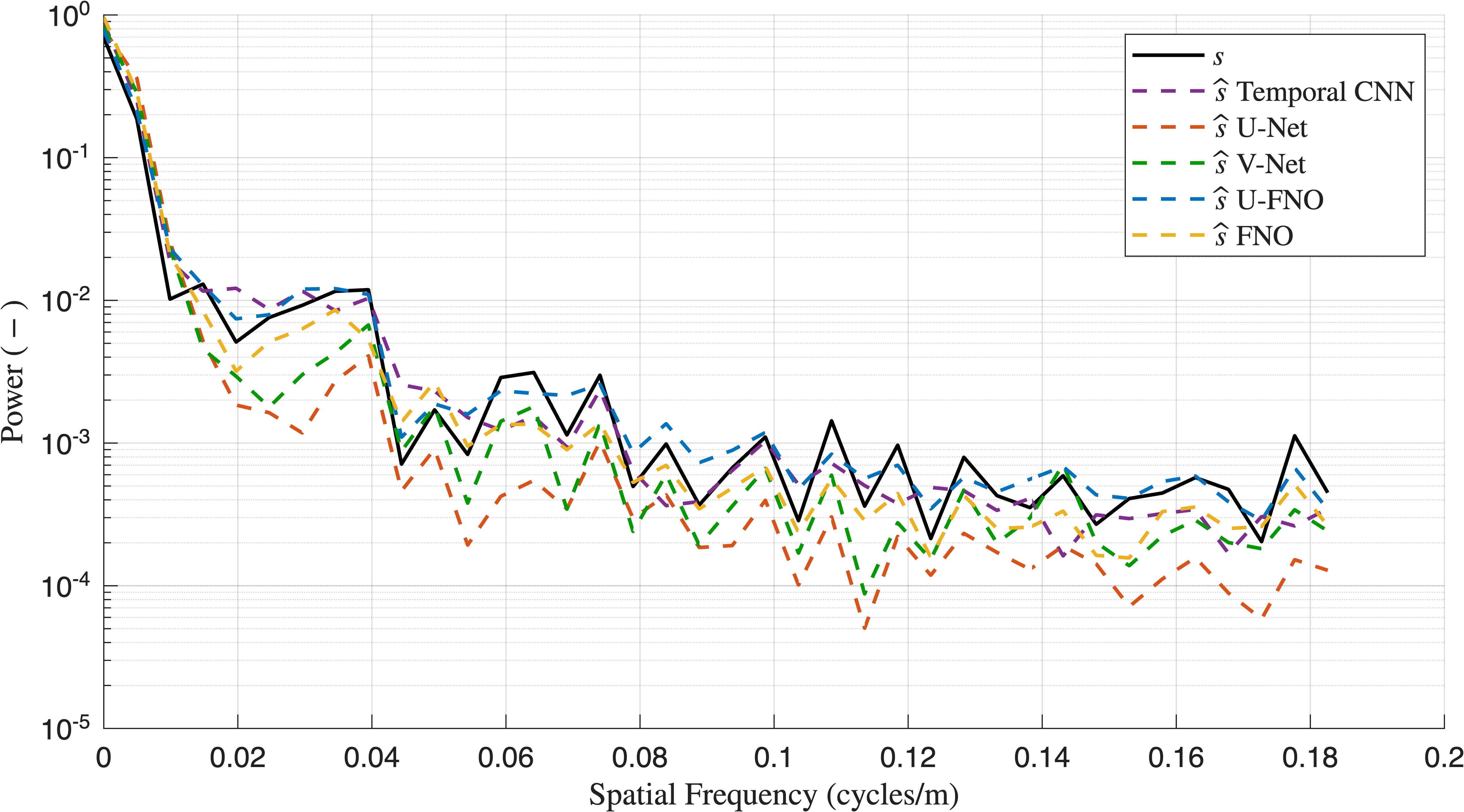}
        \caption{$z$-direction, $t = 30.0$ yr}
    \end{subfigure}
    \caption{Power spectral density of the CO$_2$ saturation field $s$ at $t = 30.0$ yr for Test Case 1.}
    \label{fig:psd_test_1}
\end{figure}

\begin{figure}[!htbp]
    \centering
    \begin{subfigure}[b]{0.45\linewidth}
        \centering
        \includegraphics[width=\linewidth]{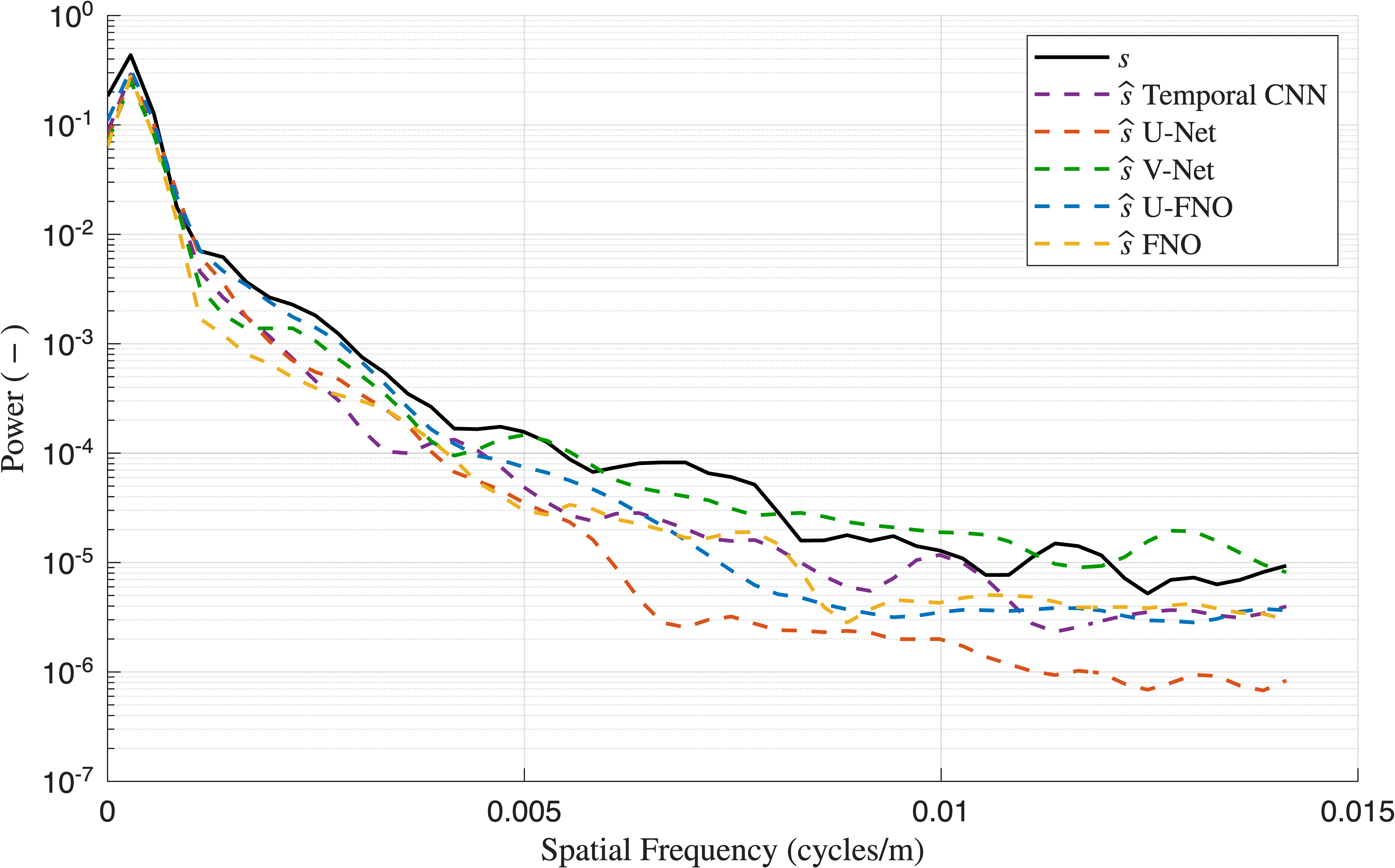}
        \caption{$r$-direction, $t = 30.0$ yr}
    \end{subfigure}
    \hfill
    \begin{subfigure}[b]{0.45\linewidth}
        \centering
        \includegraphics[width=\linewidth]{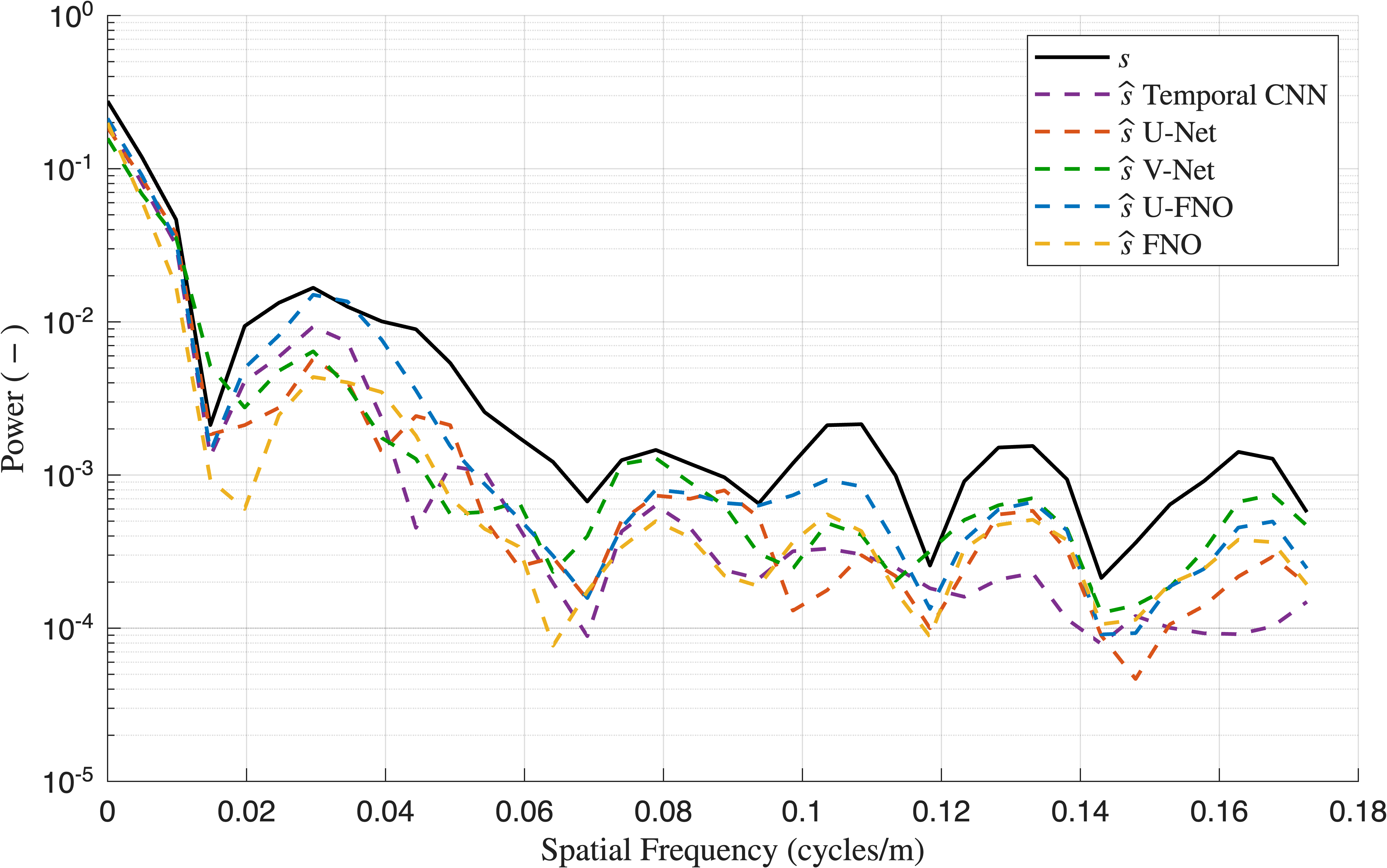}
        \caption{$z$-direction, $t = 30.0$ yr}
    \end{subfigure}
    \caption{Power spectral density of the CO$_2$ saturation field $s$ at $t = 30.0$ yr for Test Case 2.}
    \label{fig:psd_test_2}
\end{figure}

The frequency-domain behavior of each model is assessed using the Power Spectral Density (PSD) of the predicted and true fields, as shown in Figures~\ref{fig:psd_test_1} and~\ref{fig:psd_test_2}.

The PSD curves demonstrate that all models capture the low-frequency content well but diverge at higher frequencies. The U-FNO is better able to capture the high-frequency modes compared to the other architectures, as shown in Figures~\ref{fig:psd_test_1} and~\ref{fig:psd_test_2}. It should also be noted that the PSD decays more slowly in the $z$-direction than in the $r$-direction, reflecting the higher-frequency content associated with the heterogeneous and anisotropic nature of the vertical permeability field. Consequently, capturing viscous fingering and buoyancy-driven effects requires a surrogate model capable of resolving a broader spectrum of spatial frequencies.

\subsection{Pressure Build-up Field}
This section presents the predictive performance of the five surrogate model architectures for the pressure build-up field (i.e., elliptic PDE). A similar analysis to that described in Section~\ref{sec:CO2_sat} is performed, including a hyperparameter sensitivity study of the loss function, representative test case comparisons, frequency-domain analysis, and permeability sensitivity.

\subsubsection{Model Performance}
Tables~\ref{tab:pressure_results_asymmetric} and~\ref{tab:pressure_results_symmetric} summarize the performance of all trained surrogate models for the pressure build-up field over the full parametric range of the proposed loss function. The results demonstrate that the FNO achieves the highest predictive accuracy on the test dataset, followed by the Temporal CNN, U-FNO, V-Net, and U-Net. These results are consistent with the expected behaviour of each architecture, as the pressure build-up field is governed by an elliptic PDE consisting of a smooth, globally coupled field that is more diffuse and easier to approximate compared to a hyperbolic PDE. The Temporal CNN and FNO architectures produce the most accurate surrogates for this problem as their lack of skip connections propagates information globally, which leads to a more diffuse approximation. Models such as the U-FNO, U-Net, and V-Net contain skip connections, which retain higher-frequency spatial modes that are inconsistent with the low-frequency dominated structure of the pressure field. It is also observed that the performance difference between the U-Net and V-Net is marginal. This can be attributed to the smooth nature of the pressure field, which reduces the impact of the architectural differences between the learnable strided convolution downsampling of the V-Net and the fixed max pooling of the U-Net.

From these results, it is also observed that FNO-based architectures benefit from the full gradient loss function, as it provides additional context on the global behavior of the pressure build-up field. In contrast, no consistent improvement from the full gradient loss function is observed for convolutional architectures such as the Temporal CNN, U-Net, and V-Net. This is attributed to the pressure field being globally smooth, where gradients in the $z$-direction are not significant compared to the pressure gradient component associated with radial flow. Furthermore, convolutional models such as the U-Net and V-Net are not expected to perform well for this problem, as their reliance on localized high-frequency features is not well-suited to the prediction of the smooth, low-frequency pressure build-up field. Overall, the FNO trained with $\lambda_r = \lambda_z = 1.0$ achieves the best predictive accuracy with a test $R^2$ of 0.972.

The computational cost of all surrogate models for the pressure build-up field is summarized in Table~\ref{tab:computational_cost}. The training costs are consistent with those reported for the CO$_2$ saturation field, as the same model architectures and training configurations are used. It is noted that the U-Net and V-Net required longer training times for the pressure build-up field, as these models were trained using the Compute Canada Vulcan Cluster with an NVIDIA L40S GPU with 48~GB of memory, rather than the Trillium Cluster with an NVIDIA H100 GPU with 80~GB of memory.

\begin{table}[!htbp]
\scriptsize
\centering
\begin{tabular}{llccc}
\toprule
\multirow{2}{*}{\textbf{Model}} & \multirow{2}{*}{\textbf{Parameters}} &
\multirow{2}{*}{\textbf{\makecell{Forward/Backward \\ Pass Memory (MB)}}} &
\multicolumn{2}{c}{\textbf{Training Cost}} \\
\cmidrule(lr){4-5}
& & & \textbf{GPU Type} & \textbf{Time (hr)} \\
\midrule
Temporal CNN & 36,850,017 & 1,565.61 & NVIDIA H100 -- 80 GB & 16.1 \\
U-Net        & 26,894,017 & 1,152.69 & NVIDIA L40S -- 48 GB & 28.8 \\
V-Net        & 27,270,785 & 1,198.83 & NVIDIA L40S -- 48 GB & 30.2 \\
FNO          & 31,117,325 & 1,575.46 & NVIDIA H100 -- 80 GB & 28.0 \\
U-FNO        & 33,097,829 & 2,980.11 & NVIDIA H100 -- 80 GB & 63.3 \\
\bottomrule
\end{tabular}
\caption{Summary of model architecture and training cost for pressure build-up field.}
\label{tab:computational_cost}
\end{table}

\begin{figure}[!htbp]
    \centering
    \begin{subfigure}[t]{0.45\textwidth}
        \centering
        \includegraphics[width=\textwidth]{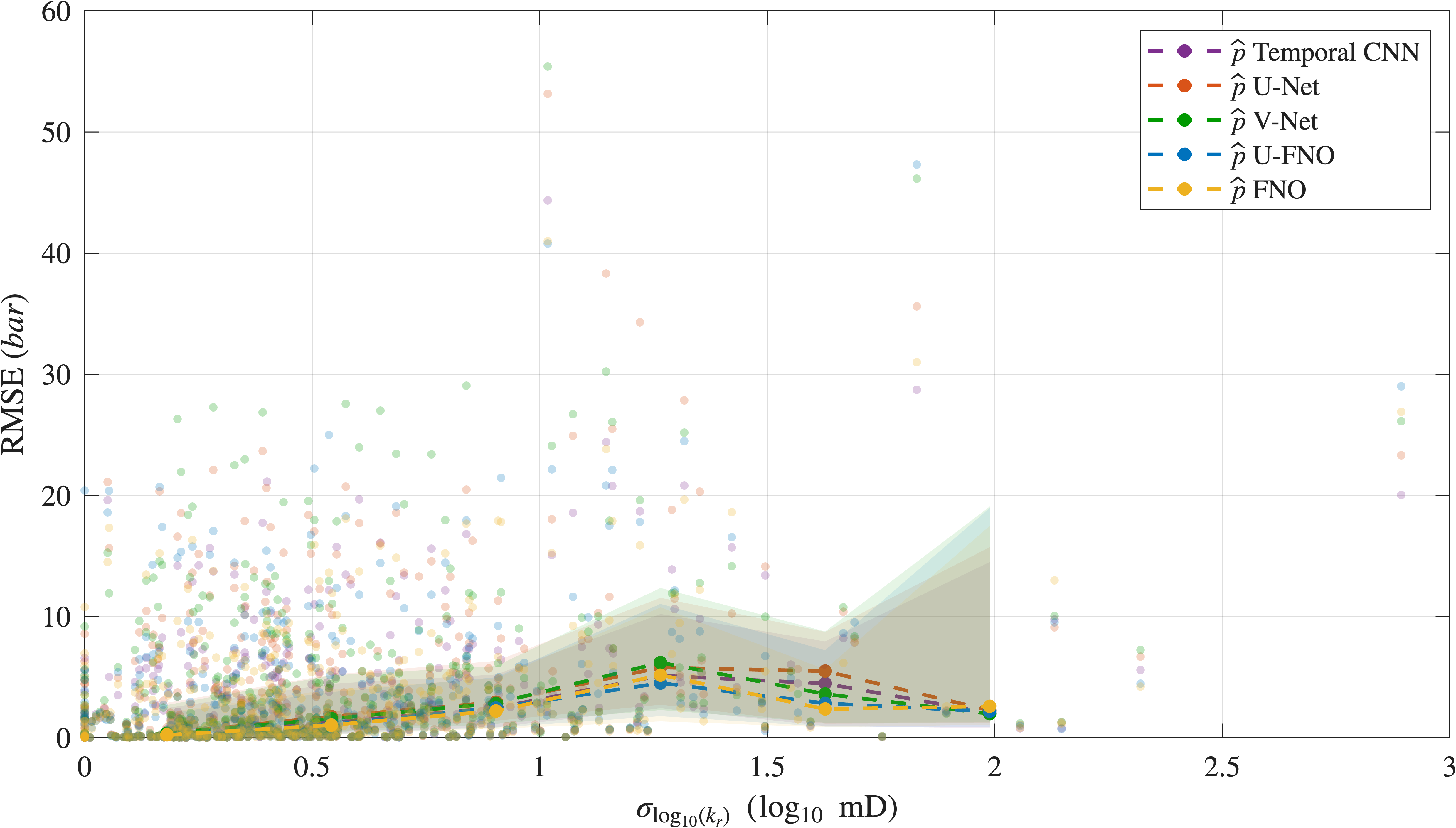}
        \caption{RMSE vs $\sigma_{\log_{10}(k_r)}$}
        \label{fig:dP_perm_r_std}
    \end{subfigure}
    \hfill
    \begin{subfigure}[t]{0.45\textwidth}
        \centering
        \includegraphics[width=\textwidth]{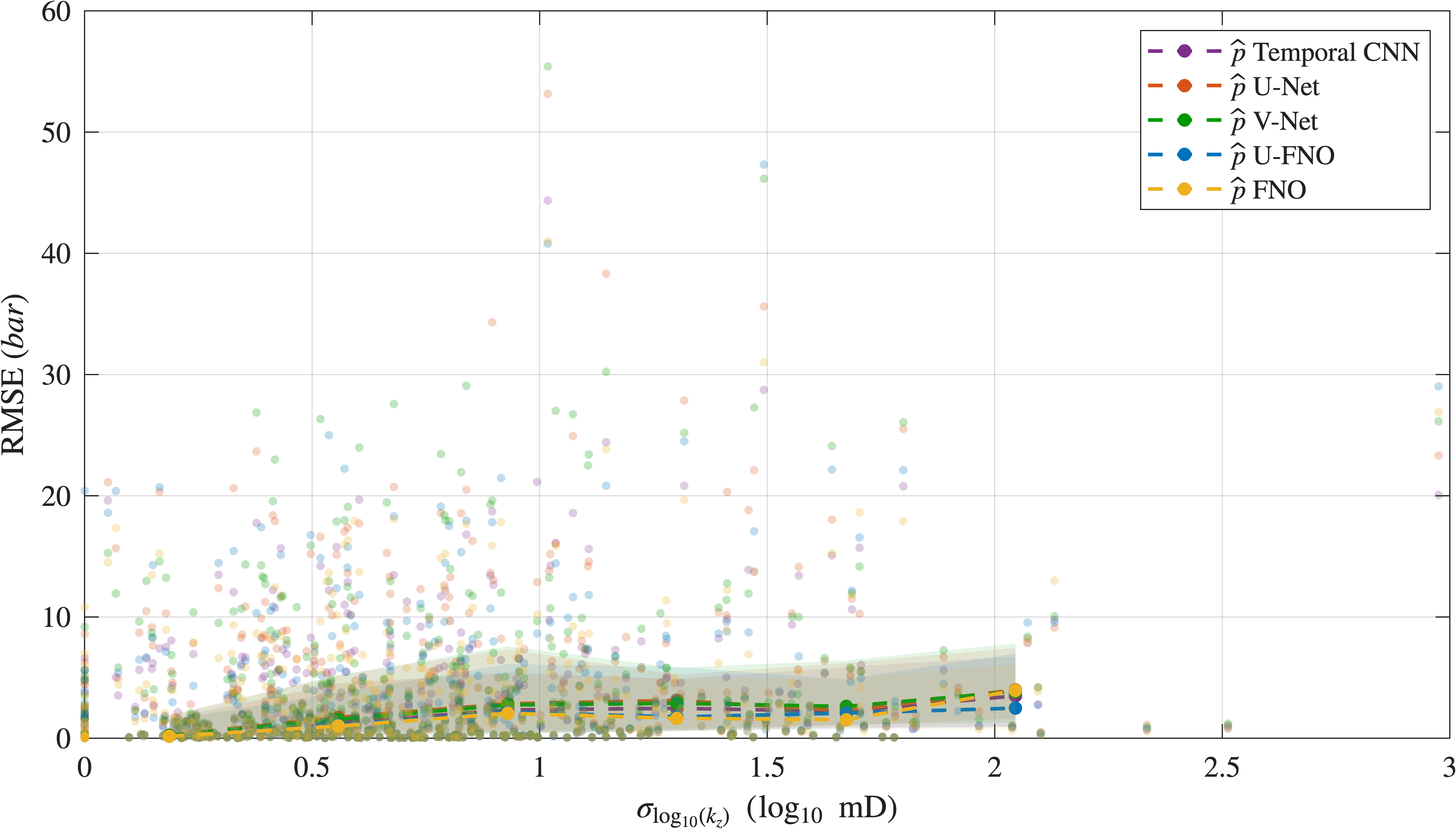}
        \caption{RMSE vs $\sigma_{\log_{10}(k_z)}$}
        \label{fig:dP_perm_z_std}
    \end{subfigure}
    \caption{Test dataset root mean square error (RMSE) vs.\ standard deviation in permeability field for all models for pressure build-up field.}
    \label{fig:dP_perm_all}
\end{figure}

The sensitivity of model performance to permeability heterogeneity for the pressure build-up field is shown in Figure~\ref{fig:dP_perm_all}. In contrast to the CO$_2$ saturation field, all models exhibit relatively uniform performance across the full range of permeability variability. This is consistent with the elliptic nature of the pressure field, which is governed by global pressure gradients and is less sensitive to local high-frequency permeability heterogeneity. The FNO and Temporal CNN maintain the lowest RMSE across the full range, while the U-Net and V-Net show higher errors, mainly associated with architectural performance.

\subsubsection{Representative Test Cases}
\begin{figure}[!htbp]
    \centering
    \begin{subfigure}[b]{0.24\linewidth}
        \centering
        \includegraphics[width=\linewidth]{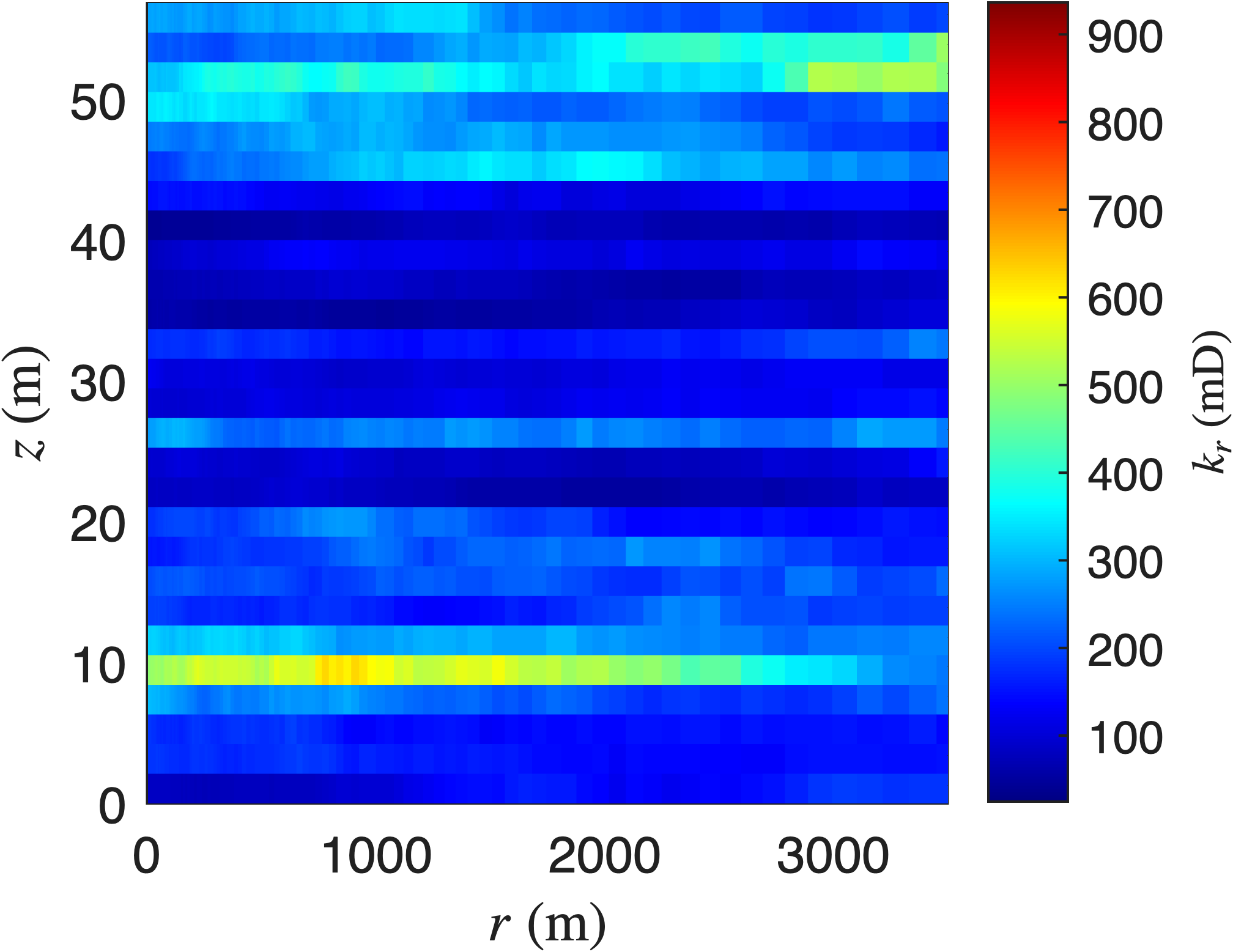}
        \caption{$k_r$ (mD)}
    \end{subfigure}
    \hfill
    \begin{subfigure}[b]{0.24\linewidth}
        \centering
        \includegraphics[width=\linewidth]{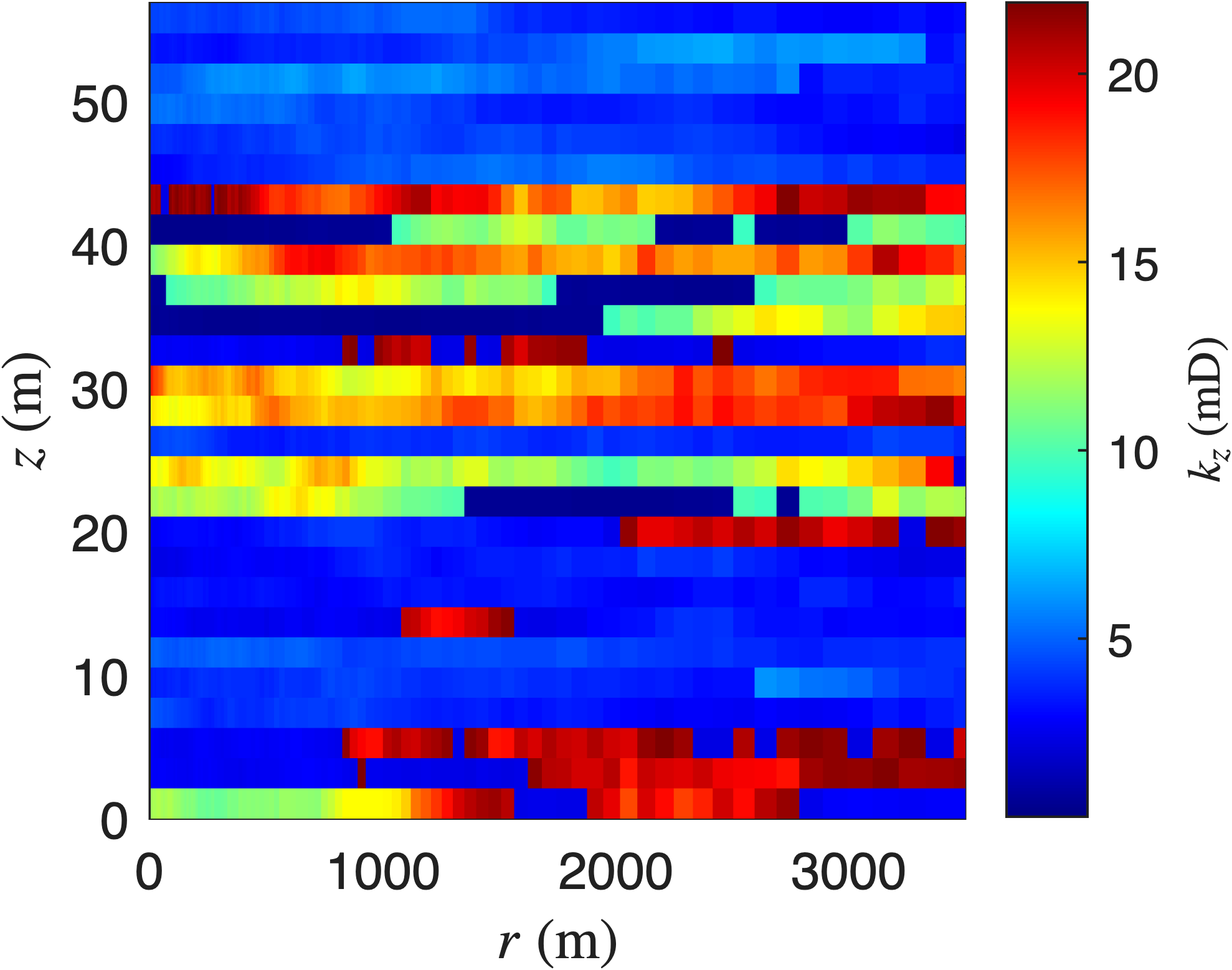}
        \caption{$k_z$ (mD)}
    \end{subfigure}
    \hfill
    \begin{subfigure}[b]{0.25\linewidth}
        \centering
        \includegraphics[width=\linewidth]{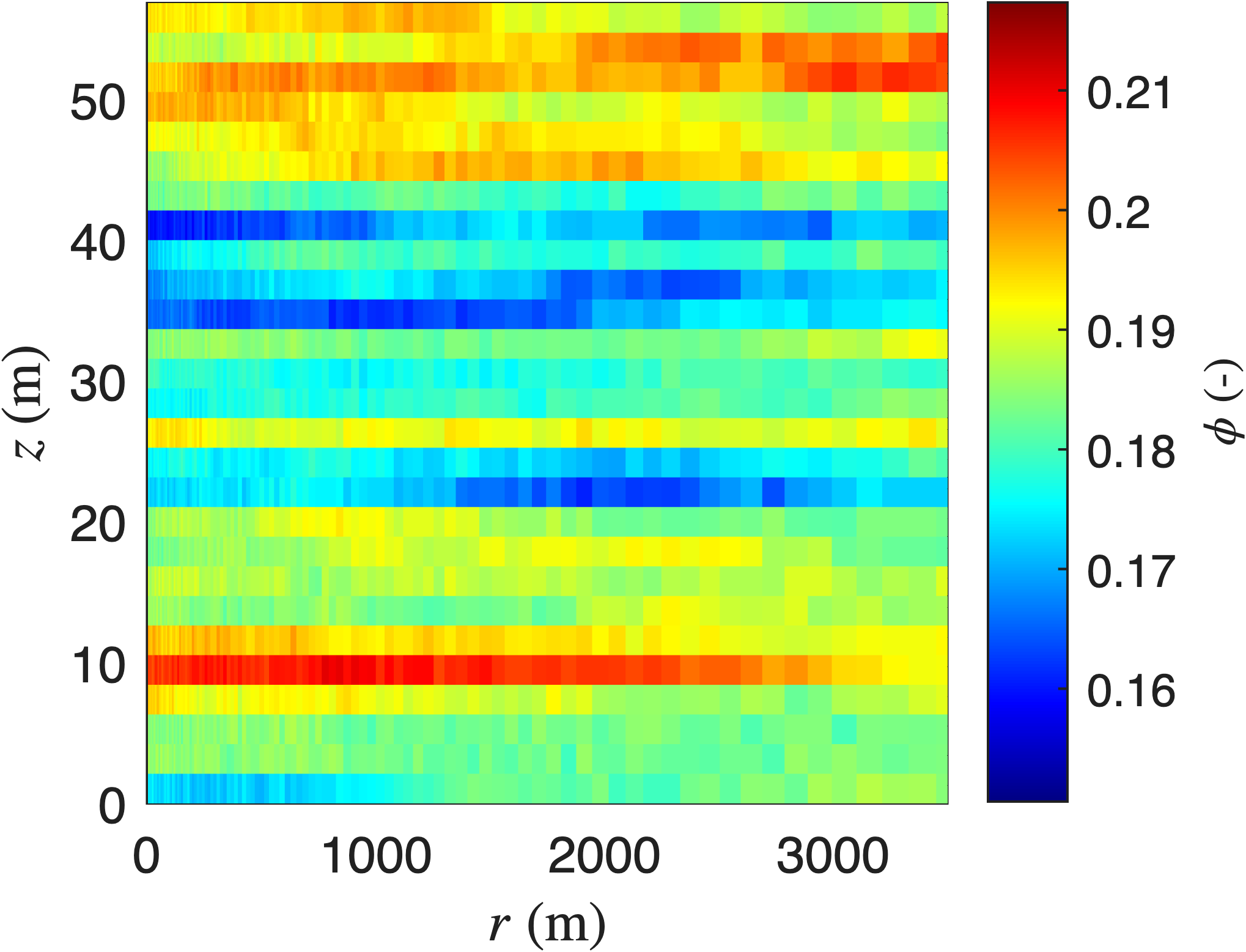}
        \caption{$\phi$ (-)}
    \end{subfigure}
    \hfill
    \begin{subfigure}[b]{0.24\linewidth}
        \centering
        \includegraphics[width=\linewidth]{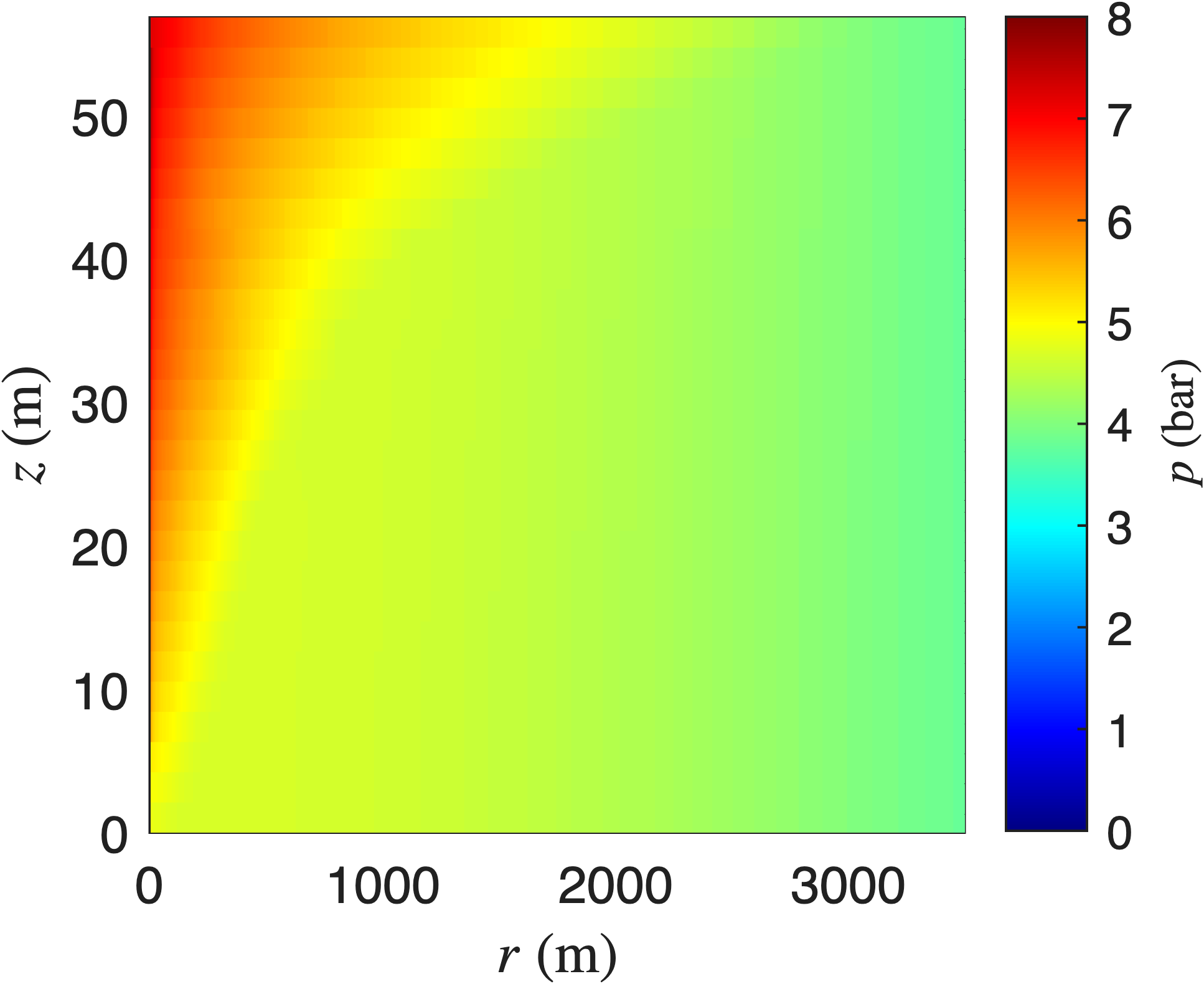}
        \caption{$p$ (bar)}
    \end{subfigure}
    \caption{Input fields and ground truth pressure build-up field ($t = 30.0$ yr) for Test Case 1. Scalar inputs: $Q_\text{inj} = 0.49$ MT/yr,\; $T = 87.9\,^\circ$C,\; $P_\text{init} = 159.0$ bar,\; $S_{wi} = 0.12$,\; $\lambda = 0.41$.}
    \label{fig:Inputs_GT_dP_Test_1}
\end{figure}

\begin{figure}[!htbp]
    \centering
    \begin{subfigure}[b]{0.24\linewidth}
        \centering
        \includegraphics[width=\linewidth]{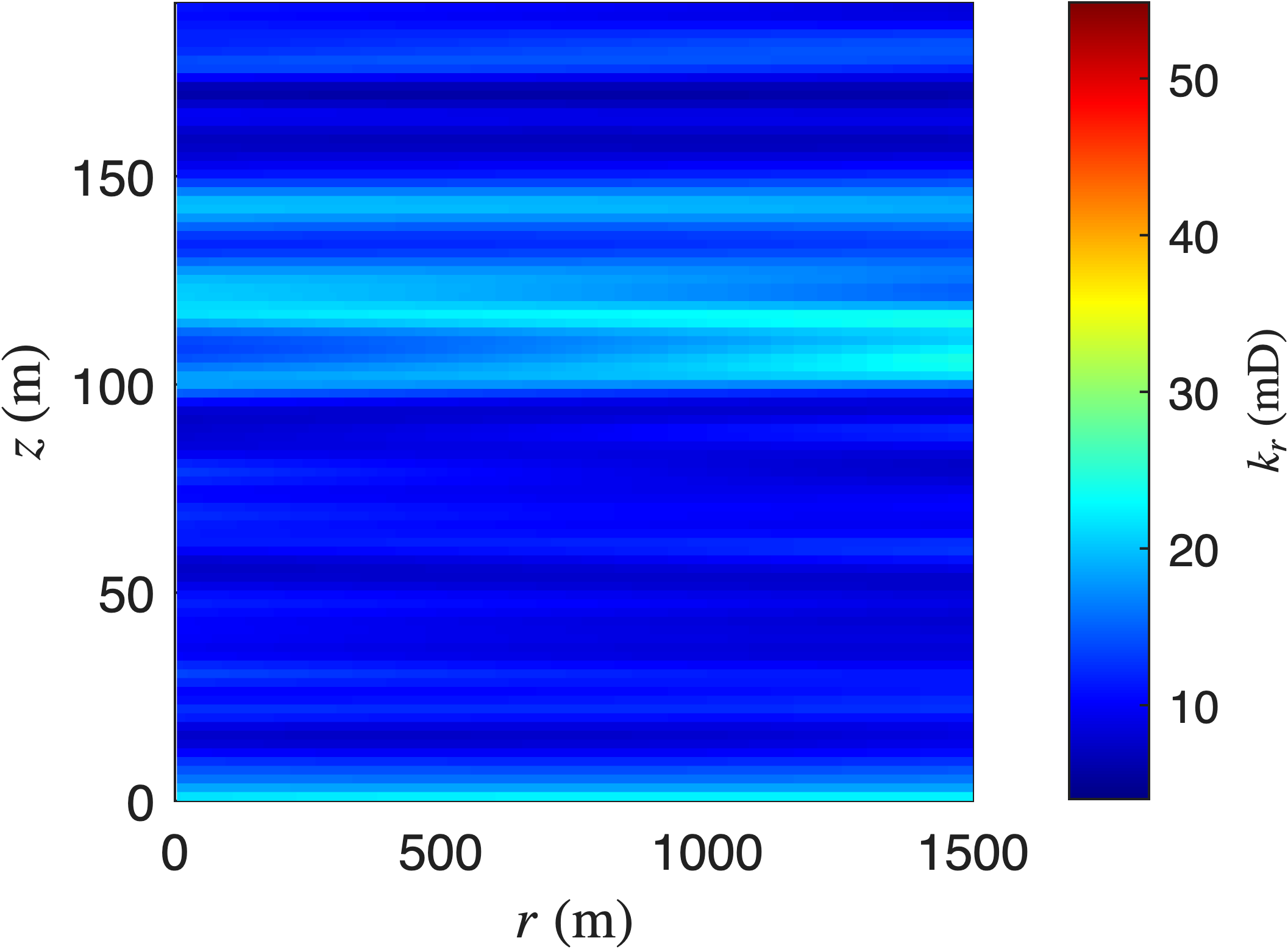}
        \caption{$k_r$ (mD)}
    \end{subfigure}
    \hfill
    \begin{subfigure}[b]{0.24\linewidth}
        \centering
        \includegraphics[width=\linewidth]{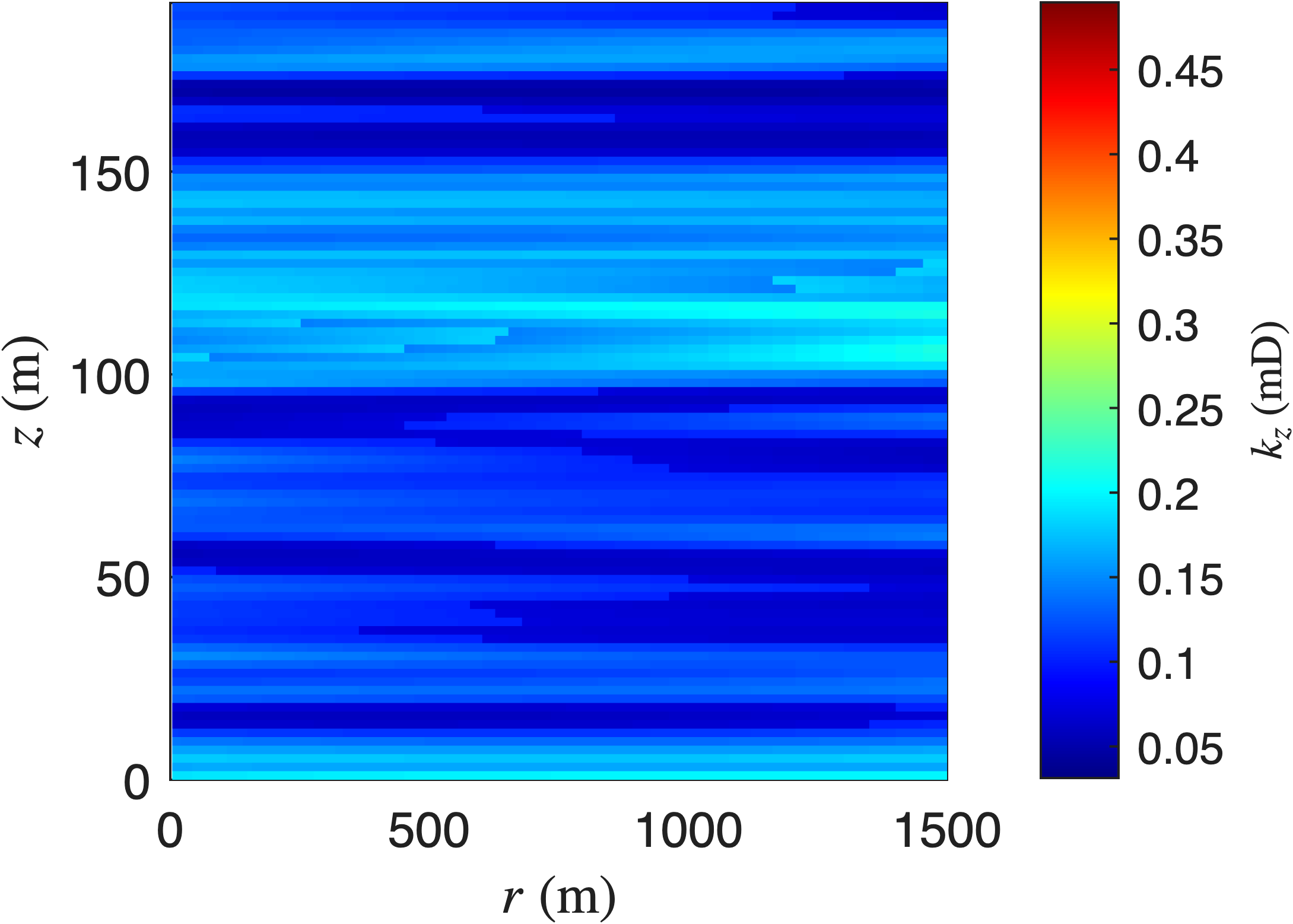}
        \caption{$k_z$ (mD)}
    \end{subfigure}
    \hfill
    \begin{subfigure}[b]{0.25\linewidth}
        \centering
        \includegraphics[width=\linewidth]{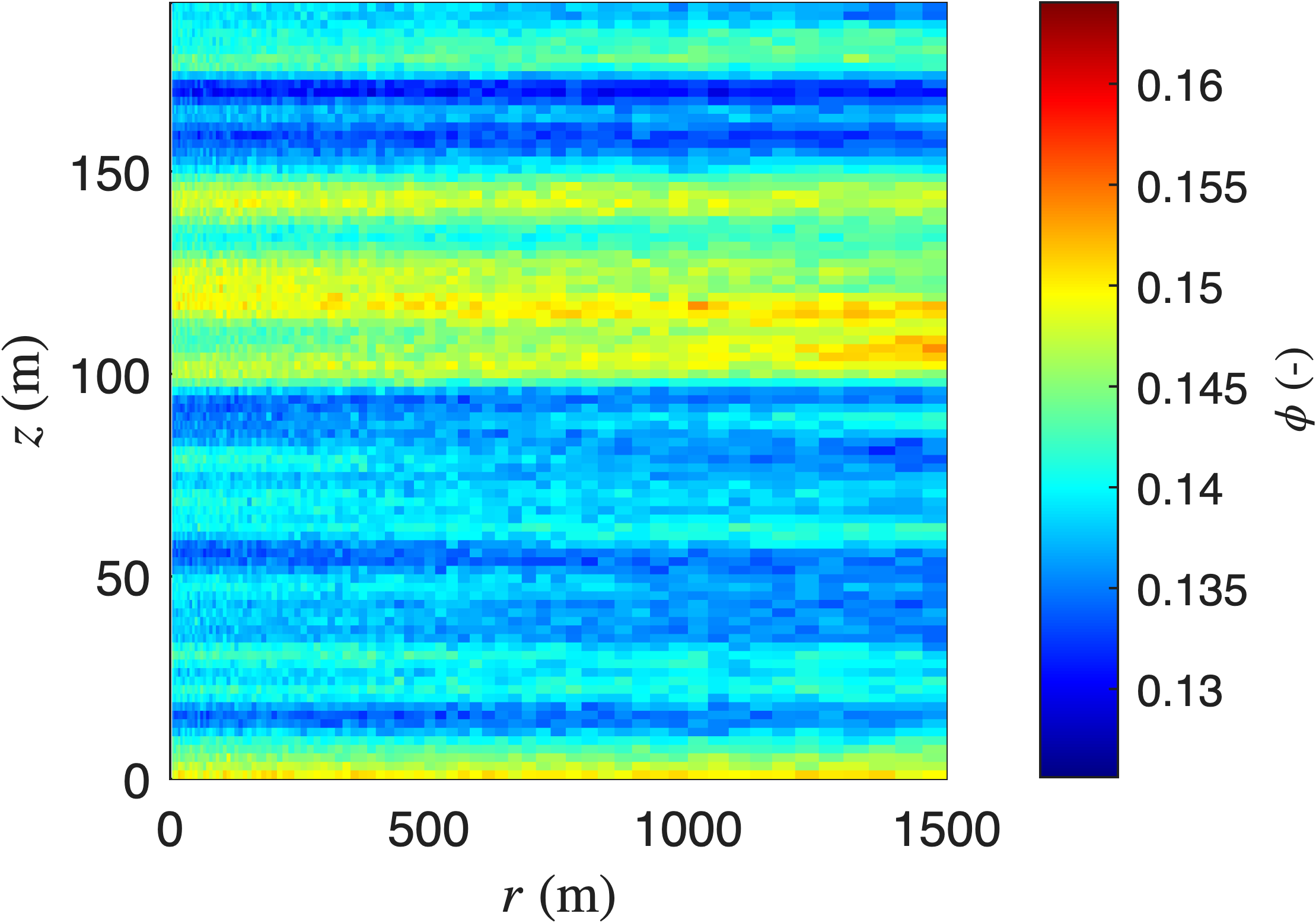}
        \caption{$\phi$ (-)}
    \end{subfigure}
    \hfill
    \begin{subfigure}[b]{0.24\linewidth}
        \centering
        \includegraphics[width=\linewidth]{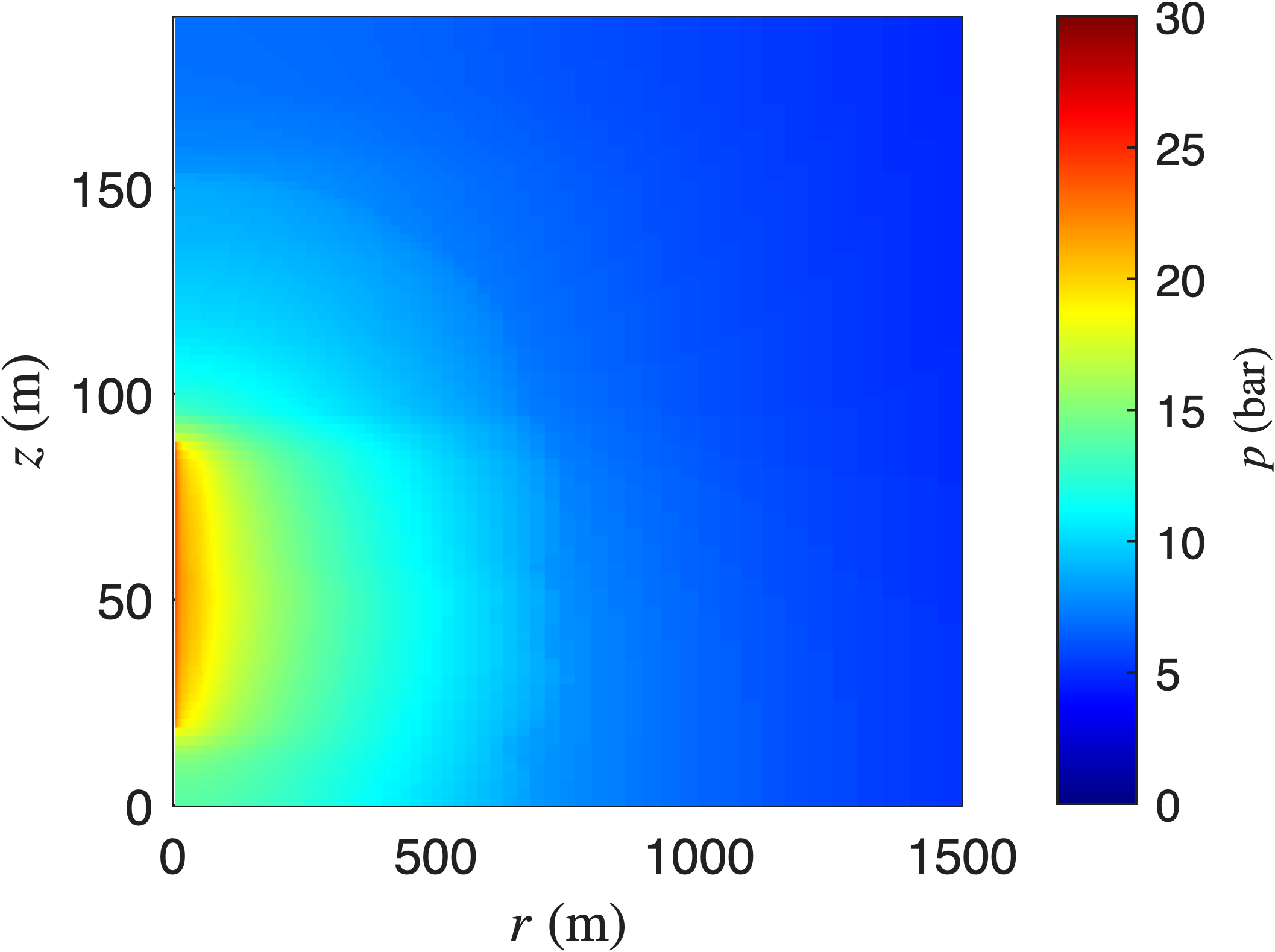}
        \caption{$p$ (bar)}
    \end{subfigure}
    \caption{Input fields and ground truth pressure build-up field ($t = 30.0$ yr) for Test Case 2. Scalar inputs: $Q_\text{inj} = 0.21$ MT/yr,\; $T = 57.0\,^\circ$C,\; $P_\text{init} = 156.4$ bar,\; $S_{wi} = 0.21$,\; $\lambda = 0.64$.}
    \label{fig:Inputs_GT_dP_Test_2}
\end{figure}

\begin{figure}[!tbp]
    \centering
    \includegraphics[width=0.74\linewidth]{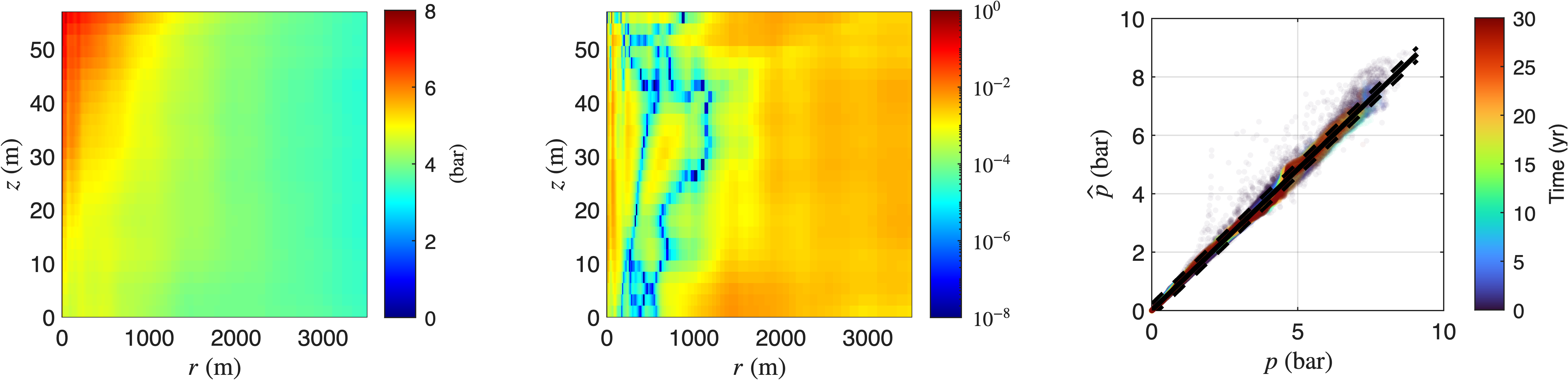}
    \modelcaption{Temporal CNN ($R^2 = 0.9954$, RMSE $= 1.46$ bar)}
    \includegraphics[width=0.74\linewidth]{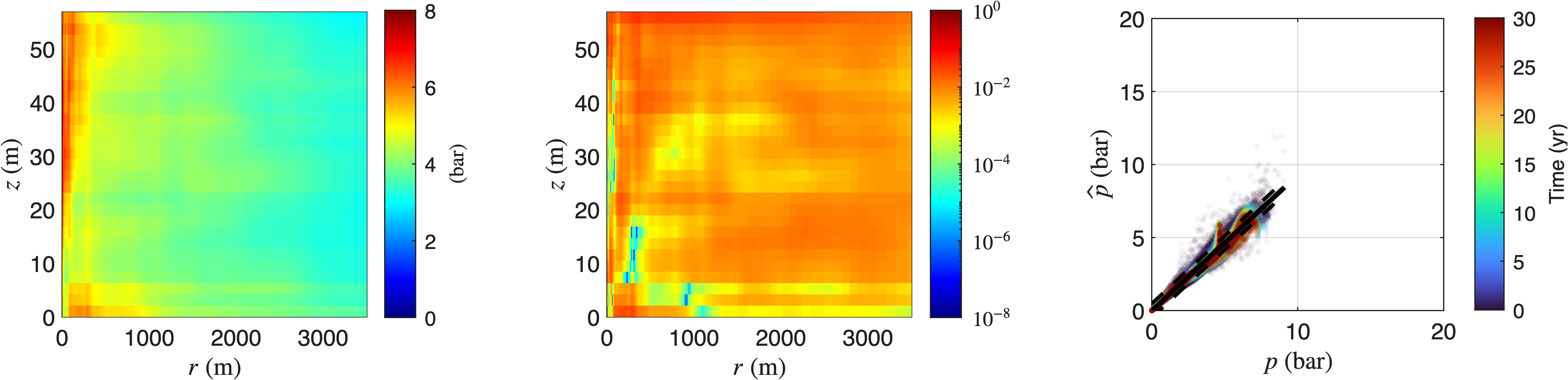}
    \modelcaption{U-Net ($R^2 = 0.9786$, RMSE $= 3.14$ bar)}
    \includegraphics[width=0.74\linewidth]{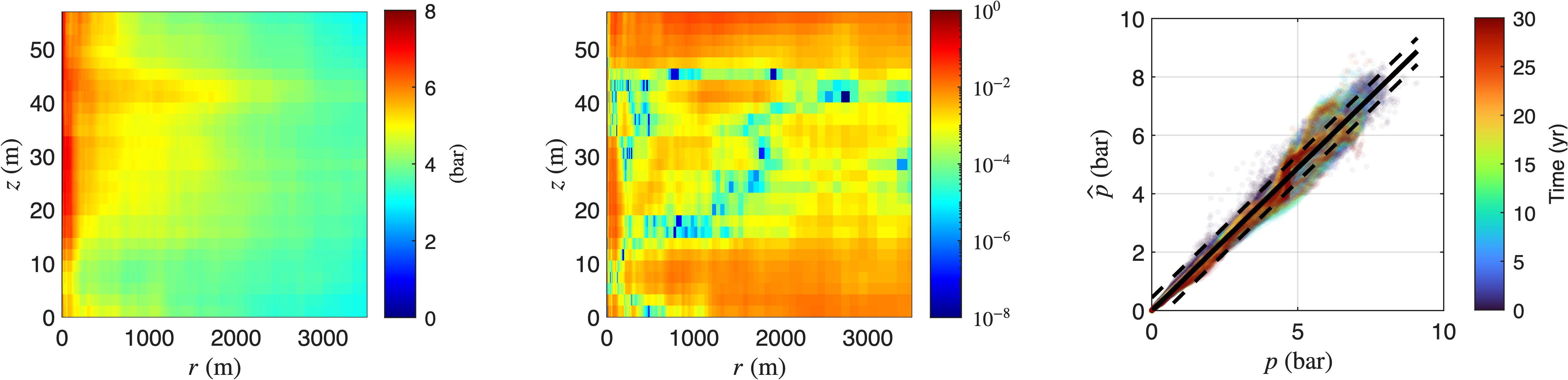}
    \modelcaption{V-Net ($R^2 = 0.9874$, RMSE $= 2.41$ bar)}
    \includegraphics[width=0.74\linewidth]{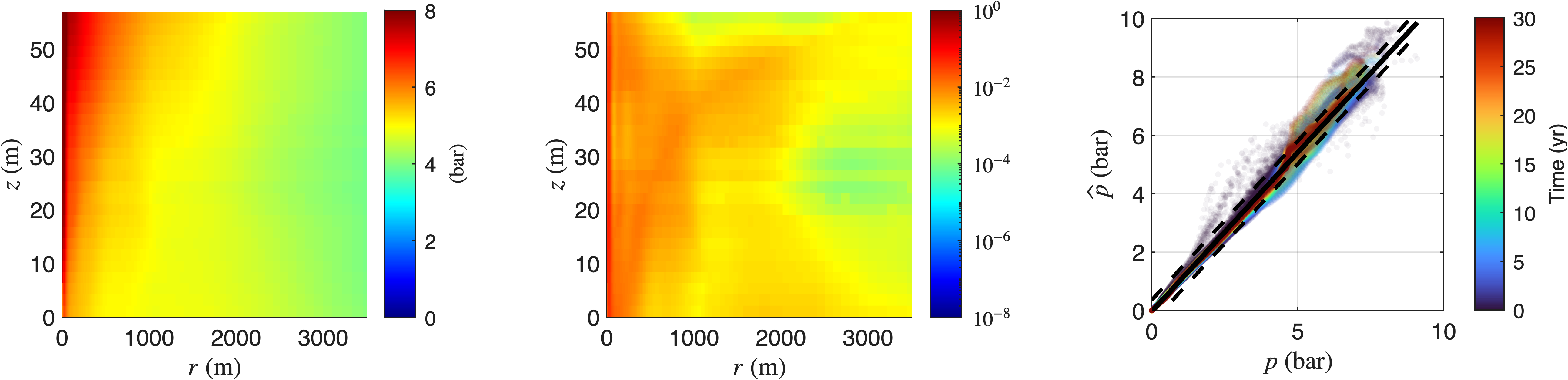}
    \modelcaption{FNO ($R^2 = 0.9802$, RMSE $= 3.02$ bar)}
    \includegraphics[width=0.74\linewidth]{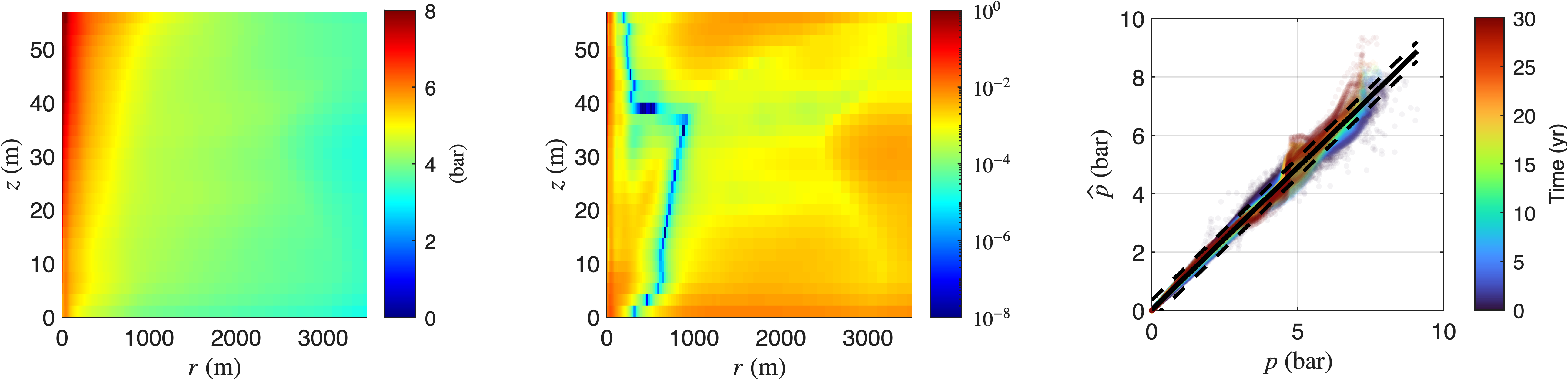}
    \modelcaption{U-FNO ($R^2 = 0.9932$, RMSE $= 1.77$ bar)}
    \caption{Pressure build-up field predictions for Test Case 1: (a) predicted field $\hat{p}$ at $t = 30.0$ yr, (b) normalized absolute error $|p - \hat{p}|/|p|_{\max}$ at $t = 30.0$ yr, and (c) parity plot over the full 30-year injection period.}
    \label{fig:dP_test_1_results}
\end{figure}

\begin{figure}[!tbp]
    \centering
    \includegraphics[width=0.74\linewidth]{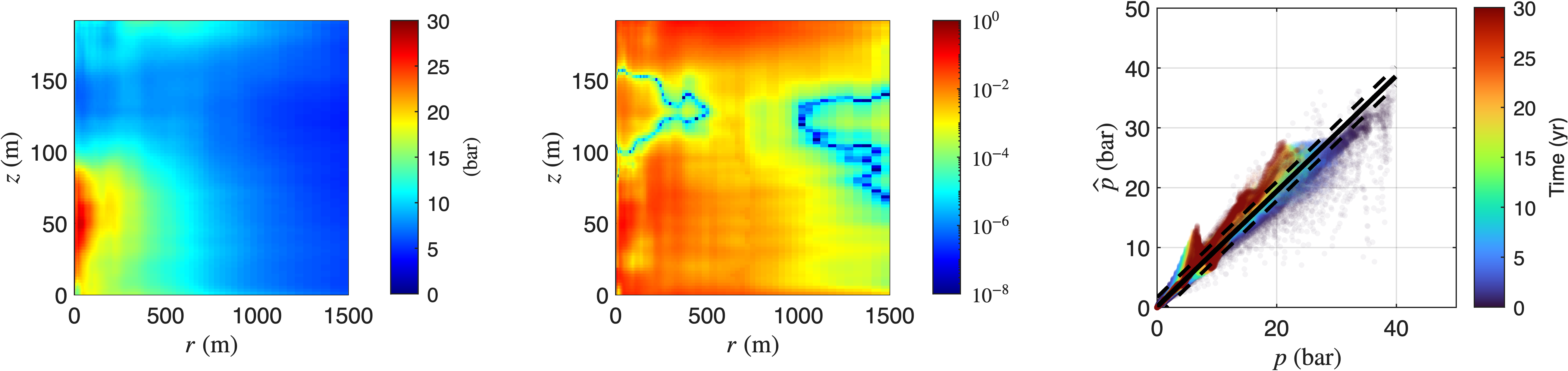}
    \modelcaption{Temporal CNN ($R^2 = 0.9721$, RMSE $= 0.827$ bar)}
    \includegraphics[width=0.74\linewidth]{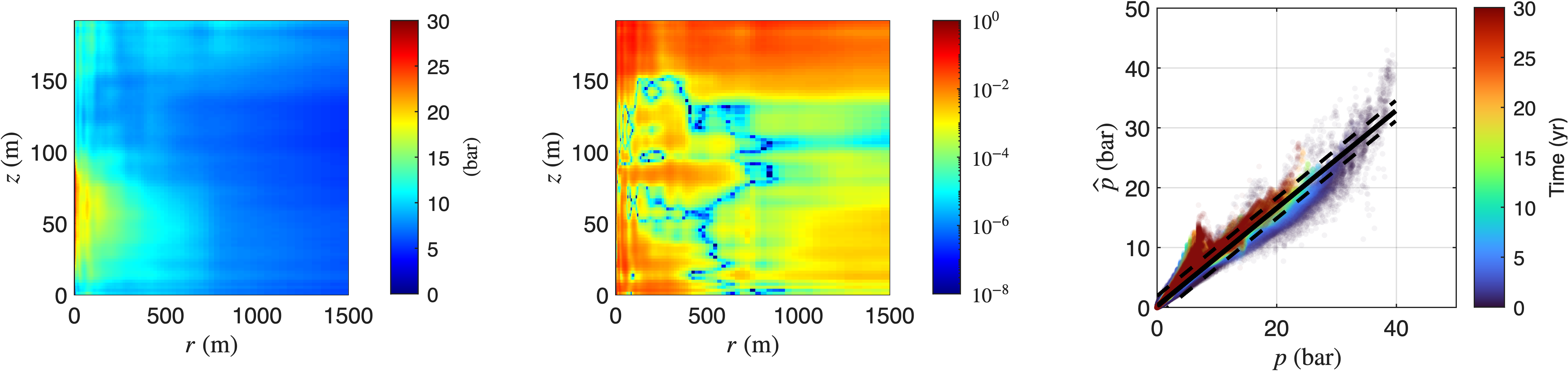}
    \modelcaption{U-Net ($R^2 = 0.9350$, RMSE $= 1.26$ bar)}
    \includegraphics[width=0.74\linewidth]{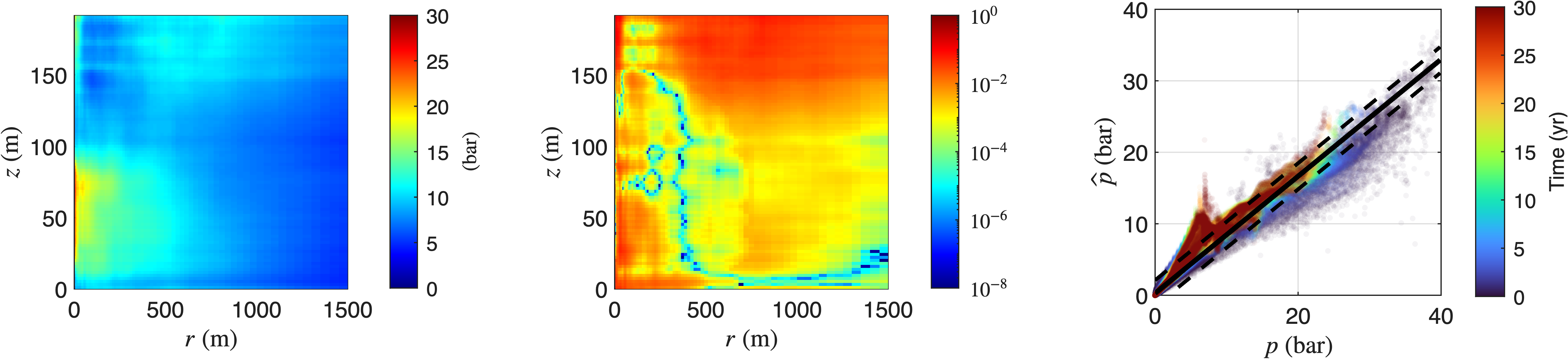}
    \modelcaption{V-Net ($R^2 = 0.9313$, RMSE $= 1.30$ bar)}
    \includegraphics[width=0.74\linewidth]{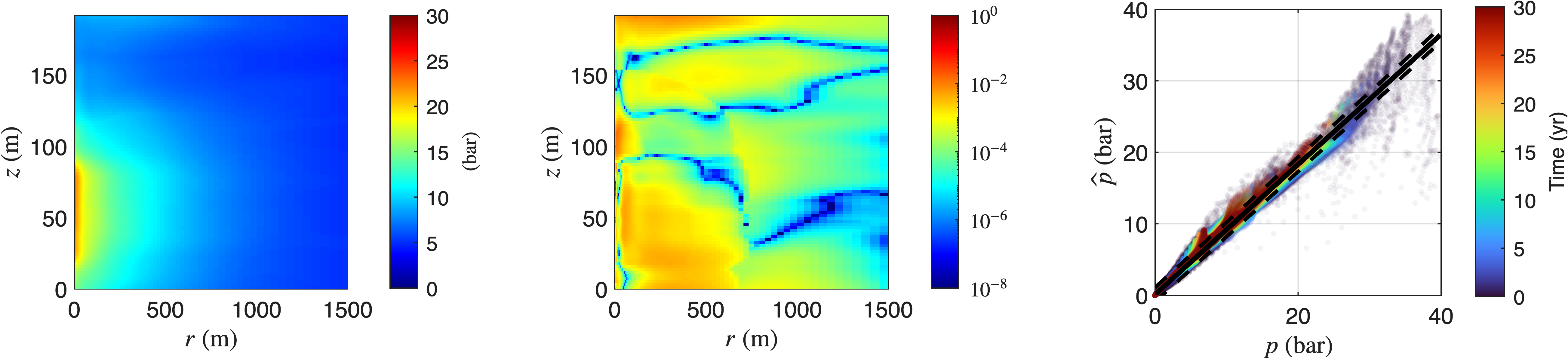}
    \modelcaption{FNO ($R^2 = 0.9809$, RMSE $= 0.685$ bar)}
    \includegraphics[width=0.74\linewidth]{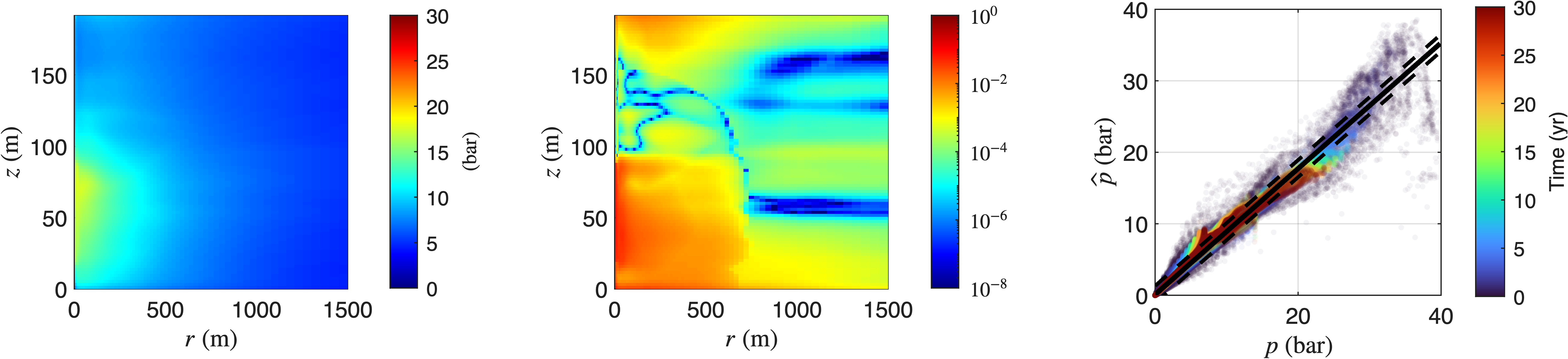}
    \modelcaption{U-FNO ($R^2 = 0.9696$, RMSE $= 0.864$ bar)}
    \caption{Pressure build-up field predictions for Test Case 2: (a) predicted field $\hat{p}$ at $t = 30.0$ yr, (b) normalized absolute error $|p - \hat{p}|/|p|_{\max}$ at $t = 30.0$ yr, and (c) parity plot over the full 30-year injection period.}
    \label{fig:dP_test_2_results}
\end{figure}

The predictive performance of each surrogate model is further evaluated on two representative test cases. The input fields and ground truth pressure build-up fields for Test Cases 1 and 2 are shown in Figures~\ref{fig:Inputs_GT_dP_Test_1} and~\ref{fig:Inputs_GT_dP_Test_2}, respectively. The predicted and true pressure build-up fields are compared at the end of the 30-year injection period in Figures~\ref{fig:dP_test_1_results} and~\ref{fig:dP_test_2_results}. The parity plots illustrate the predictive performance over the entire injection period, demonstrating each model's ability to capture the temporal dynamics of the pressure build-up field.

The Temporal CNN, FNO, and U-FNO achieve accurate predictions for both test cases, with $R^2 > 0.96$. In both cases, the FNO yields the most consistent predictions of the global pressure field and best captures the large pressure gradient near the wellbore, as reflected by the tighter parity plots over the full temporal evolution. It is also observed that convolutional architectures such as the U-Net and V-Net are not well suited for surrogate modeling of elliptic PDEs, as their reliance on skip connections introduces high-frequency spatial features that are inconsistent with the smooth, globally coupled nature of the pressure build-up field.

\subsubsection{Frequency-Domain Analysis}
\begin{figure}[!htbp]
    \centering
    \begin{subfigure}[b]{0.48\linewidth}
        \centering
        \includegraphics[width=\linewidth]{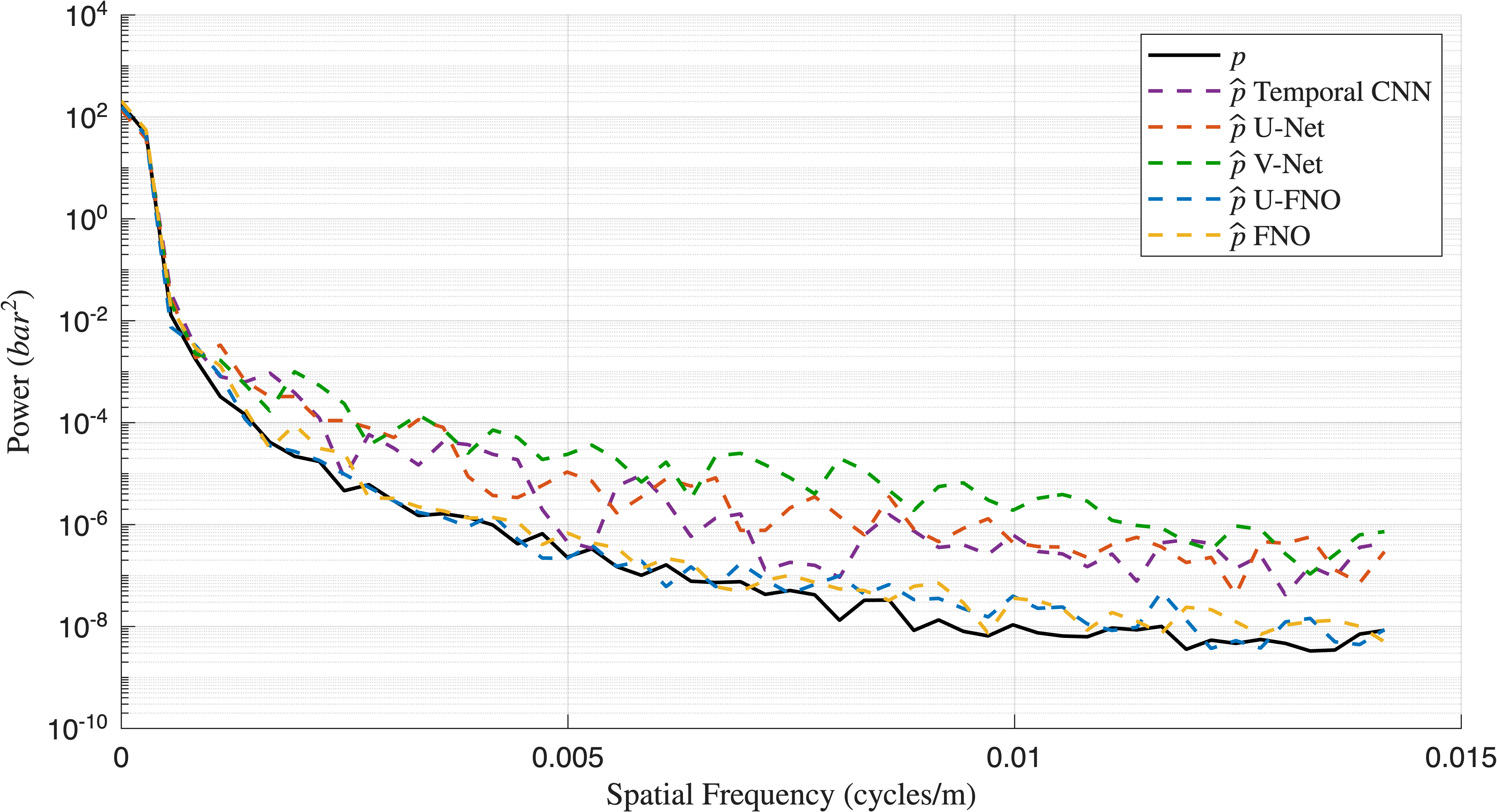}
        \caption{$r$-direction, $t = 30.0$ yr}
    \end{subfigure}
    \hfill
    \begin{subfigure}[b]{0.48\linewidth}
        \centering
        \includegraphics[width=\linewidth]{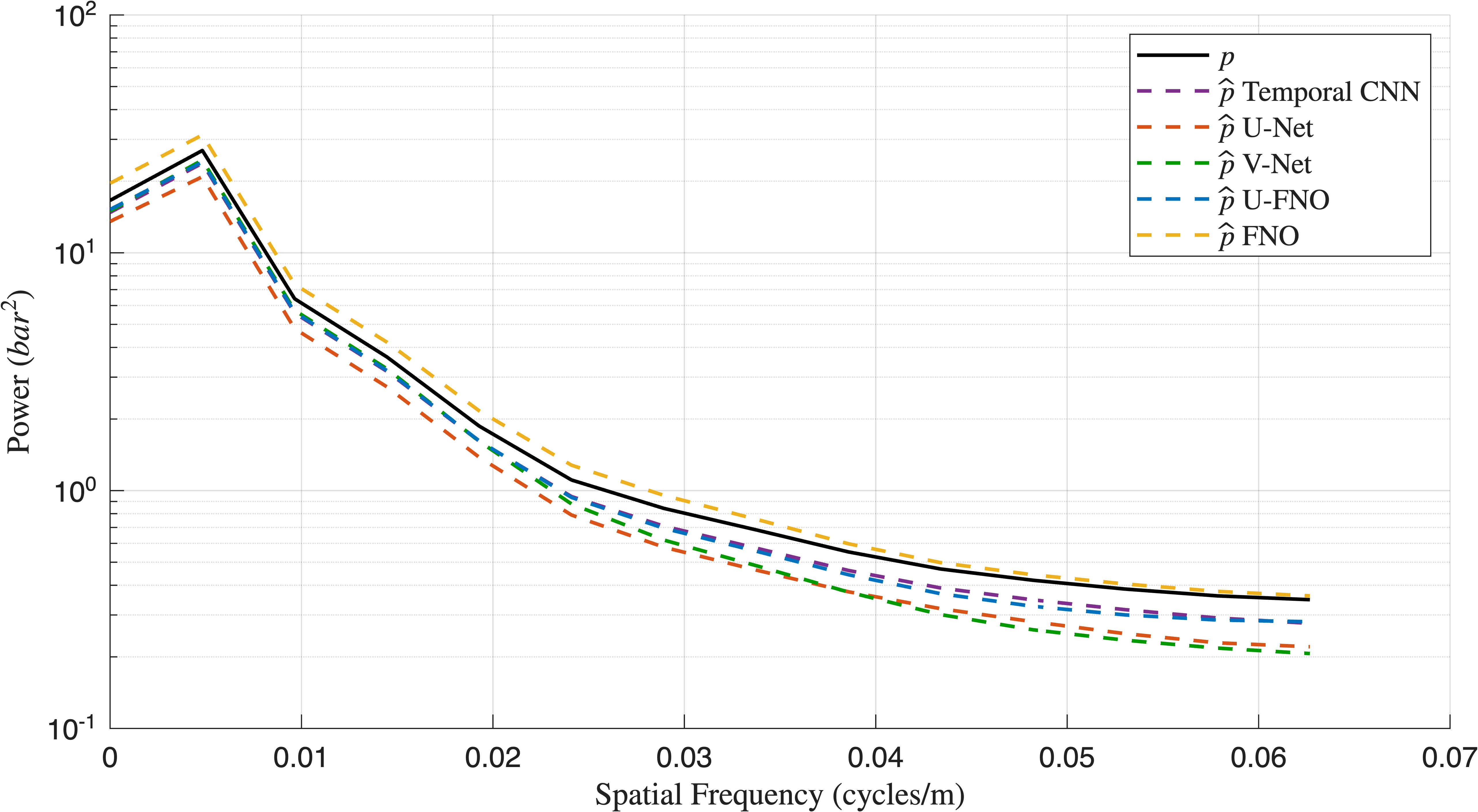}
        \caption{$z$-direction, $t = 30.0$ yr}
    \end{subfigure}
    \caption{Power spectral density of the pressure build-up field $p$ at $t = 30.0$ yr for Test Case 1.}
    \label{fig:pres_psd_test_1}
\end{figure}

\begin{figure}[!htbp]
    \centering
    \begin{subfigure}[b]{0.48\linewidth}
        \centering
        \includegraphics[width=\linewidth]{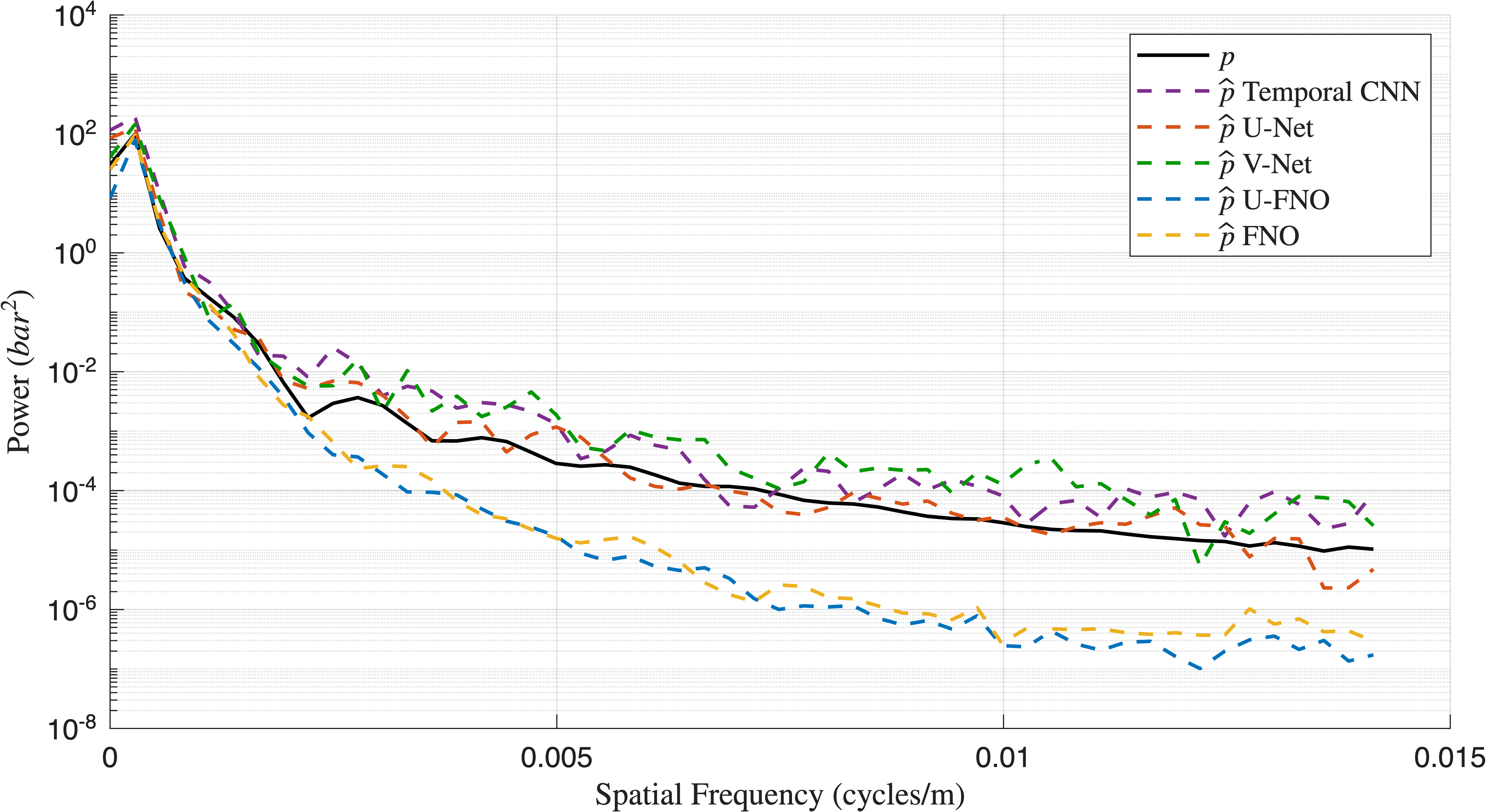}
        \caption{$r$-direction, $t = 30.0$ yr}
    \end{subfigure}
    \hfill
    \begin{subfigure}[b]{0.48\linewidth}
        \centering
        \includegraphics[width=\linewidth]{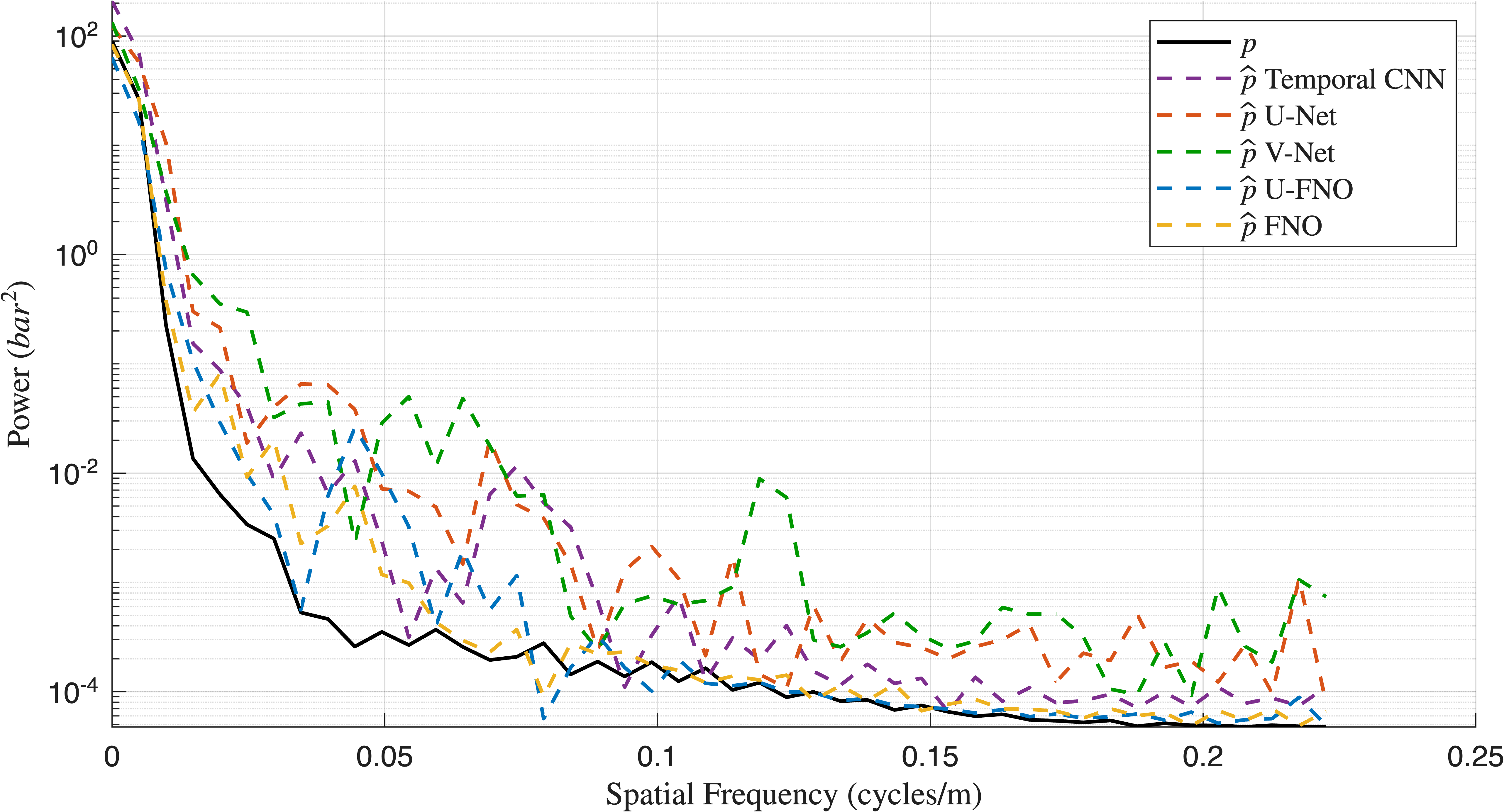}
        \caption{$z$-direction, $t = 30.0$ yr}
    \end{subfigure}
    \caption{Power spectral density of the pressure build-up field $p$ at $t = 30.0$ yr for Test Case 2.}
    \label{fig:pres_psd_test_2}
\end{figure}

The frequency-domain behaviour of each model is assessed using the Power Spectral Density (PSD) of the predicted and true fields, as shown in Figures~\ref{fig:pres_psd_test_1} and~\ref{fig:pres_psd_test_2}.

All models capture the dominant low-frequency spatial features of the pressure field. However, FNO-based architectures more accurately reproduce the full spatial spectrum, whereas convolutional models introduce spurious high-frequency noise due to their reliance on localized skip connection features. In Test Case 2, the permeability field exhibits a layered structure in the radial direction, which introduces higher-frequency content in the radial direction of the pressure field. In this case, all models show reduced accuracy at higher frequencies, as the spectral truncation in the FNO and the localized nature of convolutional features both limit the ability to resolve the broader spatial spectrum.

\section{Conclusion}
This study presents a benchmark comparison of five deep learning surrogate model architectures --- the Temporal CNN, U-Net, V-Net, FNO, and U-FNO for geological carbon sequestration modeling. Each model was trained to predict CO$_2$ saturation and pressure build-up fields for a radially symmetric, infinite-acting, anisotropic heterogeneous multi-phase flow problem, and evaluated based on its ability to capture the governing PDE type (i.e., hyperbolic vs.\ elliptic).

The results demonstrate that the choice of architecture is strongly influenced by the underlying PDE. For the CO$_2$ saturation field (i.e., hyperbolic), the U-FNO achieved the highest predictive accuracy (test $R^2 = 0.967$) by combining the global low-frequency approximation of FNO layers with the high-frequency features learned through an embedded U-Net branch. PSD analysis supported these findings, as the U-FNO more accurately preserved high-frequency components of the saturation field. For the pressure build-up field (i.e., elliptic), the FNO achieved the highest accuracy (test $R^2 = 0.972$), as spectral convolution efficiently propagates information globally, which is consistent with the smooth, diffuse solution of elliptic PDEs. Convolutional architectures with skip connections (U-Net and V-Net) performed competitively for the saturation field but introduced spurious high-frequency noise, making them unsuitable as a surrogate model for the pressure build-up field.

A modified loss function using a full spatial gradient penalty was also presented. Including both radial and vertical gradient components improved predictive accuracy for the Temporal CNN and U-FNO on the CO$_2$ saturation field, where the vertical gradient term improved the model's ability to capture buoyancy-driven flow and viscous fingering. For the pressure build-up field, only FNO-based architectures benefited from the full gradient loss, consistent with the globally smooth, radially dominated structure of the elliptic solution.

These findings highlight the strong potential of deep learning-based surrogate models for geological carbon sequestration applications. Future work should address more complex subsurface flow problems involving dynamic well controls, three-dimensional domains, and coupled thermo-hydro-mechanical processes, as well as the integration of these surrogate models within inverse modeling, history matching, uncertainty quantification, and real-time optimization workflows.

\section*{Acknowledgments}
\vspace{-\baselineskip}
\vspace{0.3cm}
\noindent
The authors gratefully acknowledge the financial support from the Natural Sciences and Engineering Research Council of Canada (NSERC) Discovery Grant program.

\section*{Code and Data Availability}
The Python scripts and dataset used in this study, originally developed by Gege Wen \cite{wen2022u} and subsequently modified by the authors, are available on \href{https://git.uwaterloo.ca/gzingaro1/comparative-study-geological-carbon-sequestration}{GitLab}.

\section*{CRediT Authorship Contribution Statement}
\vspace{-\baselineskip}
\vspace{0.3cm}
\noindent
\textbf{Giovanni Zingaro:} Conceptualization, Methodology, Software, Data Curation, Formal Analysis, Investigation, Validation, Visualization, Writing - Original Draft, Writing - Review \& Editing. \textbf{Robert Gracie:} Conceptualization, Funding Acquisition, Methodology, Resources, Supervision, Writing - Review \& Editing. \textbf{Yuri Leonenko:} Conceptualization, Funding Acquisition, Methodology, Resources, Supervision, Writing - Review \& Editing.

\bibliographystyle{elsarticle-num}
\bibliography{references}

\newpage
\appendix
\section{Model Performance Results}

\begin{table}[!htbp]
\tiny
\centering
\begin{tabular}{llcccccc}
\toprule
$\lambda_r$ & Metric & Split & Temporal CNN & U-Net & V-Net & FNO & U-FNO \\
\midrule
\multirow{15}{*}{0.01}
 & \multirow{3}{*}{$\mathcal{L}(\hat{\mathbf{s}})$}         & Training   & 1.96E-01 & 1.83E-01 & 1.88E-01 & 1.93E-01 & 1.71E-01 \\
 &                                           & Validation & 2.04E-01 & 1.94E-01 & 1.94E-01 & 1.99E-01 & 1.78E-01 \\
 &                                           & Test       & 2.05E-01 & 1.95E-01 & 1.96E-01 & 2.02E-01 & 1.81E-01 \\
 & \multirow{3}{*}{$\mathcal{L}_{\partial r}(\hat{\mathbf{s}})$} & Training   & 1.00E+00 & 9.65E-01 & 1.05E+00 & 8.66E-01 & 7.78E-01 \\
 &                                           & Validation & 1.00E+00 & 9.74E-01 & 1.05E+00 & 8.71E-01 & 7.83E-01 \\
 &                                           & Test       & 1.01E+00 & 9.85E-01 & 1.06E+00 & 8.77E-01 & 7.93E-01 \\
 & \multirow{3}{*}{$\mathcal{L}_{\partial z}(\hat{\mathbf{s}})$} & Training   & 9.91E-01 & 8.61E-01 & 9.47E-01 & 7.83E-01 & 7.11E-01 \\
 &                                           & Validation & 1.00E+00 & 8.89E-01 & 9.60E-01 & 8.07E-01 & 7.33E-01 \\
 &                                           & Test       & 9.92E-01 & 8.71E-01 & 9.51E-01 & 8.00E-01 & 7.22E-01 \\
 & \multirow{3}{*}{RMSE}                     & Training   & 3.54E-02 & 3.32E-02 & 3.38E-02 & 3.49E-02 & 3.19E-02 \\
 &                                           & Validation & 3.69E-02 & 3.51E-02 & 3.50E-02 & 3.60E-02 & 3.30E-02 \\
 &                                           & Test       & 3.73E-02 & 3.56E-02 & 3.56E-02 & 3.67E-02 & 3.39E-02 \\
 & \multirow{3}{*}{R$^2$}                    & Training   & 0.955 & 0.960 & 0.959 & 0.956 & 0.963 \\
 &                                           & Validation & 0.951 & 0.956 & 0.956 & 0.953 & 0.961 \\
 &                                           & Test       & 0.950 & 0.954 & 0.954 & 0.951 & 0.958 \\
\midrule
\multirow{15}{*}{0.1}
 & \multirow{3}{*}{$\mathcal{L}(\hat{\mathbf{s}})$}         & Training   & 1.98E-01 & 2.17E-01 & 1.98E-01 & 1.89E-01 & 1.72E-01 \\
 &                                           & Validation & 2.06E-01 & 2.26E-01 & 2.06E-01 & 1.97E-01 & 1.80E-01 \\
 &                                           & Test       & 2.07E-01 & 2.25E-01 & 2.07E-01 & 2.00E-01 & 1.82E-01 \\
 & \multirow{3}{*}{$\mathcal{L}_{\partial r}(\hat{\mathbf{s}})$} & Training   & 9.55E-01 & 9.31E-01 & 1.00E+00 & 8.57E-01 & 7.96E-01 \\
 &                                           & Validation & 9.53E-01 & 9.33E-01 & 1.01E+00 & 8.65E-01 & 8.03E-01 \\
 &                                           & Test       & 9.56E-01 & 9.47E-01 & 1.02E+00 & 8.67E-01 & 8.09E-01 \\
 & \multirow{3}{*}{$\mathcal{L}_{\partial z}(\hat{\mathbf{s}})$} & Training   & 9.73E-01 & 8.65E-01 & 9.18E-01 & 7.76E-01 & 7.49E-01 \\
 &                                           & Validation & 9.85E-01 & 8.83E-01 & 9.33E-01 & 8.02E-01 & 7.73E-01 \\
 &                                           & Test       & 9.74E-01 & 8.68E-01 & 9.25E-01 & 7.94E-01 & 7.64E-01 \\
 & \multirow{3}{*}{RMSE}                     & Training   & 3.61E-02 & 3.97E-02 & 3.56E-02 & 3.42E-02 & 3.19E-02 \\
 &                                           & Validation & 3.75E-02 & 4.16E-02 & 3.73E-02 & 3.57E-02 & 3.32E-02 \\
 &                                           & Test       & 3.79E-02 & 4.17E-02 & 3.78E-02 & 3.63E-02 & 3.41E-02 \\
 & \multirow{3}{*}{R$^2$}                    & Training   & 0.953 & 0.943 & 0.954 & 0.958 & 0.963 \\
 &                                           & Validation & 0.949 & 0.938 & 0.950 & 0.954 & 0.960 \\
 &                                           & Test       & 0.948 & 0.937 & 0.948 & 0.952 & 0.958 \\
\midrule
\multirow{15}{*}{1.0}
 & \multirow{3}{*}{$\mathcal{L}(\hat{\mathbf{s}})$}         & Training   & 2.01E-01 & 2.34E-01 & 2.30E-01 & 1.99E-01 & 1.99E-01 \\
 &                                           & Validation & 2.08E-01 & 2.41E-01 & 2.37E-01 & 2.08E-01 & 2.05E-01 \\
 &                                           & Test       & 2.08E-01 & 2.42E-01 & 2.38E-01 & 2.09E-01 & 2.08E-01 \\
 & \multirow{3}{*}{$\mathcal{L}_{\partial r}(\hat{\mathbf{s}})$} & Training   & 9.56E-01 & 1.07E+00 & 9.17E-01 & 8.19E-01 & 8.61E-01 \\
 &                                           & Validation & 9.57E-01 & 1.07E+00 & 9.17E-01 & 8.27E-01 & 8.68E-01 \\
 &                                           & Test       & 9.60E-01 & 1.08E+00 & 9.27E-01 & 8.32E-01 & 8.75E-01 \\
 & \multirow{3}{*}{$\mathcal{L}_{\partial z}(\hat{\mathbf{s}})$} & Training   & 9.18E-01 & 1.03E+00 & 9.33E-01 & 7.93E-01 & 8.26E-01 \\
 &                                           & Validation & 9.31E-01 & 1.04E+00 & 9.42E-01 & 8.19E-01 & 8.44E-01 \\
 &                                           & Test       & 9.22E-01 & 1.03E+00 & 9.24E-01 & 8.12E-01 & 8.36E-01 \\
 & \multirow{3}{*}{RMSE}                     & Training   & 3.68E-02 & 4.27E-02 & 4.16E-02 & 3.65E-02 & 3.70E-02 \\
 &                                           & Validation & 3.80E-02 & 4.41E-02 & 4.29E-02 & 3.79E-02 & 3.81E-02 \\
 &                                           & Test       & 3.83E-02 & 4.46E-02 & 4.36E-02 & 3.86E-02 & 3.92E-02 \\
 & \multirow{3}{*}{R$^2$}                    & Training   & 0.951 & 0.934 & 0.938 & 0.952 & 0.950 \\
 &                                           & Validation & 0.948 & 0.931 & 0.934 & 0.948 & 0.948 \\
 &                                           & Test       & 0.947 & 0.929 & 0.932 & 0.946 & 0.944 \\
\bottomrule
\end{tabular}
\caption*{\tiny $\mathcal{L}(\hat{\mathbf{s}}) = \frac{\|\mathbf{s} - \hat{\mathbf{s}}\|_2}{\|\mathbf{s}\|_2}$, \quad $\mathcal{L}_{\partial r}(\hat{\mathbf{s}}) = \frac{\|\partial_r \mathbf{s} - \partial_r \hat{\mathbf{s}}\|_2}{\|\partial_r \mathbf{s}\|_2}$, \quad $\mathcal{L}_{\partial z}(\hat{\mathbf{s}}) = \frac{\|\partial_z \mathbf{s} - \partial_z \hat{\mathbf{s}}\|_2}{\|\partial_z \mathbf{s}\|_2}$}
\caption{Model performance results for CO$_2$ saturation field -- radial gradient loss ($\lambda_z = 0$).}
\label{tab:results_asymmetric}
\end{table}

\begin{table}[!htbp]
\tiny
\centering
\begin{tabular}{llcccccc}
\toprule
$\lambda$ & Metric & Split & Temporal CNN & U-Net & V-Net & FNO & U-FNO \\
\midrule
\multirow{15}{*}{0.0}
 & \multirow{3}{*}{$\mathcal{L}(\hat{\mathbf{s}})$}         & Training   & 1.95E-01 & 1.92E-01 & 2.08E-01 & 1.88E-01 & 1.61E-01 \\
 &                                           & Validation & 2.03E-01 & 2.00E-01 & 2.17E-01 & 1.95E-01 & 1.68E-01 \\
 &                                           & Test       & 2.05E-01 & 2.03E-01 & 2.18E-01 & 1.98E-01 & 1.71E-01 \\
 & \multirow{3}{*}{$\mathcal{L}_{\partial r}(\hat{\mathbf{s}})$} & Training   & 1.05E+00 & 9.25E-01 & 1.06E+00 & 8.75E-01 & 7.83E-01 \\
 &                                           & Validation & 1.04E+00 & 9.28E-01 & 1.06E+00 & 8.82E-01 & 7.87E-01 \\
 &                                           & Test       & 1.05E+00 & 9.39E-01 & 1.08E+00 & 8.86E-01 & 7.96E-01 \\
 & \multirow{3}{*}{$\mathcal{L}_{\partial z}(\hat{\mathbf{s}})$} & Training   & 1.02E+00 & 8.87E-01 & 1.00E+00 & 7.86E-01 & 7.40E-01 \\
 &                                           & Validation & 1.03E+00 & 9.00E-01 & 1.02E+00 & 8.09E-01 & 7.59E-01 \\
 &                                           & Test       & 1.02E+00 & 8.87E-01 & 1.00E+00 & 8.01E-01 & 7.55E-01 \\
 & \multirow{3}{*}{RMSE}                     & Training   & 3.53E-02 & 3.50E-02 & 3.76E-02 & 3.40E-02 & 2.95E-02 \\
 &                                           & Validation & 3.68E-02 & 3.65E-02 & 3.92E-02 & 3.53E-02 & 3.09E-02 \\
 &                                           & Test       & 3.73E-02 & 3.73E-02 & 4.01E-02 & 3.60E-02 & 3.17E-02 \\
 & \multirow{3}{*}{R$^2$}                    & Training   & 0.955 & 0.956 & 0.949 & 0.958 & 0.968 \\
 &                                           & Validation & 0.951 & 0.952 & 0.945 & 0.955 & 0.965 \\
 &                                           & Test       & 0.950 & 0.950 & 0.942 & 0.953 & 0.963 \\
\midrule
\multirow{15}{*}{0.01}
 & \multirow{3}{*}{$\mathcal{L}(\hat{\mathbf{s}})$}         & Training   & 1.97E-01 & 1.87E-01 & 2.03E-01 & 1.93E-01 & 1.55E-01 \\
 &                                           & Validation & 2.04E-01 & 1.94E-01 & 2.11E-01 & 1.99E-01 & 1.62E-01 \\
 &                                           & Test       & 2.06E-01 & 1.96E-01 & 2.11E-01 & 2.02E-01 & 1.65E-01 \\
 & \multirow{3}{*}{$\mathcal{L}_{\partial r}(\hat{\mathbf{s}})$} & Training   & 1.04E+00 & 1.08E+00 & 9.50E-01 & 8.72E-01 & 7.55E-01 \\
 &                                           & Validation & 1.04E+00 & 1.09E+00 & 9.53E-01 & 8.76E-01 & 7.63E-01 \\
 &                                           & Test       & 1.05E+00 & 1.10E+00 & 9.62E-01 & 8.83E-01 & 7.74E-01 \\
 & \multirow{3}{*}{$\mathcal{L}_{\partial z}(\hat{\mathbf{s}})$} & Training   & 1.03E+00 & 9.59E-01 & 8.91E-01 & 7.90E-01 & 6.89E-01 \\
 &                                           & Validation & 1.04E+00 & 9.77E-01 & 9.06E-01 & 8.14E-01 & 7.14E-01 \\
 &                                           & Test       & 1.03E+00 & 9.62E-01 & 8.98E-01 & 8.05E-01 & 7.05E-01 \\
 & \multirow{3}{*}{RMSE}                     & Training   & 3.55E-02 & 3.37E-02 & 3.64E-02 & 3.49E-02 & 2.85E-02 \\
 &                                           & Validation & 3.68E-02 & 3.49E-02 & 3.81E-02 & 3.60E-02 & 2.97E-02 \\
 &                                           & Test       & 3.74E-02 & 3.56E-02 & 3.84E-02 & 3.66E-02 & 3.04E-02 \\
 & \multirow{3}{*}{R$^2$}                    & Training   & 0.955 & 0.959 & 0.952 & 0.956 & 0.971 \\
 &                                           & Validation & 0.951 & 0.956 & 0.948 & 0.953 & 0.968 \\
 &                                           & Test       & 0.949 & 0.955 & 0.947 & 0.951 & 0.967 \\
\midrule
\multirow{15}{*}{0.1}
 & \multirow{3}{*}{$\mathcal{L}(\hat{\mathbf{s}})$}         & Training   & 1.92E-01 & 2.08E-01 & 2.05E-01 & 1.92E-01 & 1.66E-01 \\
 &                                           & Validation & 2.01E-01 & 2.14E-01 & 2.13E-01 & 2.00E-01 & 1.74E-01 \\
 &                                           & Test       & 2.03E-01 & 2.15E-01 & 2.14E-01 & 2.01E-01 & 1.76E-01 \\
 & \multirow{3}{*}{$\mathcal{L}_{\partial r}(\hat{\mathbf{s}})$} & Training   & 9.30E-01 & 9.46E-01 & 9.15E-01 & 8.39E-01 & 7.72E-01 \\
 &                                           & Validation & 9.35E-01 & 9.48E-01 & 9.17E-01 & 8.46E-01 & 7.79E-01 \\
 &                                           & Test       & 9.43E-01 & 9.58E-01 & 9.27E-01 & 8.51E-01 & 7.87E-01 \\
 & \multirow{3}{*}{$\mathcal{L}_{\partial z}(\hat{\mathbf{s}})$} & Training   & 8.80E-01 & 8.38E-01 & 8.18E-01 & 7.70E-01 & 7.10E-01 \\
 &                                           & Validation & 9.05E-01 & 8.51E-01 & 8.34E-01 & 7.96E-01 & 7.35E-01 \\
 &                                           & Test       & 8.99E-01 & 8.36E-01 & 8.25E-01 & 7.89E-01 & 7.29E-01 \\
 & \multirow{3}{*}{RMSE}                     & Training   & 3.49E-02 & 3.79E-02 & 3.76E-02 & 3.49E-02 & 3.03E-02 \\
 &                                           & Validation & 3.64E-02 & 3.91E-02 & 3.91E-02 & 3.64E-02 & 3.17E-02 \\
 &                                           & Test       & 3.70E-02 & 3.96E-02 & 3.96E-02 & 3.68E-02 & 3.25E-02 \\
 & \multirow{3}{*}{R$^2$}                    & Training   & 0.956 & 0.948 & 0.949 & 0.956 & 0.967 \\
 &                                           & Validation & 0.952 & 0.945 & 0.945 & 0.952 & 0.964 \\
 &                                           & Test       & 0.950 & 0.943 & 0.944 & 0.951 & 0.962 \\
\midrule
\multirow{15}{*}{1.0}
 & \multirow{3}{*}{$\mathcal{L}(\hat{\mathbf{s}})$}         & Training   & 2.00E-01 & 2.14E-01 & 2.05E-01 & 1.90E-01 & 1.88E-01 \\
 &                                           & Validation & 2.07E-01 & 2.20E-01 & 2.11E-01 & 1.99E-01 & 1.95E-01 \\
 &                                           & Test       & 2.10E-01 & 2.21E-01 & 2.14E-01 & 2.01E-01 & 1.94E-01 \\
 & \multirow{3}{*}{$\mathcal{L}_{\partial r}(\hat{\mathbf{s}})$} & Training   & 9.52E-01 & 9.93E-01 & 8.58E-01 & 8.17E-01 & 7.86E-01 \\
 &                                           & Validation & 9.53E-01 & 9.94E-01 & 8.61E-01 & 8.27E-01 & 7.96E-01 \\
 &                                           & Test       & 9.62E-01 & 1.00E+00 & 8.71E-01 & 8.34E-01 & 8.00E-01 \\
 & \multirow{3}{*}{$\mathcal{L}_{\partial z}(\hat{\mathbf{s}})$} & Training   & 8.52E-01 & 8.13E-01 & 7.28E-01 & 7.56E-01 & 6.86E-01 \\
 &                                           & Validation & 8.76E-01 & 8.32E-01 & 7.51E-01 & 7.89E-01 & 7.13E-01 \\
 &                                           & Test       & 8.71E-01 & 8.19E-01 & 7.45E-01 & 7.83E-01 & 7.08E-01 \\
 & \multirow{3}{*}{RMSE}                     & Training   & 3.65E-02 & 3.86E-02 & 3.80E-02 & 3.40E-02 & 3.51E-02 \\
 &                                           & Validation & 3.77E-02 & 3.97E-02 & 3.91E-02 & 3.57E-02 & 3.63E-02 \\
 &                                           & Test       & 3.85E-02 & 4.05E-02 & 4.01E-02 & 3.62E-02 & 3.65E-02 \\
 & \multirow{3}{*}{R$^2$}                    & Training   & 0.952 & 0.946 & 0.948 & 0.958 & 0.955 \\
 &                                           & Validation & 0.949 & 0.944 & 0.945 & 0.954 & 0.952 \\
 &                                           & Test       & 0.947 & 0.941 & 0.942 & 0.952 & 0.952 \\
\bottomrule
\end{tabular}
\caption*{\tiny $\mathcal{L}(\hat{\mathbf{s}}) = \frac{\|\mathbf{s} - \hat{\mathbf{s}}\|_2}{\|\mathbf{s}\|_2}$, \quad $\mathcal{L}_{\partial r}(\hat{\mathbf{s}}) = \frac{\|\partial_r \mathbf{s} - \partial_r \hat{\mathbf{s}}\|_2}{\|\partial_r \mathbf{s}\|_2}$, \quad $\mathcal{L}_{\partial z}(\hat{\mathbf{s}}) = \frac{\|\partial_z \mathbf{s} - \partial_z \hat{\mathbf{s}}\|_2}{\|\partial_z \mathbf{s}\|_2}$}
\caption{Model performance results for CO$_2$ saturation field -- full gradient loss ($\lambda_r = \lambda_z$).}
\label{tab:results_symmetric}
\end{table}

\begin{table}[!htbp]
\tiny
\centering
\begin{tabular}{llcccccc}
\toprule
$\lambda_r$ & Metric & Split & Temporal CNN & U-Net & V-Net & FNO & U-FNO \\
\midrule
\multirow{15}{*}{0.01}
 & \multirow{3}{*}{$\mathcal{L}(\hat{\mathbf{p}})$}         & Training   & 1.20E-01 & 1.82E-01 & 1.59E-01 & 1.38E-01 & 1.25E-01 \\
 &                                           & Validation & 1.31E-01 & 1.88E-01 & 1.66E-01 & 1.42E-01 & 1.29E-01 \\
 &                                           & Test       & 1.36E-01 & 1.96E-01 & 1.75E-01 & 1.43E-01 & 1.34E-01 \\
 & \multirow{3}{*}{$\mathcal{L}_{\partial r}(\hat{\mathbf{p}})$} & Training   & 6.92E-01 & 1.19E+00 & 9.84E-01 & 9.99E-01 & 8.59E-01 \\
 &                                           & Validation & 6.96E-01 & 1.20E+00 & 9.76E-01 & 1.01E+00 & 8.59E-01 \\
 &                                           & Test       & 7.02E-01 & 1.18E+00 & 9.80E-01 & 9.82E-01 & 8.52E-01 \\
 & \multirow{3}{*}{$\mathcal{L}_{\partial z}(\hat{\mathbf{p}})$} & Training   & 1.13E+00 & 2.13E+00 & 1.86E+00 & 1.24E+00 & 1.10E+00 \\
 &                                           & Validation & 1.06E+00 & 1.95E+00 & 1.70E+00 & 1.19E+00 & 1.04E+00 \\
 &                                           & Test       & 9.47E-01 & 1.75E+00 & 1.55E+00 & 1.24E+00 & 9.61E-01 \\
 & \multirow{3}{*}{RMSE}                     & Training   & 3.23E+00 & 5.10E+00 & 4.67E+00 & 4.13E+00 & 3.75E+00 \\
 &                                           & Validation & 3.58E+00 & 5.16E+00 & 4.87E+00 & 4.25E+00 & 3.83E+00 \\
 &                                           & Test       & 4.20E+00 & 6.30E+00 & 5.92E+00 & 4.64E+00 & 4.54E+00 \\
 & \multirow{3}{*}{R$^2$}                    & Training   & 0.980 & 0.952 & 0.960 & 0.965 & 0.974 \\
 &                                           & Validation & 0.976 & 0.951 & 0.955 & 0.965 & 0.973 \\
 &                                           & Test       & 0.971 & 0.937 & 0.944 & 0.965 & 0.967 \\
\midrule
\multirow{15}{*}{0.1}
 & \multirow{3}{*}{$\mathcal{L}(\hat{\mathbf{p}})$}         & Training   & 1.28E-01 & 1.55E-01 & 1.78E-01 & 1.31E-01 & 1.53E-01 \\
 &                                           & Validation & 1.36E-01 & 1.63E-01 & 1.84E-01 & 1.35E-01 & 1.57E-01 \\
 &                                           & Test       & 1.45E-01 & 1.70E-01 & 1.91E-01 & 1.39E-01 & 1.59E-01 \\
 & \multirow{3}{*}{$\mathcal{L}_{\partial r}(\hat{\mathbf{p}})$} & Training   & 6.95E-01 & 1.05E+00 & 1.03E+00 & 9.10E-01 & 1.26E+00 \\
 &                                           & Validation & 6.95E-01 & 1.04E+00 & 1.03E+00 & 9.14E-01 & 1.24E+00 \\
 &                                           & Test       & 6.98E-01 & 1.05E+00 & 1.02E+00 & 9.09E-01 & 1.25E+00 \\
 & \multirow{3}{*}{$\mathcal{L}_{\partial z}(\hat{\mathbf{p}})$} & Training   & 1.11E+00 & 1.81E+00 & 1.95E+00 & 1.09E+00 & 1.94E+00 \\
 &                                           & Validation & 1.05E+00 & 1.69E+00 & 1.83E+00 & 1.05E+00 & 1.78E+00 \\
 &                                           & Test       & 9.61E-01 & 1.50E+00 & 1.58E+00 & 1.10E+00 & 1.54E+00 \\
 & \multirow{3}{*}{RMSE}                     & Training   & 3.51E+00 & 4.24E+00 & 5.61E+00 & 3.64E+00 & 4.68E+00 \\
 &                                           & Validation & 3.79E+00 & 4.53E+00 & 5.66E+00 & 3.79E+00 & 4.67E+00 \\
 &                                           & Test       & 4.61E+00 & 5.36E+00 & 6.91E+00 & 4.43E+00 & 5.27E+00 \\
 & \multirow{3}{*}{R$^2$}                    & Training   & 0.977 & 0.966 & 0.944 & 0.974 & 0.955 \\
 &                                           & Validation & 0.973 & 0.962 & 0.943 & 0.972 & 0.958 \\
 &                                           & Test       & 0.965 & 0.953 & 0.926 & 0.968 & 0.956 \\
\midrule
\multirow{15}{*}{1.0}
 & \multirow{3}{*}{$\mathcal{L}(\hat{\mathbf{p}})$}         & Training   & 1.41E-01 & 1.66E-01 & 1.68E-01 & 1.27E-01 & 1.43E-01 \\
 &                                           & Validation & 1.53E-01 & 1.70E-01 & 1.75E-01 & 1.31E-01 & 1.51E-01 \\
 &                                           & Test       & 1.59E-01 & 1.79E-01 & 1.84E-01 & 1.34E-01 & 1.50E-01 \\
 & \multirow{3}{*}{$\mathcal{L}_{\partial r}(\hat{\mathbf{p}})$} & Training   & 6.73E-01 & 8.49E-01 & 8.30E-01 & 7.87E-01 & 8.00E-01 \\
 &                                           & Validation & 6.95E-01 & 8.47E-01 & 8.29E-01 & 7.86E-01 & 8.01E-01 \\
 &                                           & Test       & 6.84E-01 & 8.52E-01 & 8.38E-01 & 7.82E-01 & 8.04E-01 \\
 & \multirow{3}{*}{$\mathcal{L}_{\partial z}(\hat{\mathbf{p}})$} & Training   & 1.16E+00 & 1.95E+00 & 1.89E+00 & 1.11E+00 & 1.84E+00 \\
 &                                           & Validation & 1.09E+00 & 1.74E+00 & 1.75E+00 & 1.05E+00 & 1.69E+00 \\
 &                                           & Test       & 1.02E+00 & 1.61E+00 & 1.57E+00 & 1.06E+00 & 1.50E+00 \\
 & \multirow{3}{*}{RMSE}                     & Training   & 3.85E+00 & 5.18E+00 & 5.06E+00 & 3.55E+00 & 4.36E+00 \\
 &                                           & Validation & 4.13E+00 & 5.20E+00 & 5.11E+00 & 3.68E+00 & 4.61E+00 \\
 &                                           & Test       & 4.80E+00 & 6.33E+00 & 6.44E+00 & 4.44E+00 & 4.85E+00 \\
 & \multirow{3}{*}{R$^2$}                    & Training   & 0.972 & 0.950 & 0.950 & 0.976 & 0.961 \\
 &                                           & Validation & 0.966 & 0.951 & 0.950 & 0.974 & 0.958 \\
 &                                           & Test       & 0.962 & 0.933 & 0.925 & 0.969 & 0.961 \\
\bottomrule
\end{tabular}
\caption*{\tiny $\mathcal{L}(\hat{\mathbf{p}}) = \frac{\|\mathbf{p} - \hat{\mathbf{p}}\|_2}{\|\mathbf{p}\|_2}$, \quad $\mathcal{L}_{\partial r}(\hat{\mathbf{p}}) = \frac{\|\partial_r \mathbf{p} - \partial_r \hat{\mathbf{p}}\|_2}{\|\partial_r \mathbf{p}\|_2}$, \quad $\mathcal{L}_{\partial z}(\hat{\mathbf{p}}) = \frac{\|\partial_z \mathbf{p} - \partial_z \hat{\mathbf{p}}\|_2}{\|\partial_z \mathbf{p}\|_2}$}
\caption{Model performance results for pressure build-up field -- radial gradient loss ($\lambda_z = 0$).}
\label{tab:pressure_results_asymmetric}
\end{table}

\begin{table}[!htbp]
\tiny
\centering
\begin{tabular}{llcccccc}
\toprule
$\lambda$ & Metric & Split & Temporal CNN & U-Net & V-Net & FNO & U-FNO \\
\midrule
\multirow{15}{*}{0.0}
 & \multirow{3}{*}{$\mathcal{L}(\hat{\mathbf{p}})$}         & Training   & 1.20E-01 & 1.59E-01 & 1.56E-01 & 1.32E-01 & 1.26E-01 \\
 &                                           & Validation & 1.27E-01 & 1.67E-01 & 1.61E-01 & 1.36E-01 & 1.32E-01 \\
 &                                           & Test       & 1.36E-01 & 1.74E-01 & 1.69E-01 & 1.36E-01 & 1.34E-01 \\
 & \multirow{3}{*}{$\mathcal{L}_{\partial r}(\hat{\mathbf{p}})$} & Training   & 6.96E-01 & 1.02E+00 & 9.82E-01 & 8.28E-01 & 9.61E-01 \\
 &                                           & Validation & 6.97E-01 & 1.02E+00 & 9.71E-01 & 8.25E-01 & 9.64E-01 \\
 &                                           & Test       & 7.00E-01 & 1.02E+00 & 9.70E-01 & 8.11E-01 & 9.46E-01 \\
 & \multirow{3}{*}{$\mathcal{L}_{\partial z}(\hat{\mathbf{p}})$} & Training   & 1.11E+00 & 1.86E+00 & 1.77E+00 & 1.20E+00 & 1.70E+00 \\
 &                                           & Validation & 1.05E+00 & 1.72E+00 & 1.63E+00 & 1.16E+00 & 1.56E+00 \\
 &                                           & Test       & 9.60E-01 & 1.63E+00 & 1.51E+00 & 1.22E+00 & 1.37E+00 \\
 & \multirow{3}{*}{RMSE}                     & Training   & 3.31E+00 & 4.69E+00 & 4.70E+00 & 3.80E+00 & 3.66E+00 \\
 &                                           & Validation & 3.44E+00 & 4.85E+00 & 4.87E+00 & 3.89E+00 & 3.82E+00 \\
 &                                           & Test       & 4.38E+00 & 5.81E+00 & 5.77E+00 & 4.32E+00 & 4.35E+00 \\
 & \multirow{3}{*}{R$^2$}                    & Training   & 0.979 & 0.957 & 0.960 & 0.970 & 0.974 \\
 &                                           & Validation & 0.978 & 0.955 & 0.958 & 0.971 & 0.971 \\
 &                                           & Test       & 0.968 & 0.944 & 0.947 & 0.969 & 0.969 \\
\midrule
\multirow{15}{*}{0.01}
 & \multirow{3}{*}{$\mathcal{L}(\hat{\mathbf{p}})$}         & Training   & 1.27E-01 & 2.38E-01 & 1.97E-01 & 1.35E-01 & 1.32E-01 \\
 &                                           & Validation & 1.34E-01 & 2.46E-01 & 2.05E-01 & 1.38E-01 & 1.37E-01 \\
 &                                           & Test       & 1.44E-01 & 2.49E-01 & 2.05E-01 & 1.38E-01 & 1.39E-01 \\
 & \multirow{3}{*}{$\mathcal{L}_{\partial r}(\hat{\mathbf{p}})$} & Training   & 6.75E-01 & 1.41E+00 & 1.11E+00 & 9.35E-01 & 1.06E+00 \\
 &                                           & Validation & 6.78E-01 & 1.41E+00 & 1.11E+00 & 9.40E-01 & 1.06E+00 \\
 &                                           & Test       & 6.77E-01 & 1.39E+00 & 1.10E+00 & 9.06E-01 & 1.05E+00 \\
 & \multirow{3}{*}{$\mathcal{L}_{\partial z}(\hat{\mathbf{p}})$} & Training   & 1.03E+00 & 2.21E+00 & 2.18E+00 & 1.13E+00 & 1.54E+00 \\
 &                                           & Validation & 9.96E-01 & 2.08E+00 & 2.02E+00 & 1.10E+00 & 1.43E+00 \\
 &                                           & Test       & 9.06E-01 & 1.86E+00 & 1.70E+00 & 1.06E+00 & 1.29E+00 \\
 & \multirow{3}{*}{RMSE}                     & Training   & 3.34E+00 & 6.31E+00 & 5.21E+00 & 3.89E+00 & 4.01E+00 \\
 &                                           & Validation & 3.55E+00 & 6.32E+00 & 5.42E+00 & 4.00E+00 & 4.13E+00 \\
 &                                           & Test       & 4.45E+00 & 7.53E+00 & 6.13E+00 & 4.33E+00 & 4.68E+00 \\
 & \multirow{3}{*}{R$^2$}                    & Training   & 0.979 & 0.925 & 0.946 & 0.969 & 0.967 \\
 &                                           & Validation & 0.978 & 0.925 & 0.942 & 0.968 & 0.967 \\
 &                                           & Test       & 0.968 & 0.909 & 0.935 & 0.969 & 0.966 \\
\midrule
\multirow{15}{*}{0.1}
 & \multirow{3}{*}{$\mathcal{L}(\hat{\mathbf{p}})$}         & Training   & 1.25E-01 & 1.57E-01 & 1.62E-01 & 1.29E-01 & 1.24E-01 \\
 &                                           & Validation & 1.33E-01 & 1.63E-01 & 1.67E-01 & 1.32E-01 & 1.30E-01 \\
 &                                           & Test       & 1.41E-01 & 1.74E-01 & 1.73E-01 & 1.35E-01 & 1.33E-01 \\
 & \multirow{3}{*}{$\mathcal{L}_{\partial r}(\hat{\mathbf{p}})$} & Training   & 6.24E-01 & 8.77E-01 & 9.29E-01 & 7.83E-01 & 7.97E-01 \\
 &                                           & Validation & 6.35E-01 & 8.77E-01 & 9.21E-01 & 7.93E-01 & 7.93E-01 \\
 &                                           & Test       & 6.30E-01 & 8.72E-01 & 9.30E-01 & 7.69E-01 & 7.96E-01 \\
 & \multirow{3}{*}{$\mathcal{L}_{\partial z}(\hat{\mathbf{p}})$} & Training   & 9.49E-01 & 1.29E+00 & 1.41E+00 & 9.11E-01 & 1.10E+00 \\
 &                                           & Validation & 9.16E-01 & 1.20E+00 & 1.35E+00 & 8.87E-01 & 1.05E+00 \\
 &                                           & Test       & 8.48E-01 & 1.12E+00 & 1.20E+00 & 8.70E-01 & 9.74E-01 \\
 & \multirow{3}{*}{RMSE}                     & Training   & 3.44E+00 & 4.60E+00 & 4.54E+00 & 3.64E+00 & 3.93E+00 \\
 &                                           & Validation & 3.67E+00 & 4.69E+00 & 4.68E+00 & 3.79E+00 & 4.06E+00 \\
 &                                           & Test       & 4.45E+00 & 5.83E+00 & 5.68E+00 & 4.36E+00 & 4.64E+00 \\
 & \multirow{3}{*}{R$^2$}                    & Training   & 0.978 & 0.959 & 0.960 & 0.974 & 0.969 \\
 &                                           & Validation & 0.975 & 0.958 & 0.959 & 0.973 & 0.968 \\
 &                                           & Test       & 0.968 & 0.944 & 0.947 & 0.969 & 0.965 \\
\midrule
\multirow{15}{*}{1.0}
 & \multirow{3}{*}{$\mathcal{L}(\hat{\mathbf{p}})$}         & Training   & 1.31E-01 & 1.75E-01 & 1.80E-01 & 1.22E-01 & 1.39E-01 \\
 &                                           & Validation & 1.39E-01 & 1.77E-01 & 1.86E-01 & 1.26E-01 & 1.44E-01 \\
 &                                           & Test       & 1.46E-01 & 1.89E-01 & 1.93E-01 & 1.28E-01 & 1.49E-01 \\
 & \multirow{3}{*}{$\mathcal{L}_{\partial r}(\hat{\mathbf{p}})$} & Training   & 7.67E-01 & 7.55E-01 & 8.02E-01 & 7.28E-01 & 7.52E-01 \\
 &                                           & Validation & 7.75E-01 & 7.53E-01 & 8.03E-01 & 7.33E-01 & 7.58E-01 \\
 &                                           & Test       & 7.68E-01 & 7.47E-01 & 7.98E-01 & 7.13E-01 & 7.51E-01 \\
 & \multirow{3}{*}{$\mathcal{L}_{\partial z}(\hat{\mathbf{p}})$} & Training   & 7.79E-01 & 9.51E-01 & 1.01E+00 & 7.63E-01 & 8.56E-01 \\
 &                                           & Validation & 7.68E-01 & 9.19E-01 & 9.66E-01 & 7.58E-01 & 8.24E-01 \\
 &                                           & Test       & 7.27E-01 & 8.64E-01 & 8.91E-01 & 7.21E-01 & 7.60E-01 \\
 & \multirow{3}{*}{RMSE}                     & Training   & 3.76E+00 & 5.71E+00 & 5.58E+00 & 3.33E+00 & 4.32E+00 \\
 &                                           & Validation & 4.03E+00 & 5.65E+00 & 5.62E+00 & 3.50E+00 & 4.44E+00 \\
 &                                           & Test       & 4.77E+00 & 7.09E+00 & 6.94E+00 & 4.06E+00 & 5.24E+00 \\
 & \multirow{3}{*}{R$^2$}                    & Training   & 0.974 & 0.942 & 0.945 & 0.978 & 0.962 \\
 &                                           & Validation & 0.970 & 0.944 & 0.944 & 0.976 & 0.959 \\
 &                                           & Test       & 0.964 & 0.919 & 0.925 & 0.972 & 0.955 \\
\bottomrule
\end{tabular}
\caption*{\tiny $\mathcal{L}(\hat{\mathbf{p}}) = \frac{\|\mathbf{p} - \hat{\mathbf{p}}\|_2}{\|\mathbf{p}\|_2}$, \quad $\mathcal{L}_{\partial r}(\hat{\mathbf{p}}) = \frac{\|\partial_r \mathbf{p} - \partial_r \hat{\mathbf{p}}\|_2}{\|\partial_r \mathbf{p}\|_2}$, \quad $\mathcal{L}_{\partial z}(\hat{\mathbf{p}}) = \frac{\|\partial_z \mathbf{p} - \partial_z \hat{\mathbf{p}}\|_2}{\|\partial_z \mathbf{p}\|_2}$}
\caption{Model performance results for pressure build-up field -- full gradient loss ($\lambda_r = \lambda_z$).}
\label{tab:pressure_results_symmetric}
\end{table}

\end{document}